\pdfoutput=1
\documentclass{aa}  
\usepackage{textgreek}
\usepackage{graphicx}
\usepackage{comment}
\usepackage{mathtools}
\usepackage{amsmath}
\usepackage{fancyref}
\usepackage{longtable}
\usepackage{rotating}
\usepackage{pdflscape}
\usepackage{enumerate}
\usepackage{xtab,booktabs}
\usepackage{textcomp}
\usepackage[citecolor=blue, urlcolor=blue, linkcolor=blue, colorlinks=true]{hyperref}
\usepackage{txfonts}
\usepackage{soul}
\usepackage{natbib}
\newcommand{\subrv}{\textnormal{\tiny \textsc{rv}}}
\usepackage{afterpage}
\usepackage{capt-of}
\usepackage{float}
\usepackage{lipsum}
\usepackage{ulem}
\usepackage[dvipsnames]{xcolor}

\begin{document}

   \title{Search for associations containing young stars (SACY)}

   \subtitle{VIII. An updated census of spectroscopic binary systems showing hints of non-universal multiplicity among these associations}

  \author{S. Z\'u\~niga-Fern\'andez 
          \inst{1, 2, 3} \thanks{sebastian.zuniga@postgrado.uv.cl}
          \and A. Bayo 
          \inst{3, 1}
          \and P. Elliott 
          \inst{4}
          \and C. Zamora 
          \inst{1, 3}
          \and G. Corval\'an 
          \inst{1, 3}
          \and{X. Haubois}
          \inst{2}
          \and{J. M. Corral-Santana}
          \inst{2}
          \and{J. Olofsson}
          \inst{3, 1}
          \and{N. Hu\'elamo}
          \inst{5}
          \and{M. F. Sterzik}
          \inst{7}
          \and{C. A. O. Torres}
          \inst{6}
          \and{G. R. Quast}
          \inst{6}
          \and{C. H. F. Melo}
          \inst{2}
          }
   \institute{
   N\'ucleo Milenio de Formaci\'on Planetaria (NPF), Valpara\'iso, Chile
   \and
    European Southern Observatory, Alonso de C\'ordova 3107, Vitacura, Casilla 19001, Santiago de Chile, Chile
   \and
    Universidad de Valpara\'iso, Instituto de F\'isica y Astronom\'ia (IFA), Avenida Gran Breta\~na 1111, Casilla 5030, Valpara\'iso, Chile
   \and Department of Physics and Astronomy, York University, Toronto, ON M3J 1P3, Canada
   \and Centro de Astrobiolog\'{\i}a (CSIC-INTA), ESAC Campus, Camino del Castillo s/n, E-28692 Villanueva de la Ca\~nada, Madrid, Spain
   \and Laborat\'orio Nacional de Astrof\'isica / MCTIC, Rua Estados Unidos 154, 37504-364 Itajub\'a (MG), Brazil
   \and European Southern Observatory, Karl-Schwarzschild-Str. 2, D-85748, Garching bei Munchen, Germany}

   \date{Received ; accepted }

\abstract{Nearby young associations offer one of the best opportunities to study in detail the properties of young stellar and substellar objects thanks to their proximity ($< 200$\,pc) and age ($\sim 5-150$\,Myr). Previous works have identified spectroscopic ($< 5$\,au) binaries, close ($5-1000$\,au) visual binaries and wide or extremely wide ($1,000-100,000$\,au) binaries in the young associations. In most of the previous analyses, single-lined spectroscopic binaries  (SB1) were identified based on radial velocities variations. However, this apparent variation can also be caused by mechanisms unrelated to multiplicity.}
{We seek to update the spectroscopy binary fraction of the SACY (Search for Associations Containing Young stars) sample taking in consideration all possible biases in our identification of binary candidates, such as activity and rotation.}
{Using high-resolution spectroscopic observations we have produced $\sim$1300 cross-correlation functions (CCFs) to disentangle the previously mentioned sources of contamination. The radial velocity values obtained were cross-matched with the literature and were used to revise and update the spectroscopic binary (SB) fraction in each of the SACY association. In order to better describe the CCF profile, we calculated a set of high-order cross-correlation features to determine the origin of the variations in radial velocities.}
{We identified 68 SB candidates from our sample of 410 objects. Our results hint that the youngest associations have a higher SB fraction. Specifically, we found sensitivity-corrected SB fractions of $22 \substack{+15 \\ -11}  \%$ for \textepsilon~Cha , $31 \substack{+16 \\ -14} \%$ for TW Hya and $32 \substack{+9 \\ -8} \%$ for \textbeta~Pictoris, in contrast with the five oldest associations we have sampled ($\sim 35-125$ Myr) which are $\sim 10\%$ or lower. This result seems independent of the methodology used to asses membership to the associations.}
{The new CCF analysis, radial velocity estimates and SB candidates are particularly relevant for membership revision of targets in young stellar associations. These targets would be ideal candidates for follow-up campaigns using high-resolution techniques in order to confirm binarity, resolve the orbits, and ideally calculate dynamical masses. Additionally, if the results on SB fraction in the youngest associations are confirmed, it could hint of non-universal multiplicity among SACY associations.}


   \maketitle
%

\section{Introduction}
\label{sec:intro}

Since the first nearby young moving group of stars was identified around $30$ years ago \citep[TW Hya association,][]{delaReza1989, Kastner1997} extensive research has been conducted on these stellar associations: from identifying new ones and their members, to characterizing their chemical composition, dynamics, ages and multiplicity fractions \citep[see][among others]{Zuckerman2004,Torres2008,Shkolnik2012,Malo2014,Elliott2016b,Gagne2018A}. These nearby populations, given their age ($\sim$ 5 -- 150\,Myr) and proximity ($<$ 200\,pc), are great laboratories for studying the properties of young stellar and substellar objects.

Recent studies have used youth signatures (such as the presence of $\rm H{\rm \alpha}$ in emission or the detection of the Li \textlambda~6707 \AA~ line) and 6D kinematics (i.e. Galactic position and Galactic velocity in the 6 parameter space, XYZ and UVW) to estimate membership \citep{Scheneider2019,Lee2019}. In this context, multiplicity studies (particularly the search for tight binaries) play an important role since age diagnostics, velocity determinations, and astrometry are often affected by the use of single-star models on blended multiple systems.

More generally speaking, stellar multiplicity is important in a broad range of fields (e.g. supernova rates), but we will focus here on its impact in the star formation processes. Works on multiplicity as a function of environment, and detailed studies of composition and orbital parameters, provide valuable empirical data to improve our understanding of stellar evolution and unresolved stellar populations. These empirical estimates are of particular interest at younger ages and close separations where the theoretical models remain still only loosely constrained \citep{Duchene2007,Connelley2008,Tobin2016}, and the literature is still far from the more exhaustive work done for main sequence (MS) stars with volume limited samples \citep{Tokovinin2014a,Tokovinin2019a,Tokovinin2019b,Sperauskas2019,GaiaESO2020}. 

It is widely accepted that almost half of solar-type stars spend their time in the MS as multiple systems \citep{Tokovinin2014a,Raghavan2010}. There is also increasing evidence that multiplicity is even higher at very young ages \citep{Tobin2016}, possibly indicating the primordial nature of multiplicity in the processes of star formation. Observational studies suggest an overall decrease of the binary fraction from pre-MS ages to field ages \citep{Ghez1997,Kouwenhoven2007,Raghavan2010}. This decrease could be a consequence of disruption process in long period systems due to interactions with other systems \citep{Raghavan2010} or due to the dynamical evolution of wide companions in triple or higher order systems \citep{Sterzik2002,Reipurth2012,Elliott2016b}. In contrast with wide binaries, tight binaries are expected to ``last" longer given their larger binding energy. A number of observational results on tight binaries have indicated that the overall SB fraction remains unchanged after $1$\,Myr  \citep{Nguyen2012,Tokovinin2014b,Elliott2014}. However, recently, \cite{Jaehnig2017} suggested that some SBs (periods $\approx10^2 - 10^4$\,days) in pre-MS clusters ($\approx 1-10$ Myr) can be dynamical disrupted prior to reaching the MS. The evolution and the formation channel of multiple stellar systems can not be easily determined by field stars, where billions of years of dynamical evolution have already occurred. Therefore it is necessary to devote specific studies of the stellar multiplicity from star-forming regions (SFRs) to the young associations ($1-100$ Myr).

The multiplicity studies for the youngest stars ($\leq 100$ Myr) are still dominated by low number statistics. This is particularly critical in the case of SBs (sub-au separation scales) where high-resolution techniques are mandatory  \citep{Melo2003,Nguyen2012,VianaAlmeida2012}, but some of these techniques can be contaminated by phenomena such as activity and rotation, inherent to the young ages involved (see Section~\ref{sec:ccf_calc}). In principle, the preferred mechanism to form some of these close binaries ($\lesssim 100$\,au) is disk fragmentation, where the disk fragments due to gravitational instabilities \citep{Bonell1994,Zhu2012}. However, the formation mechanisms could be affected by environment conditions. In particular, \cite{Bate2019} found an apparent trend for multiple systems to be preferentially tighter when formed at lower metallicity environments. On the other hand, the tightest systems ($\lesssim10$\,au) cannot form directly neither via turbulent nor disk fragmentation, and the emerging consensus is that some processing must dynamically evolve the initial separations to closer  ones \citep{Bate2002}. In particular, \cite{Tokovinin2006} found that $\sim63\%$ of MS SBs were members of high-order multiple systems (see \citealt{Elliott2016b} for a similar result focused on the $\beta$ Pictoris moving group). Interestingly, $\sim 98\%$ of SBs with orbital periods shorter than 3 days have additional companions. This result seems to provide observational support to the dynamical evolution hypothesis commented before. Further SB studies in younger population ($\leq$ 100\,Myr) are, in any case, still needed to provide improved statistics on more pristine populations.

This work is the continuation of a series of studies of multiplicity in young associations over a wide range of orbital parameters \citep[$a\sim 0.1-10^{4}$\,au:][]{Elliott2014, Elliott2015, Elliott2016a, Elliott2016b}. In particular, this work focuses on SB identification within SACY via cross-correlation function (CCF), not only using the radial velocity (RV) variations with time as a sign of multiplicity, but also incorporating high-order features as a complementary tool to establish the origin of the variation. After modelling and applying observational bias corrections, we present the results on SB fraction in each association within the SACY sample and the list of SB candidates, including notes on individual objects.  

\section{Sample}
\label{sec:sample}

The sample presented in this work is drawn from our database of young association members, as in \cite{Elliott2016a}, mainly collected from \cite{Torres2006, Torres2008, Zuckerman2011, Malo2014, Kraus2014, Elliott2014} and \cite{Murphy2015}. The membership of each object to the different associations was assessed using the convergence method described in \cite{Torres2006} and \cite{Torres2008} with the updated distances from the second Gaia data release \citep[Gaia DR2,][]{GaiaDR2}. The full membership study and further analysis will be presented in Torres et al. (in prep.).

In addition, the targets selected for this work have to fulfil at least one of the following selection criteria:
\begin{itemize}
    \item The objects have at least one high-resolution spectrum in our database, from which a CCF can be calculated.
    \vspace{0.2cm}
    \item The target has at least one RV measurement (with uncertainty $\leq$\,3\,km~s$^{-1}$) and one $v~\mathrm{sin}~i$ value in the literature (with uncertainty $\leq$\,5\,km~s$^{-1}$).
\end{itemize}

We will be referred as ``the sample" to which was obtained with the SACY convergence method unless otherwise indicated. Our sample covers an approximate mass range of 0.1 -- 1.5\,M$_\odot$, with the majority of objects having an estimated mass around 1\,M$_\odot$. Masses were estimated from the 2MASS near-infrared magnitudes and parallactic distances using the evolutionary tracks from \cite{Baraffe2015}. Our final sample size is 410 objects, 303 of which have two or more epochs of high-resolution spectra. Further details on the literature's measurements used in our sample are summarised in Sec.~\ref{sec:lit_values}, and all relevant parameters for this work are listed in Table \ref{table:summary_tab}.

\section{Observations and additional data}
\label{sec:obs}
We obtained spectra taken  with the Ultraviolet and Visual Echelle Spectrograph (UVES; \textlambda/\textDelta\textlambda $ $ $\sim 40,000$ with 1\arcsec~slit, \citealt{UVES2000}) at Paranal, Chile.  These observations came from three of our observing campaigns, taken between 2015 and 2016.
We also added data retrieved from the ESO phase 3 public archive \footnote{\url{http://archive.eso.org/wdb/wdb/adp/phase3_main/form}}. Our data were taken with a 1\arcsec~slit width in the wavelength range $3250-6800$\,\AA. The time separation between different observing epochs of a given source ranges from 1 day to $\sim$1\,month.

The data were reduced with the EsoRex\footnote{\url{https://www.eso.org/sci/software/cpl/esorex.html}} pipeline of UVES, using the {\it uves\_obs\_redchain} recipe (bias corrected, dark current corrected, flat-fielded, wavelength-calibrated and extracted). This provides three spectra from the two arms of the instrument (BLUE and REDL/REDU, with wavelength coverage $3250-4500$\,\AA, $4800-5800$\,\AA, and $5800-6800$\,\AA, respectively).  For the calculation of CCF in this work, we combined all three spectra if the average signal-to-noise ratio (S/N) for the BLUE spectrum is $>$ 10. Otherwise, we combined the REDU and REDL spectra only. In total, we present 998 individual CCFs from UVES observations.

\subsection{Archival high-resolution spectra}

In order to maximise the time baseline and available spectral information for each target, we used the publicly available Phase 3 data taken with the Fibre-fed Extended Range \'Echelle Spectrograph (FEROS/2.2\,m, \citealt{FEROS1999}) and the High Accuracy Radial velocity Planet Searcher (HARPS/3.6\,m, \citealt{HARPS2003}). 

FEROS is a high-resolution \'Echelle spectrograph ($\textrm{\textlambda/\textDelta\textlambda} \approx 50,000$) installed at the MPG/ESO 2.2-m telescope located at ESO’s La Silla Observatory, Chile.  The wavelength range of the reduced spectra is $3527-9217$\,\AA.  The one dimensional Phase 3 spectra are given in the barycentric reference frame.

HARPS is also a high-resolution \'Echelle spectrograph ($\textrm{\textlambda/\textDelta\textlambda} \approx 115,000$), mounted on the 3.6\,m telescope, also located at La Silla Observatory in Chile.  The wavelength range is $3781-6912$\,\AA~and the Phase 3 spectra are given in the barycentric reference frame.  

We searched for any available science spectra for targets in common with our database of young moving group members. From all archival spectra we successfully calculated CCFs for 167 observations taken with FEROS and 97 CCFs for observations taken with HARPS. These data are also included in the analysis presented in this work. 

\subsection{Previously published quantities}
\label{sec:lit_values}

Table~\ref{tab:mjd_estimations} lists the references used in this work for both the RV and $v~\mathrm{sin}~i$ values. As mentioned previously, we only include values that have uncertainties $\leq 3$\,km~s$^{-1}$ and $\leq 5$\,km~s$^{-1}$ for RV and $v~\mathrm{sin}~i$, respectively.  The table is split into two sections: the top one shows values that do not have associated Modified Julian Dates (MJD) values for each RV. The bottom section corresponds to surveys that do have individual MJD values for each observation. 

{
\begin{table}
\tiny
\begin{center}
\caption{Previous catalogues of RV and $v~\mathrm{sin}~i$ values used in this work.  The bottom section shows those values with associated MJDs, while the top section show values for which MJDs have been estimated from the respective MJD-range.}
\begin{tabular}{p{3.2cm}p{1.1cm}  p{1.5cm} p{1.5cm}}
\hline\hline\\
  \multicolumn{1}{l}{Ref.} &             
  \multicolumn{1}{l}{Values} &               
  \multicolumn{1}{l}{MJD-range} &             
  \multicolumn{1}{l}{Ref. code}  \\         
  \multicolumn{1}{l}{} &             
\multicolumn{1}{l}{} &             
  \multicolumn{1}{l}{} &         
  \multicolumn{1}{l}{}  \\      
  \hline\\
  \multicolumn{4}{c}{MJD estimated from observation range}  \\     
  \hline\\    
\cite{Schlieder2012} & RV, $v~\mathrm{sin}~i$ & 54718-55685 & SC12 \\
\cite{Shkolnik2012} & RV \tablefootmark{a} & 53725-54455 & SH12 \\
\cite{Torres2006}& RV, $v~\mathrm{sin}~i$  & 51179-53826 & TO06 \\
\cite{Lopez2006}& RV \tablefootmark{b} & 51910-52796  & LO06 \\
\cite{Rodriguez2013} & RV & 56171-56230 & RO13 \\
\cite{Maldonado2010}& RV & 53552-54771 & MA10 \\
\cite{Moor2013}& RV & 55013-55669 & MO13 \\
\cite{Reiners2009}& RV & 54475-54835 & RE09 \\
\cite{Gontcharov2006}& RV & 47892-52275 & GO06 \\
  \hline\\
  \multicolumn{4}{c}{Exact MJD values available for each observation}  \\     
  \hline\\    
\cite{Malo2014}& RV, $v~\mathrm{sin}~i$  & 54996-56532 & MA14 \\
\cite{Kraus2014}& RV \tablefootmark{c}  & 56124-56327 & KR14 \\
\cite{Montes2001b}& RV & 51384-51566 & MO01b \\
\cite{Mochnacki2002}& RV & 51082-52003 & MO02\\
\cite{Bailey2012} & RV, $v~\mathrm{sin}~i$ & 53327-54963 & BA12 \\
\cite{Desidera2015}& RV, $v~\mathrm{sin}~i$  &53102-55399 & DE15 \\
\hline\\[-0.4ex]
\end{tabular}
\tablefoot{\tablefoottext{a}{Extended from \cite{Shkolnik2010}}, \tablefoottext{b}{Stars added to the initial sample of \cite{Zuckerman2004}}, \tablefoottext{c}{$v~\mathrm{sin}~i$ values not used from \cite{Kraus2014} as these values are the standard deviation of the broadening function, not calibrated $v~\mathrm{sin}~i$ values. }}
\label{tab:mjd_estimations}
\end{center}
\end{table}}

\subsection{Gaia Data Release 2}
\label{sec:GaiaDR2}

The second Gaia data release\footnote{\url{https://www.cosmos.esa.int/web/gaia/dr2}} (hereafter: Gaia DR2) was issued on 25 April 2018, providing accurate proper motions and parallaxes (among other astrophysical parameters) for more than a billion sources. In particular, this Gaia data release also includes for the first time RV values \citep{Katz2019} for objects with a mean G magnitude between $\sim$4 and $\sim$13 and effective temperatures ($T_{\rm eff}$) between 3550 and 6900\,K.

The overall precision of the RV at the bright-end is in the order of $200-300$\,m~s$^{-1}$ while at the faint-end it deteriorates to $\sim1.2$\,km~s$^{-1}$ for a $T_{\rm eff}$ of 4750\,K and $\sim2.5$\,km~s$^{-1}$ for a $T_{\rm eff}$ of 6500\,K.

Stars identified as double-lined spectroscopic binaries are not reported in Gaia DR2, while variable single-lined, variable star, and non detected double-lined spectroscopic binaries have been treated as single stars in the same release \citep{Sartoretti2018}.

\begin{figure}[h]
\begin{center}
\includegraphics[width=0.45\textwidth]{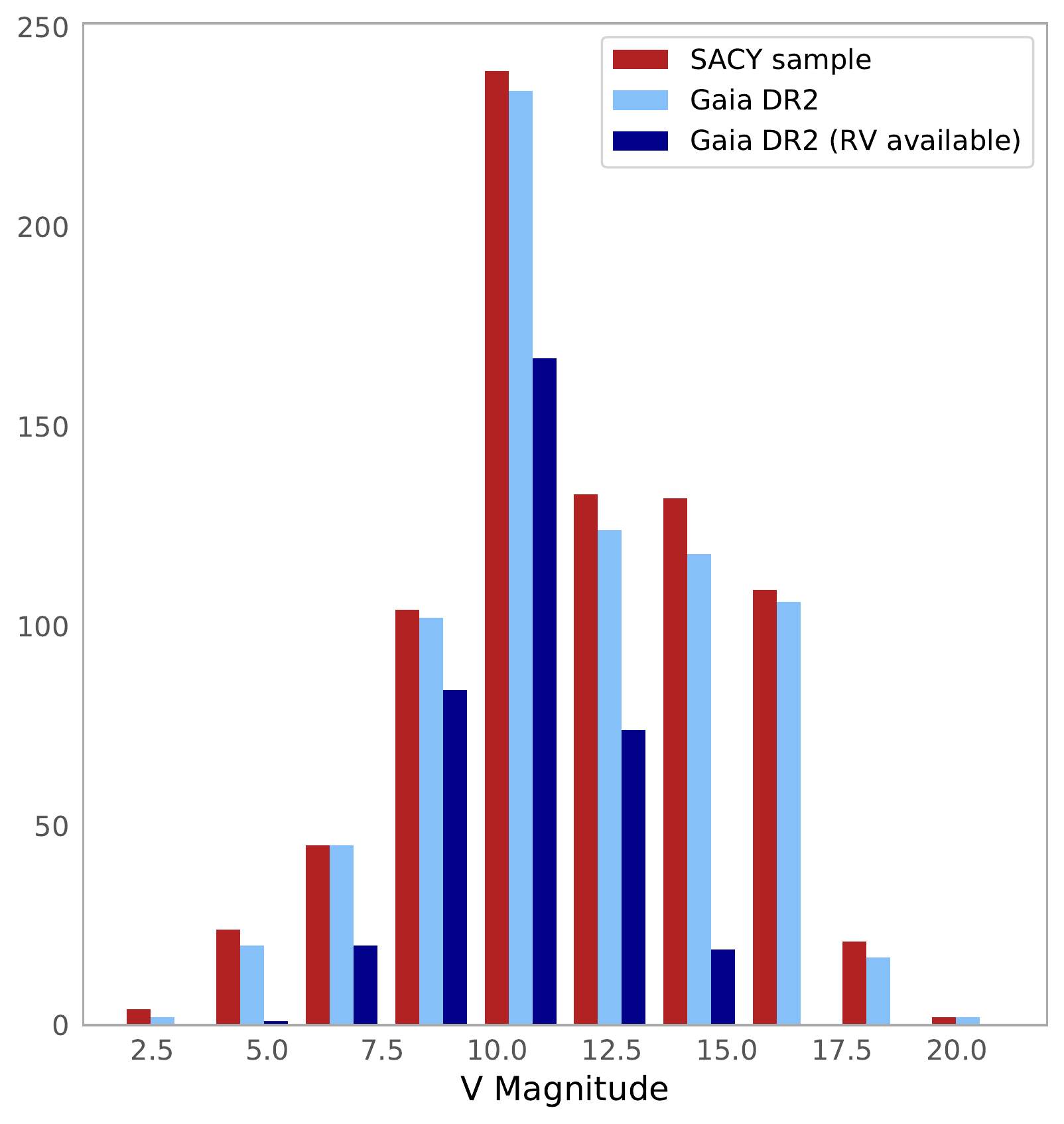}
\vspace{-0.2cm}
\caption{V-magnitude distribution of all members of the SACY sample along with those with counterpart in the Gaia DR2. We reach a completeness of $96.2\%$ where $44.7 \%$ of the objects count with a Gaia RV estimate.}
\label{fig:gaia_sacy}
\end{center}
\end{figure}

We retrieved Gaia DR2 data for all the objects in the SACY sample using the \texttt{astroquery Vizier} package
\footnote{\url{https://astroquery.readthedocs.io/en/latest/vizier/vizier.html}}. We updated our local database to use identifiers resovable by the Sesame service and the Gaia DR2 queries were based on those identifiers. Objects not resolved by identifiers were instead searched by coordinates. In both cases we ran an initial query with a $10$\arcsec~radius and used the proper motions of the closest Gaia source, within the radius, to derive its J2000 coordinates (that are those originally included in our local database). Those J2000 coordinates were then matched to the coordinates in our local database with a 1\arcsec~radius. Objects outside of this 1\arcsec~radius were individually inspected (see Fig. \ref{fig:GDR2appendix} in the Appendix) by cross validating using Simbad, Vizier and the TESS input catalogue \citep[TIC-8, ][]{Stassun2019}. We recovered Gaia DR2 counterparts for $805$ out of $837$ targets in our local database, corresponding to a completeness of $96.2 \%$ (see Fig. \ref{fig:gaia_sacy}). From these $805$ objects, $374$ have RV measurements from Gaia, which were used in this work as an additional epoch of data.

Our database comprises $2379$ RV measurements and $1515$ $v~\mathrm{sin}~i$ values, $1151$ of which come from our CCF calculation of high-resolution spectra. All these values together with other additional properties can be found in Table \ref{tab:RV_indiv} and \ref{tab:vsini_indiv}.

\subsection{Assessing membership using BANYAN $\Sigma$}
\label{sec:BANYAN}
In order to asses any possible bias throughout this work with the use of the convergence method to build the census of the different associations, we have followed an independent path, utilising the BANYAN $\Sigma$ tool \footnote{\url{https://github.com/jgagneastro/banyan_sigma}} for young association membership.

Accurate RV, distances and proper motion values are key ingredients in the accuracy of our convergence method \citep{Torres2006,Torres2008}. Similarly, the recovery rate of BANYAN $\Sigma$ is 68\% when proper motion and RV are used and 90\% when parallaxes are included \citep{Gagne2018A}. Therefore, as we did for the convergence method, we fed the RV measurements collected in this work plus the Gaia DR2 proper motion and parallaxes to the BANYAN $\Sigma$ tool for membership assessment. 

\begin{figure}[h]
\begin{center}
\includegraphics[width=0.46\textwidth]{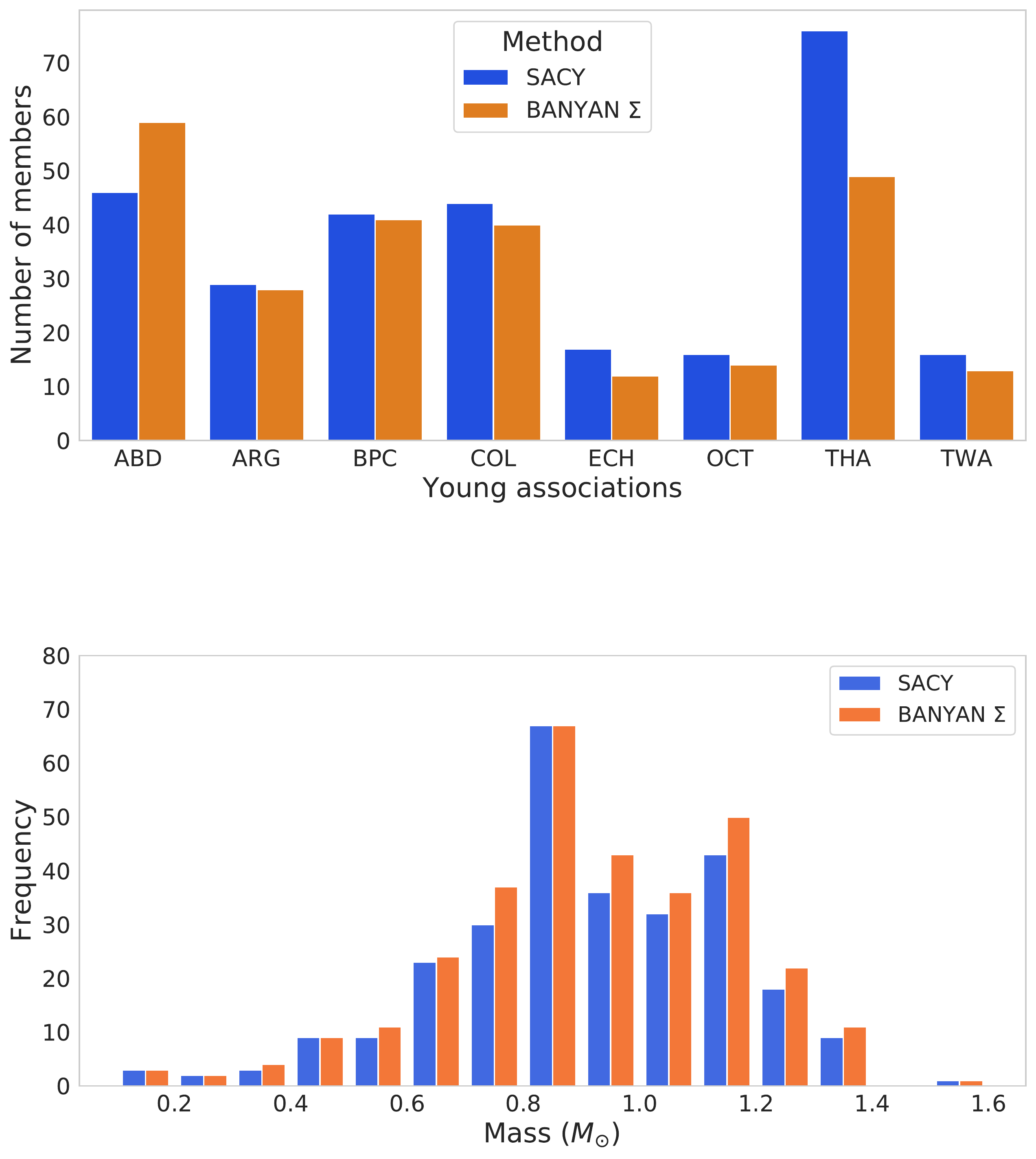}
\vspace{-0.2cm}
\caption{\textit{Top:} Number of targets belonging to each young associations identified by our convergence method (SACY) and BANYAN $\Sigma$. \textit{Bottom:} Mass function of the census built with the convergence method and BANYAN $\Sigma$ for membership assessment.}
\label{fig:sacy_ban_members}
\end{center}
\end{figure}

It is out of the scope of this work to develop or establish a metric to compare in details the outcome of the two methodologies. However, the two resulting censuses, allow us to test the robustness of our results against moderate changes in membership (see Sec. \ref{sec:results} for further details). The membership results for the SACY convergence method and BANYAN $\Sigma$ are available in Table \ref{table:summary_tab} and summarised in Fig.~\ref{fig:sacy_ban_members}. The mass distributions of the samples analysed throughout this work (using either our convergence method or BANYAN $\Sigma$ tool) are shown in the bottom panel of Fig.~\ref{fig:sacy_ban_members}. As it can be seen, the only associations with noticeable differences regarding total number of members are ABD and THA.

\subsection{Rotational periods from light curves}
\label{sec:LC}
In order to estimate the rotational periods of the objects in the sample, we queried two of the main missions delivering  precise light curves: K2 and TESS \citep{K22014,TESS2015}. We proceeded in the following manner:
\begin{enumerate}
    \item We queried the archives of both missions via the MAST API (via the \texttt{astroquery} package within \texttt{astropy}) with the J2000 coordinates of each object and a search radius of 0.002\,$\deg$ ($\sim$7$\arcsec$). We obtained light curves for 272 out of 410 objects ($\sim65$\% of the sample). In particular, 266 were taken with TESS (across different sectors) and six with K2.
    \item In all cases we chose the Pre-search Data Conditioned Simple Aperture Photometry (PCDSAP) fluxes and characterised the variability of the sources via their Lomb-Scargle (LS) periodograms \citep[calculated with \texttt{astropy.timeseries.lombscargle},][]{VanderPlas2015}.
    \item Even though the false alarm probabilities (FAPs) of the peaks identified in the LS periodogram were extremely low (typically well below  $10^{-4}$), we performed a simple quality check for the identified periods in the following way: we folded each light curve to the period with the highest intensity in the LS periodogram and model the modulation by calculating the median, binning the phased curve in $100$ bins. Such trend was subtracted from the phased light curve and the median absolute deviation (MAD) of those residuals was compared to the MAD of the original phased light curve. 
   \item In the case of TESS data, additional checks need to be done to account for the large pixel size of its detector. In order to estimate the contamination that could affect each of the light curves, we modified the existing python package \texttt{tpfplotter} \citep{Aller2020} that, in short, provides the number of Gaia sources within a $\Delta$G mag (Gaia G mag, this $\Delta$ is defined by the user) of the science target that fall in the pipeline aperture of TESS. We modified the code in order to take into account both, the proper motion of our targets and the cross-match with Gaia DR2 explained in Sec.~\ref{sec:GaiaDR2}. 
   We chose a $\Delta$Gmag of $5$ magnitudes and in Table~\ref{table:summary_tab} we include notes on the minimum $\Delta$Gmag found within the aperture. We note that a number ($27$ to be precise) of our Gaia cross-match identifications are not recovered in Simbad. Even though we stand by those identifications, we have identified them in the column \textit{LC notes} of Table~\ref{table:summary_tab}. 
  \item We classified a period as ``good quality" if the MAD of the residuals is at least three times smaller than the MAD of the original phased light curve and if there are no Gaia sources that fall in the aperture with $\Delta$G mag $< 5$. Periods which fulfil the criteria based on the MAD of the residuals but have contaminants in the aperture with  $2.5 \leq \Delta$G mag $\leq 5$ should be considered with caution. For periods that present contaminants in the TESS aperture and do not fulfil the criteria based on the MAD are not considered as reliable for the rest of the analysis and  are flagged as ``bad quality''. For an example of clearly contaminated light curve (rotational periods not to be trusted) see appendix \ref{sec:LC_qflag}.
\end{enumerate}
 Our estimated periods as a function of median $v~\mathrm{sin}~i$ from our work are presented in Fig. \ref{fig:LC_vs_vsini} (see the details regarding $v~\mathrm{sin}~i$ estimation in appendix.~\ref{sec:vsini_calib}). This relation was used throughout our analysis as a complementary source to evaluate the nature of SB candidates.
 
\begin{figure}[h]
\begin{center}
\includegraphics[width=0.48\textwidth]{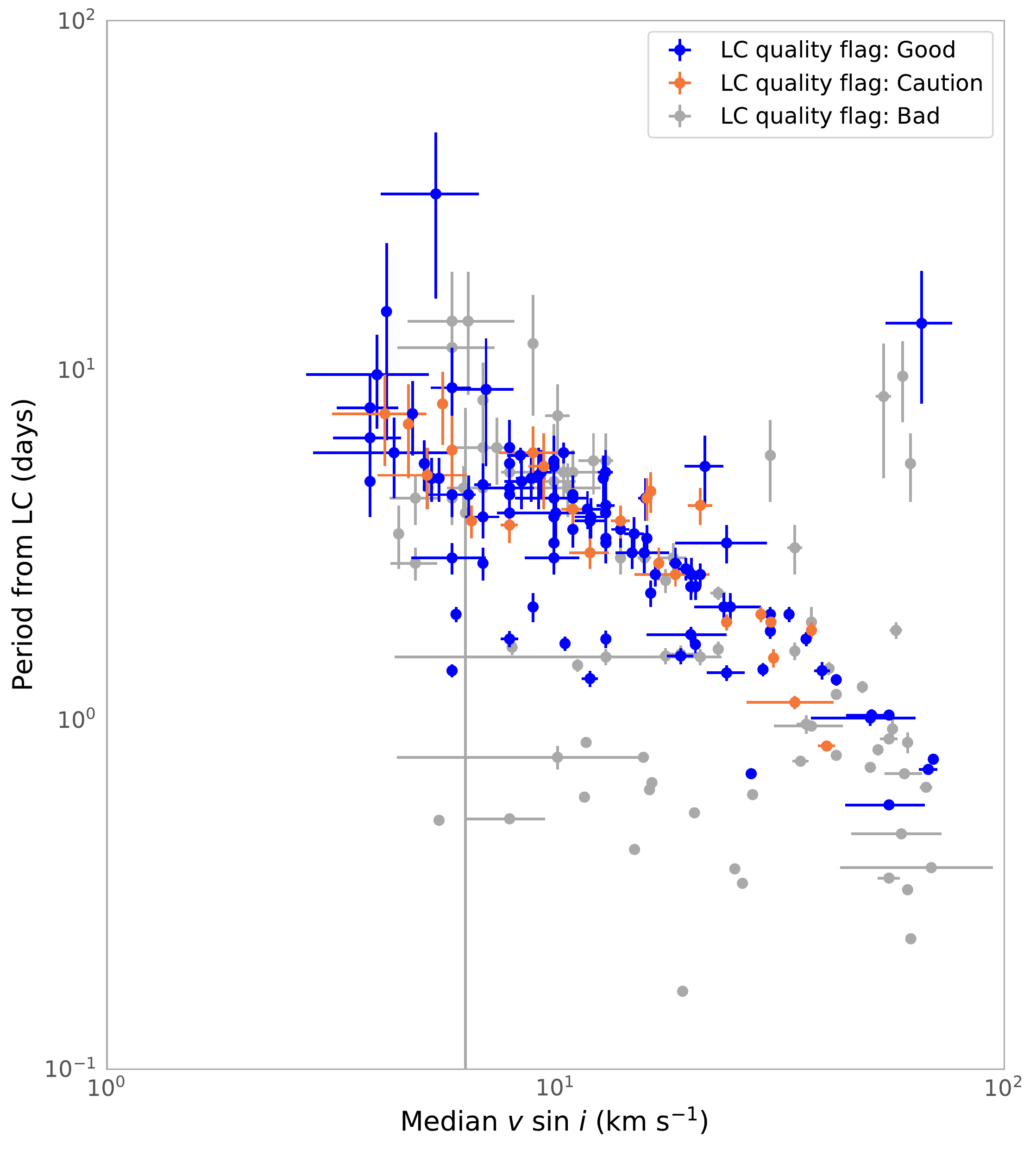}
\vspace{-0.2cm}
\caption{Rotational periods estimated from the light curves versus median $v~\mathrm{sin}~i$ from our work. The quality flag of the period defined in Sec. \ref{sec:LC}, is color-coded as grey, orange and blue for bad, caution and good, respectively.}
\label{fig:LC_vs_vsini}
\end{center}
\end{figure}

\begin{figure*}[h]
\begin{center}
\includegraphics[width=0.9\textwidth]{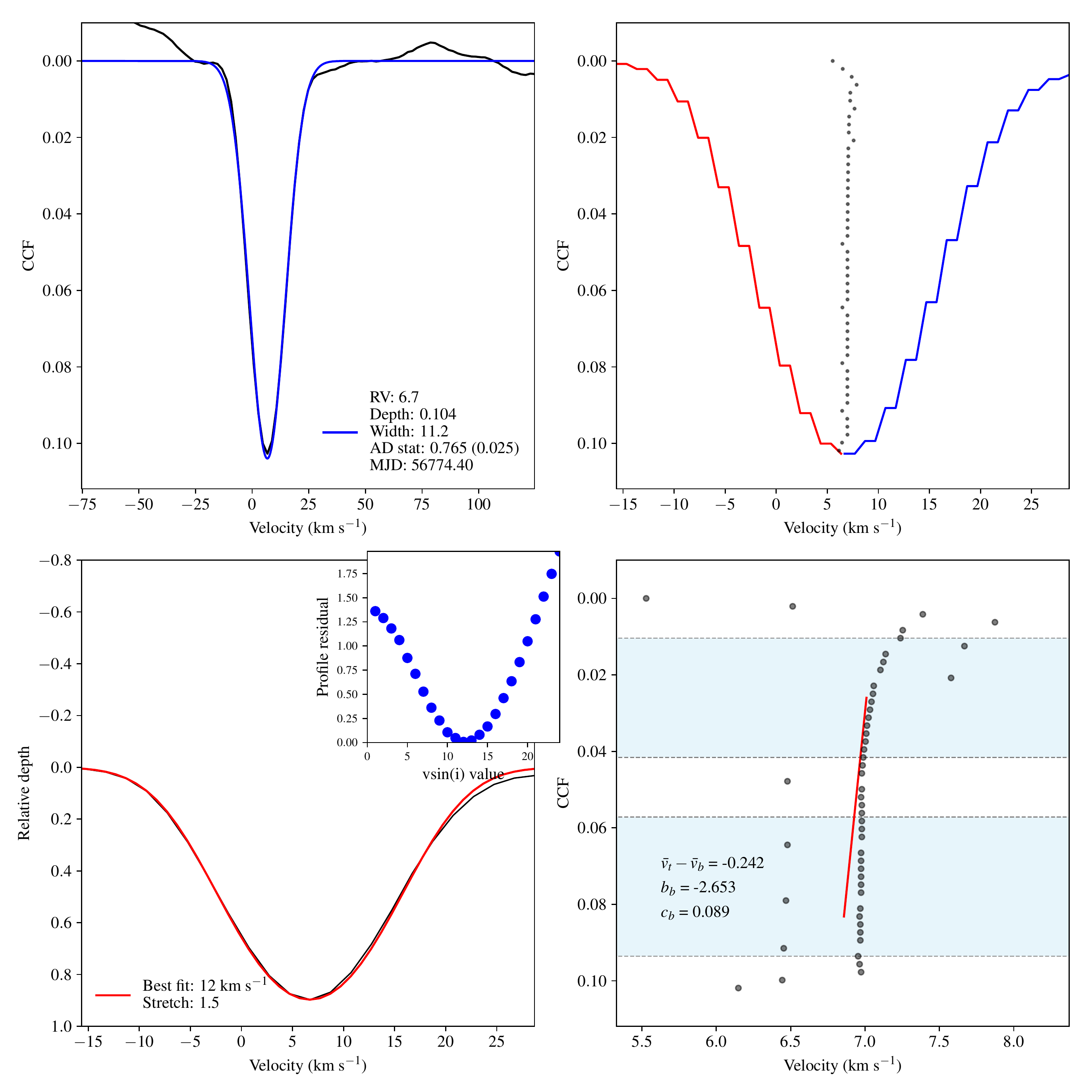}
\vspace{-0.2cm}
\caption{An example of the graphical output from our CCF calculation code for one target. {\it Top left}: The CCF profile. The quantities shown in the lower left are the peak of the fitted Gaussian profile (RV), the depth of the CCF, the width ($\sigma$) of the Gaussian profile, the Anderson-Darling statistic for normality between $-\sigma$ and +$\sigma$ with its respective significance level and the MJD of the observation. {\it Top right}: The 2\,$\sigma$ region of the CCF profile and the bisector (grey dots).  {\it Bottom left}: The normalised CCF fitted with the best-fit rotational profile (from profiles in the $v~\mathrm{sin}~i$ range 1--200\,km~s$^{-1}$).  The residuals of fits are shown in the inset. {\it Bottom right}: The bisector slope along with three metrics of its shape ($b_b$, $c_b$ and $BIS$). See text in Section~\ref{sec:ccf_properties} and ~\ref{sec:ccf_calc} for further details. }
\label{fig:ccf_output}
\end{center}
\end{figure*}

\section{Properties and calculation of CCF profiles}
\label{sec:ccf_properties}
There are two main ways of calculating CCFs from high-resolution spectra, using either observations of RV standard stars or using a numerical mask, acting as a standard star. In this analysis we used a CORAVEL-type numerical mask which was convoluted with the observed spectrum for each observation \citep[for further details see][]{Queloz1995}. For the sake of homogeneity and given the relatively narrow range of spectral types in our sample (see Table
~\ref{table:summary_tab}), we use a single K0 mask in our analysis. 

Only in the cases where the K0 mask completely failed in the CCF calculation (assessed by the goodness of fit of the Gaussian profile to the CCF), we used other available masks (F0 or M4, depending on the spectral type of the star). However, for consistency, the CCF profiles and respective properties of such objects are not included in the statistical analysis of our measurements.

The CCFs analysis and the SB update presented in this work follows up what was presented by \cite{Elliott2014}. However, here we do not only enlarge our database of observations, but we have also chosen to use a much more detailed approach in calculating the CCFs for each observation; by introducing high-order features of the CCFs, we can distinguish between apparent RV variation caused by poor fitting of the CCF and variation produced by bound companions and/or stellar activity.

\subsection{Sources of uncertainty}

The uncertainty in RV calculation using a numerical mask ($\sigma_\mathrm{meas.}$) can be derived with the following equation \citep{Baranne1996}:

\begin{equation}
\label{eq:rv_uncertainty}
\sigma_\mathrm{meas.} = \frac{C(T_\mathrm{eff})}{D \times S/N}\frac{1+0.2\omega}{3}~\mathrm{km~s^{-1}}
\end{equation}

\noindent where $C(T_\mathrm{eff}$) is a constant that depends on both the spectral type of the star and the mask used, which is typically 0.04; $\omega$ is the (noiseless) full width at half maximum (in km~s$^{-1}$) of the CCF; $D$ is the (noiseless) relative depth, and S/N is the mean signal-to-noise ratio.

This uncertainty is relevant to one measurement of RV from a single observation and a single instrument. Given the high S/N of our data, typically $\sim50 - 100$, the calculated uncertainty is almost negligible. A more empirical approach can be taken by studying RV from different epochs and gauging the level of intrinsic variation of the star.  As these stars are often variable, the CCF profiles are not always completely symmetric \citep{Lagrange2013} and, therefore the uncertainty calculated using Equation~\ref{eq:rv_uncertainty} is underestimated.  Thus, following the analysis presented in \cite{Elliott2014}, we use an empirical approach to estimate RV uncertainties (see Sec. \ref{sec:rv_std_versus_vsini} for further details).

\subsection{Cross-correlation features}
In order to better describe the CCF profile we calculate a set of high-order cross-correlation features: \\
\begin{itemize}
\item {\it Bisector}:  The bisector is calculated from the midpoint of the line for each element of intensity that defines the CCF profile,  shown by the grey dots in upper right panel in Fig.~\ref{fig:ccf_output}.\\

\item {\it Bisector inverse slope}: Here we adopt the bisector inverse slope (BIS) as defined by \cite{Queloz2001}:
\begin{equation}
BIS=\bar{v}_t - \bar{v}_b
\end{equation}

where $\bar{v}_t$ is the mean bisector velocity in the region between 10\% and 40\% of the line depth and $\bar{v}_b$ is the mean bisector velocity between 55\% and 90\% of the line depth. These two regions are highlighted in the bottom right panel of Fig.~\ref{fig:ccf_output}.\\

\item {\it Bisector slope ($b_b$)}: This is defined as the inverse slope from a linear fit (shown by the red line in the bottom right panel of Fig.~\ref{fig:ccf_output}) for the region between 25\%-80\% of the CCF's depth \citep{Dall2006}.  \\

\item {\it Curvature ($c_b$)}: The curvature of the CCF's profile is defined as:

\begin{equation}
c_b=(v_3 - v_2) - (v_2 - v_1)
\end{equation}

where $v_1$, $v_2$, and  $v_3$ are the mean bisector velocity on the 20-30\%, 40-55\%, and 75-100\% of the CCF's depth. This definition is from \cite{Dall2006} which is a slightly modified version of the curvature presented in \cite{Povich2001}. \\

\begin{figure*}[h]
\begin{center}
\hspace{-0.44cm}
\includegraphics[width=0.85\textwidth]{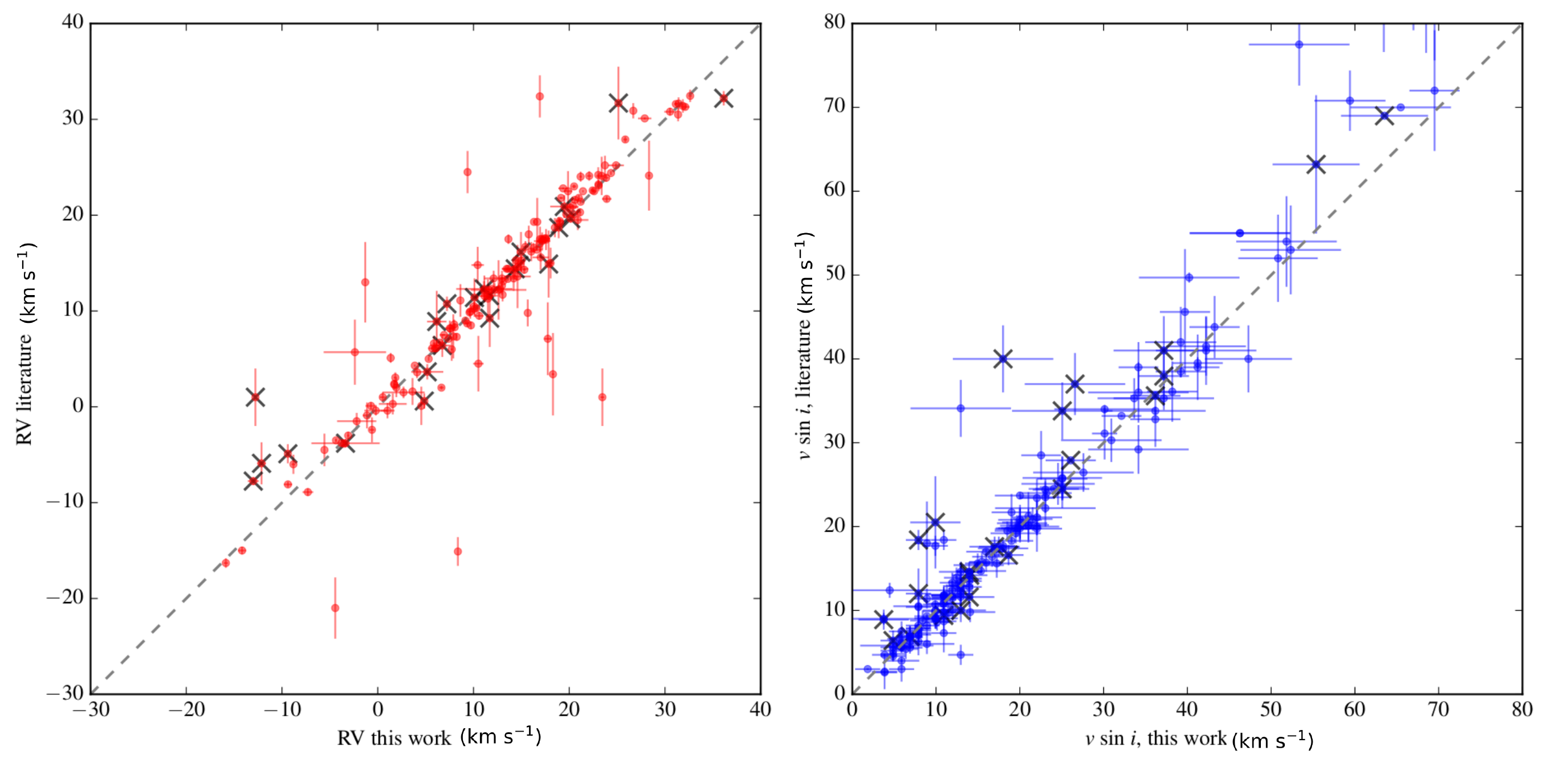}
\vspace{-0.3cm}
\caption{{\it Left panel}: RV values calculated in this work versus values from the literature.  Crosses represent previously identified spectroscopic multiple systems.  The 1:1 relation is shown by the dashed line.  {\it Right panel}: Same as upper panel, but for $v~\mathrm{sin}~i$ values. }
\label{fig:literature_comp}
\end{center}
\end{figure*}

\item {\it Anderson-Darling statistic (AD)}: We use the AD statistic around the peak of the CCF profile as a test for normality, i.e. how Gaussian-like the profile is.  We perform this test around the 1\,$\sigma$ region around the peak of the CCF profile.  The AD statistic and its significance are shown in the upper left panel of Figure~\ref{fig:ccf_output}, i.e. the null hypothesis, that the function is not Gaussian, cannot be rejected at a significant level. \\ 

\item {\it Profile residual}: The CCF profile is fitted by a set of rotational profiles \citep{Gray1976} to determine its $v~\mathrm{sin}~i$ value.  In order to quantify the validity of this fit we calculated the overall residual for each $v~\mathrm{sin}~i$ profile (from 1 - 200\,km~s$^{-1}$).  The minimum of this set of residuals is used to determine the best-fit profile for each observation, but also the absolute value is retained. That way we can compare the absolute residuals as a function of other properties in our sample.

\end{itemize}

\section{Estimates of radial and rotational velocities}
\label{sec:ccf_calc}

To calculate all the properties defined in the previous section from the available high-resolution optical spectra, we wrote a series of functions\footnote{Code is available at \url{https://github.com/szunigaf}}. Those functions compute the CCFs, and return these properties as a ``digest" of the information contained in the CCFs.

Figure~\ref{fig:ccf_output} shows the summary graphical output from the master function described before. 
The CCF is shown in the top left panel of Fig. \ref{fig:ccf_output}, i.e. the resulting profile of the star's spectrum with the numerical mask (in black) and the Gaussian profile fitted to the data (in blue). The grey dots in the right panel of Fig. \ref{fig:ccf_output} represent the bisector of the profile whereas the red and blue parts show the two separate sides of the $2 \sigma$ region of the star's CCF profile. Another relevant output from our functions is the star's normalised CCF profile with the best-fit rotational profile (bottom left in Fig.~\ref{fig:ccf_output} from a series of profiles with $v~\mathrm{sin}~i$ from 1 - 200\,km~s$^{-1}$).  The legend shows the best fitting profile value and the {\it stretch factor} which is a measure of how much the best-fit $v~\mathrm{sin}~i$ profile was stretched to achieve the fit.  The inset in the upper right shows an area around the minimum of the residuals from different $v~\mathrm{sin}~i$ profile fitting, highlighting in this case that 7\,km~s$^{-1}$ is clearly the best fit.  Note that these $v~\mathrm{sin}~i$ values are ``raw", see Appendix~\ref{sec:vsini_calib} for details on calibration. The three metrics of the bisector are also given by our functions (see bottom right panel in Fig.~\ref{fig:ccf_output}). Namely the BIS ($\bar{v}_t - \bar{v}_b$), the slope ($b_b$) and the curvature ($c_b$) which help to quantitatively characterise the properties of the bisector.

We visually inspected each of the CCF outputs and removed any observations where the CCF calculation had clearly failed (or a different mask had to be used), mostly due to low S/N. This left $1375$ CCFs for further analysis. 

Several broadening mechanisms can contribute to the width of the CCF, these can either be inherent to the star (surface gravity, effective temperature, rotation, turbulence) or arise from the instrument used to obtain the observations. Therefore to accurately measure rotational velocities we have to account for non-rotational broadening mechanisms, both physical and instrumental. The details for our calibration approach can be found in Appendix~\ref{sec:vsini_uncertainties}.  

With our calibrated $v~\mathrm{sin}~i$ values and barycentric RVs, we were able to look at the overall properties of our targets by combining individual observations.  We were also able to identify any clear double-lined spectroscopic binaries from their double-peaked CCF profiles, see Appendix \ref{sec:individ_sources}.

\subsection{Cross-match with literature}
For each object, we compared the median RVs and $v~\mathrm{sin}~i$ from our database with previously published values (see Table~\ref{tab:mjd_estimations} for references) to ensure there was no significant offset. Figure~\ref{fig:literature_comp} shows the results of this comparison.  The error bars for each quantity represent the standard deviation from multiple observations. 

\begin{figure*}[h]
\begin{center}
\includegraphics[width=0.85\textwidth]{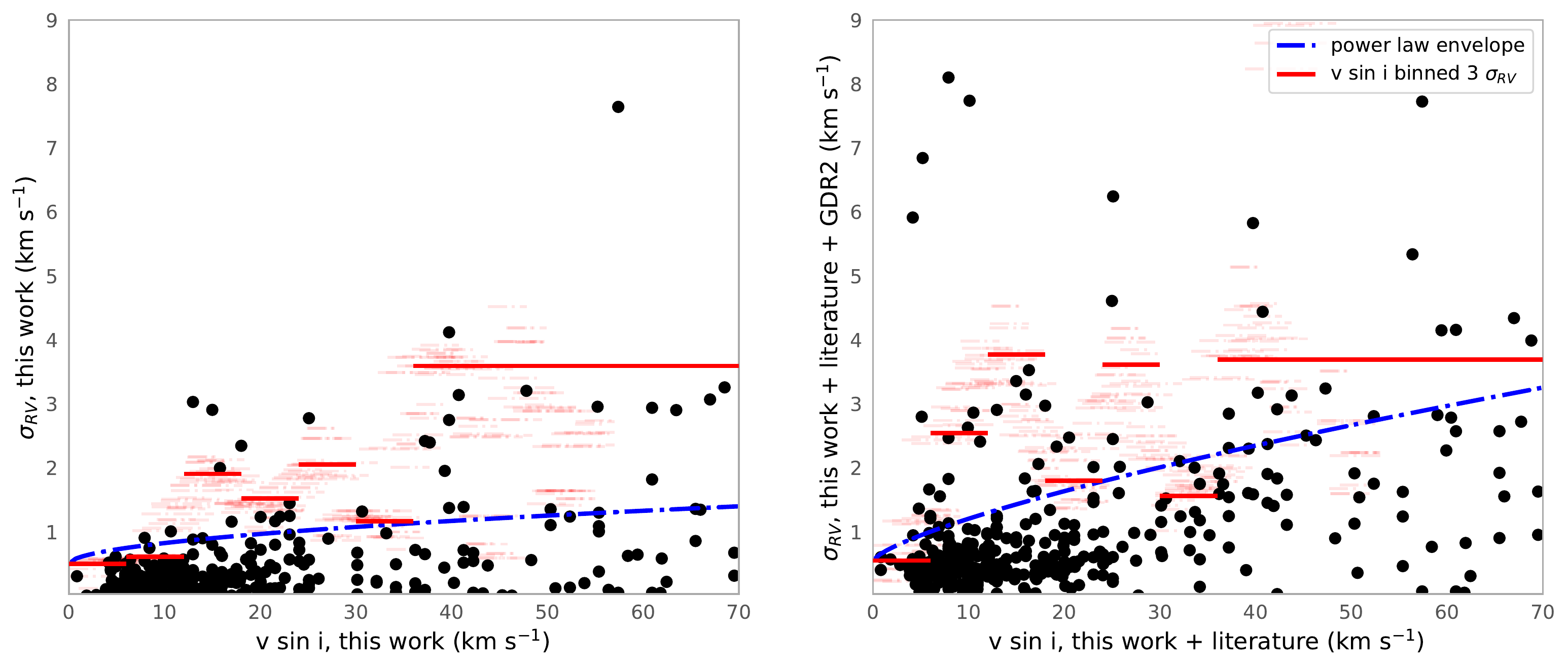}
\vspace{-0.2cm}
\caption{{\it Left panel}: The standard deviation in RV as a function of  $v~\mathrm{sin}~i$ for measurements calculated in this work. The 3\,$\sigma$ value from binning in $6$\,km~s$^{-1}$ bins are represented by the red solid lines. The power-law envelope is represented by dash-dotted blue line. {\it Right panel}: Same as left panel but including values from literature and Gaia DR2 for the standard deviation estimates.}
\label{fig:rv_std_vs_vsini}
\end{center}
\end{figure*}

Black crosses represent objects previously identified as multiple systems, i.e. those not likely to follow the 1:1 relation. We also note that for $v~\mathrm{sin}~i \gtrsim 50$\,km~s$^{-1}$, the broader CCF profile translates into a larger uncertainty on the estimate of this quantity (see Appendix \ref{sec:vsini_calib}). With all of this into account, the 1:1 relation describes adequately the comparison of both sets of values for objects considered as a single stars, demonstrating that our new functions calculating CCF properties are working correctly. 

\section{Using multiple measurements to identify single-lined spectroscopic binaries}

Most previous studies identifying  SB1 solely rely on the analysis of multi-epoch RV values. However, in this work we use the high-order CCF features, if possible, when investigating any potential RV variation to better conclude on the true nature of the object. We made an initial list of systems to be further investigated by looking at both RV and $v~\mathrm{sin}~i$ variation as a function of $v~\mathrm{sin}~i$.

\subsection{Distinguishing RV variation as a function of rotation}
\label{sec:rv_std_versus_vsini}

Typically, the variation in RV ($\sigma_\subrv$) is used to flag potential SB1.  However, this apparent variation can also be caused by mechanisms unrelated to multiplicity. \cite{Elliott2014} used a single value (global $\sigma_{\subrv}=$ 2.7\,km~s$^{-1}$) to flag potential SB1, irrespective of their $v~\mathrm{sin}~i$ values.  However, in this work we show that $\sigma_\mathrm{\subrv}$ is a function of $v~\mathrm{sin}~i$, i.e. the apparent radial velocity variation is intrinsically related to the target's $v~\mathrm{sin}~i$. This was also demonstrated in \cite{Bailey2012} using near-infrared radial velocities. The relationship can be explained by the peak of the CCF being less well-defined the broader the profile is. We can exploit this relationship to revisit the spectroscopic multiplicity of stars in our sample.

Fig.~\ref{fig:rv_std_vs_vsini} shows $\sigma_\subrv$ versus $v~\mathrm{sin}~i$ for stars in our sample that are not double- or triple-lined spectroscopic binaries, and that have observations for at least two different epochs.   
The left panel shows the estimates from this work while the right panel presents our values together with those compiled from literature and Gaia DR2. For the sake of homogeneity, to be considered, the literature data also has to fulfil the criteria of having an uncertainty on RV and $v~\mathrm{sin}~i$ lower than $3$ and $5$\,km~s$^{-1}$, respectively (Sec. \ref{sec:sample}).

Considering only our measurements, we note that the dispersion in RV is relatively low for slow rotators. For example, 3\,$\sigma$ variation of $0.7$ and $1.1$\,km~s$^{-1}$ for $v~\mathrm{sin}~i$ of $\approx 5$ and $10$\,km~s$^{-1}$, respectively (shown by the solid red line in Fig.~\ref{fig:rv_std_vs_vsini}). Only at $v~\mathrm{sin}~i \approx$40\,km~s$^{-1}$ more than 3\,km~s$^{-1}$ RV variations are observed.  When measurements from the literature are considered, on average, RV variations increase which is expected from combining observations from different instruments, heterogeneity in the procedure to perform the estimates, and a longer time-span between observations.

As mentioned before, a relationship between $v~\mathrm{sin}~i$ and $\sigma_\mathrm{\subrv}$ is expected. In order to obtain a general and empirical description this relation, we calculated the 3$\sigma$ interval for $\sigma_\subrv$ using an array of binned $v~\mathrm{sin}~i$ values.  We ran a Monte Carlo simulation using the 3$\sigma$ statistics for different bin size and phase (the starting point of the binning). The bin size range was between 3  and 7 km~s$^{-1}$. This range was estimated from the three most common-used bin size estimation method: \cite{Freedman1981}, \cite{Scott1979} and \cite{Sturges1926}. The selected initial phase range covers from $0$ to $4$\,km~s$^{-1}$. This exercise allowed us to address the dispersion in the results that can be explained solely in therms of the choice of phase and bin size. Each realisation is represented by a light red line in Fig. \ref{fig:rv_std_vs_vsini}. It is out of the scope of this work to characterise in details the underlining physical structure between the $\sigma_\subrv$ values as a function of $v~\mathrm{sin}~i$. The only purpose of the simple analysis presented here is to have a first order estimate of the effect of the rotation velocity in the RV determination and, consequently, in its variation. The final adopted thresholds to be used as ``caution" flags when assessing multiplicity are those resulting from a $6$\,km~s$^{-1}$ step between 0 and $42$\,km~s$^{-1}$ (solid red line, Fig.~\ref{fig:rv_std_vs_vsini}). This bin size was selected by taking in consideration the better compromise between sampling and the minimum number of points in each bin. Beyond $43$\,km~s$^{-1}$ on $v~\mathrm{sin}~i$, the number of points in each bin is $\lesssim$10, and therefore the statistics become less reliable. However, we can assume that a very rough positive correlation is maintained or at the very least that it does not invert, i.e. the higher the $v~\mathrm{sin}~i$, the larger the RV variation is. 

As an alternative method to identify SB candidates, we fit a power-law of the form $\sigma_\subrv = m~{(v~\mathrm{sin}~i)}^b$ and then we scale it up to keep a conservative envelope that leave about $85\%$ of the points below it. The fit is obtained using a Huber loss function (\citealp{Huber1964}), which is more robust to outliers than squared loss function \citep{Ivezic2014}, and is shown as a dashed-dotted blue line in Fig.\,\ref{fig:rv_std_vs_vsini}. We identified SB candidates using both selection criteria and further investigated the nature of any targets with RV variation lying above either of those thresholds. We investigated the true SB nature of any targets with RV variations above those thresholds (see Table \ref{tab:flagged_variable_sources} and Appendix \ref{sec:individ_sources}).

\subsection{Distinguishing fast rotators from blended binaries}

\begin{figure}[h]
\begin{center}
\includegraphics[width=0.46\textwidth]{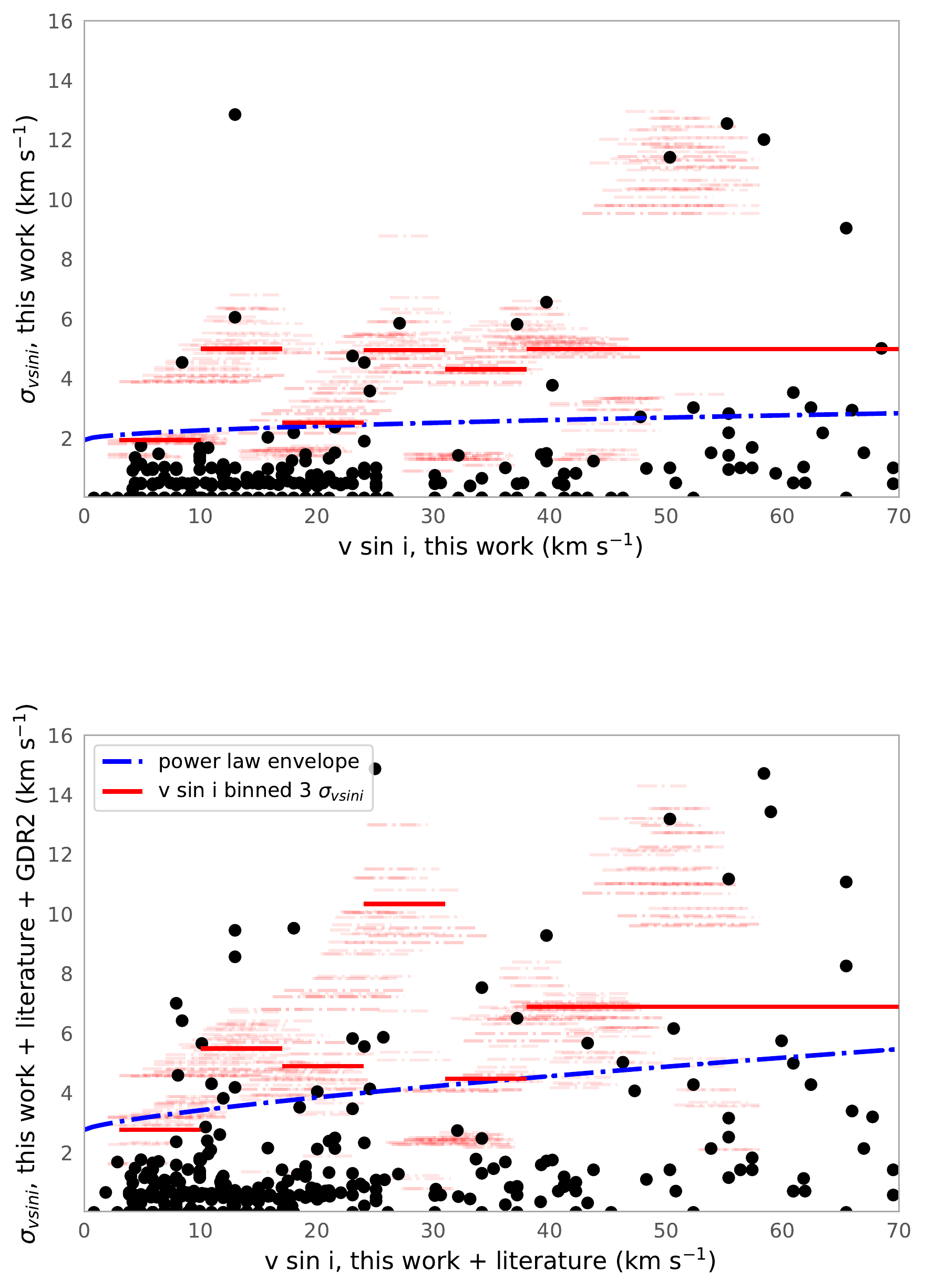}
\vspace{-0.2cm}
\caption{{\it Top panel}: $v~\mathrm{sin}~i$ versus the standard deviation in $v~\mathrm{sin}~i$ for measurements calculated in this work. The  3\,$\sigma$ values, from $3$ to $45$\,km~s$^{-1}$  binned in $7$\,km~s$^{-1}$ bins,  are shown by red solid lines. The power-law envelope is represented by dash-dotted blue line. }{\it Bottom panel}: Same as upper panel, however, including values from the literature.
\label{fig:vsini_std_vs_vsini}
\end{center}
\end{figure}

Large projected rotational velocity values could not only result from a single fast rotator, but also from a blended profile of two slower rotators.  If the latter is the case, one would expect $v~\mathrm{sin}~i$ values varying in time depending on the system's phase at the time of the observations. To investigate any potential systems of this kind, similarly to Fig.~\ref{fig:rv_std_vs_vsini}, we looked at the typical variations in $v~\mathrm{sin}~i$ as a function of median $v~\mathrm{sin}~i$.  These results are shown in Fig.~\ref{fig:vsini_std_vs_vsini}. Note that as our $v~\mathrm{sin}~i$ values are calculated from a grid of rotational profiles with 1\,km~s$^{-1}$ step, we are insensitive to smaller variations and therefore many objects appear to be constant. Following a similar approach to the one of the previous subsection, we calculated an upper envelopes to the variations in $v~\mathrm{sin}~i$ and flagged systems above those levels for further inspection.

\subsection{Using the BIS versus RV relation}
\label{sec:BIS_vs_RV}

Another way to validate whether a RV variation is induced by a bound companion is to include the BIS as an additional source of information. \cite{Lagrange2013} used this technique searching for giant planets in a sample of $26$ stars, some of which are in the young associations studied here. Significant anti-correlation between the BIS and RV suggests that the RV jitter is most likely due to stellar activity \citep{Desort2007}.
This technique relies on a large number of measurements per target and therefore in this work we are limited to a small number of stars in our sample. Therefore, in our case, this technique allowed us to rule out a few potential SBs rather than to identify new systems. The BIS and RV values are listed in Table \ref{tab:RV_indiv}.

\subsection{Spectroscopic binaries from the literature}

We searched the literature to identify formerly flagged SBs from our sample to assess the robustness of our method. For all previously identified spectroscopic binaries, we recover a very large fraction of them ($84\%^{+11}_{-8}$). Most of the non-recovered SBs correspond to objects or systems with very few observations in our local database, but for a few of the objects, our analysis contradicts the ``SB flag" found in the literature (see Appendix~\ref{sec:individ_sources} for comments on the individual sources). 

\subsection{Close visual binaries from the literature}

Some multiple systems have the right configuration and are located at the right distance for them to be resolvable with direct imaging techniques (with adaptive optics, AO hereafter) and, in addition, display RV variations of the primary. A good example of such system is V343\,Nor \citep{Nielsen2016}.  Looking for similar cases, we compiled a list of targets from the literature that have AO-discovered known companions (typically, with estimated periods of $\approx$1000\,days, Table~\ref{tab:vb_targets}).  

Unfortunately, within this AO sample of four close visual binaries, none of them had sufficient time coverage in our database of high-resolution spectra to achieve\st{d} the sensitivity needed to detect any companion-induced RV changes. However, the orbits of all four systems have been determined in previous works, as noted in Table~\ref{tab:vb_targets}.

{
\begin{table}
\tiny
\begin{center}
\caption{Kinematic properties of previously identified close visual binaries within our sample.}
\begin{tabular}{p{1.3cm}p{0.7cm}  p{0.6cm} p{0.6cm} p{0.6cm}p{0.6cm}p{0.5cm} }
\hline\hline\\
  \multicolumn{1}{l}{ID} &             
  \multicolumn{1}{l}{$\sigma_\mathrm{\subrv}$} &               
  \multicolumn{1}{l}{ $v~\mathrm{sin}~i$} &     
  \multicolumn{1}{l}{Time span} &                 
  \multicolumn{1}{l}{Num. obs}  &        
  \multicolumn{1}{l}{P} &   
  \multicolumn{1}{l}{Ref.} \\       
  \multicolumn{1}{l}{} &             
\multicolumn{1}{l}{(km~s$^{-1}$)} &             
  \multicolumn{1}{l}{(km~s$^{-1}$)} &         
  \multicolumn{1}{l}{(day)} &           
  \multicolumn{1}{l}{}  &
  \multicolumn{1}{l}{(year)}  &
  \multicolumn{1}{l}{}  \\    
  \hline\\
TWA 22 & 0.19 & 9.9 & 64 & 3 & 5.15 & a \\ 
HD 98800 & 0.07 & $< 5$ & 4 & 2 & 0.86 & b, g \\ 
HD 16760  & \ldots & $< 5$ & \ldots & 1 & 1.27 & c, d \\ 
HD 36705 & \ldots & $\approx$75 & \ldots & 1 & 11.74 & e, f \\ 
\hline\\[-0.4ex]
\end{tabular}
\tablefoot{a: \cite{Bonnefoy2009}, b: \cite{Malkov2012}, c: \cite{Bouchy2009}, d: \cite{Sato2009}, e: \cite{Close2005}, f: \cite{Nielsen2005}, g: \cite{Torres1995b}}
\label{tab:vb_targets}
\end{center}
\end{table}}

\subsection{Detection of SBs candidates}

The final list of SB candidates identified in this work is presented in Table~\ref{tab:flagged_variable_sources}. In a few cases, our analysis contradicts previous claims of multiplicity from the literature, while in some other cases, we do not recover the SB nature of some candidates, which we attribute to the sampling of the data available to us (see details on Appendix \ref{sec:individ_sources}).

Out of the $381$ objects from the compilation of our work, the literature (Table \ref{tab:mjd_estimations}) and Gaia DR2, we identified $68$ SB candidates. For each candidate, we compiled all the information available regarding RV and $v~\mathrm{sin}~i$ both from our work and the literature and used those values to establish a final classification regarding their multiplicity. The conclusion ({\it Conc.}) column of Table~\ref{tab:flagged_variable_sources} presents the summary of this analysis, where the values ``Y'', ``N'' or ``?'' correspond to 
``multiple system", ``not a multiple system according to the data available", or ``inconclusive".

While specific comments for particularly interesting or challenging candidates can be found in  Appendix~\ref{sec:individ_sources}, there were a number of cases where the variable flag of $v~\mathrm{sin}~i$ turned out to be a misleading diagnostic. In these cases, a closer inspection of the CCF profiles revealed that the variability was not real and just induced by a poor fitting of the rotational profile. In such cases, it is still possible that the candidate is an unresolved SB, but, since we do not have sufficient evidence to support that conclusion, we flagged those candidates as inconclusive.

\onecolumn{
\centering
\tiny{
\LTcapwidth=\textwidth
\begin{longtable}[l]{p{3.4cm} p{0.8cm} p{0.8cm} p{0.8cm}p{0.8cm} p{0.8cm} p{0.8cm} p{0.8cm} p{0.8cm} p{0.9cm} p{0.1cm} p{0.3cm} p{0.3cm}}
\caption{Properties of targets flagged as potential SB1 systems in the analysis presented in this work. Standard deviation are calculated for targets with two or more epochs. Targets previously flagged but not recovered in this work are available in Appendix~\ref{sec:individ_sources}.  The new and recovered SB2/SB3 targets are available in Appendix~\ref{sec:individ_sources} and Table \ref{table:summary_tab}.
}\label{tab:flagged_variable_sources}\\
\hline\hline\\
  \multicolumn{1}{l}{ID} &             
  \multicolumn{4}{c}{Values calculated in this work} &             
  \multicolumn{4}{c}{Values calculated in this work + literature} &               
  \multicolumn{1}{l}{\# obs} &               
  \multicolumn{1}{l}{Flag}  &    
  \multicolumn{1}{l}{Conc.}  \\[1ex]
  \multicolumn{1}{l}{} &             
  \multicolumn{1}{l}{$\mathrm{RV}_{\mathrm{median}}$} &             
  \multicolumn{1}{l}{$\sigma_\mathrm{\subrv}$} &               
  \multicolumn{1}{l}{$\mathrm{vsini}_{\mathrm{median}}$} &        
  \multicolumn{1}{l}{$\sigma_\mathrm{vsini}$} &                    
  \multicolumn{1}{l}{$\mathrm{RV}_{\mathrm{median}}$} &             
  \multicolumn{1}{l}{$\sigma_\mathrm{\subrv}$} &               
  \multicolumn{1}{l}{$\mathrm{vsini}_{\mathrm{median}}$} &        
  \multicolumn{1}{l}{$\sigma_\mathrm{vsini}$} &                    
  \multicolumn{1}{l}{} &               
  \multicolumn{1}{l}{}  &
  \multicolumn{1}{l}{}  \\         
  \multicolumn{1}{l}{} &             
  \multicolumn{1}{l}{}  &  
  \multicolumn{1}{l}{}  &  
    \multicolumn{1}{l}{}  &
    \multicolumn{1}{l}{}  &
    \multicolumn{1}{l}{}  &
    \multicolumn{1}{l}{}  &
    \multicolumn{1}{l}{}  &                
    \multicolumn{1}{l}{}  &
    \multicolumn{1}{l}{}  &        
    \multicolumn{1}{l}{}  &            
    \multicolumn{1}{l}{}  &          
    \multicolumn{1}{c}{}  \\
  \hline\\
      \multicolumn{13}{c}{Potential SB1 systems from variable RV and/or $v~\mathrm{sin}~i$ values}  \\  
  \hline\\      
CD-46 644 & 23.70 & 0.03 & 34.16 & 0.0 & 24.22 & 0.96 & 34.16 & 7.54 & 2 (4) & & N \\ 
HD 17332 A &  4.62 & 0.75 & 8.41 & 4.55 & 4.20 & 0.66 & 8.41 & 4.55 & 2 (4) & & ? \\ 
CD-56 1032A &  31.87 & 4.12 & 39.72 & 6.56 & 31.87 & 5.83 & 39.72 & 9.28 & 2 (2) & & Y \\ 
CPD-19 878 &  25.59 & 1.32 & 30.63 & 0.51 & 25.59 & 1.32 & 30.63 & 0.51 & 4 (4) & & ? \\ 
TYC 7627-2190-1 & 21.94 & 3.03 & 12.95 & 12.85 & 21.94 & 3.71 & 24.98 & 14.88 & 3 (4) & & Y \\ 
 V*PXVir &  -12.99 & 0.52 & 4.17 & 0.31 & -12.39 & 5.81 & 4.16 & 0.35 & 4 (8) & SB1 & Y \\ 
 HD 159911 &  21.77 & 0.63 & 58.4 & 12.02 & 21.77 & 0.63 & 58.4 & 12.02 & 3 (3) & & Y \\ 
 CD-43 3604 &  17.5 & 2.35 & 18.0 & 2.19 & 17.43 & 2.66 & 18.0 & 9.52 & 4 (5) & & Y \\ 
V* V379 Vel &  14.645 & 0.045 & 7.9 & 1.5 & 14.6 & 1.49 & 7.9 & 1.5 & 2 (3) &  & ? \\ 
 TYC 8594-58-1 &  11.03 & 0.650 & 12.95 & 0.0 & 11.03 & 0.75 & 12.95 & 9.45 & 4 (5) & & N \\ 
 2MASS J12203437-7539286 &  4.86 & 0.02 & 7.9 & 1.5 & 4.86 & 2.47 & 7.90 & 2.37 & 2 (3) & & Y \\ 
HD 129496 &  -6.07 & 3.07 & 66.99 & 1.51 & -6.07 & 3.07 & 66.99 & 1.51 & 2 (2) & & N \\ 
 V*AFLep &  20.89 & 1.11 & 50.32 & 11.42 & 21.39 & 1.25 & 50.32 & 11.42 & 4 (5) &  & N \\ 
 HD 139084 &  5.17 & 1.99 & 15.77 & 0.56 & 5.10 & 1.76 & 15.88 & 0.55 & 9 (11) & SB1 & Y \\ 
 HD 139084 B &  4.55 & 0.01 & 15.98 & 1.50 & 2.32 & 3.14 & 15.98 & 1.5 & 1 (2) & & N \\ 
 HD 164249 A & -0.14 & 1.17 & 21.54 & 2.37 & -0.09 & 1.06 & 21.03 & 2.25 & 8 (11) &  & N \\ 
 HD 164249 B & -0.6 & 0.28 & 12.95 & 6.06 & -0.88 & 0.88 & 12.95 & 6.06 & 2 (3) & & N \\ 
 CD-31 16041 & -8.81 & 0.20 & 40.22 & 3.78 & -8.73 & 1.25 & 43.25 & 4.92 & 3 (4) & & N \\ 
 V*PZTel &  -2.99 & 2.96 & 55.23 & 12.55 & -3.54 & 2.71 & 58.99 & 12.81 & 10 (12) & & N \\ 
 HD 199143 &  -22.73 & \ldots & 58.40 & \ldots & -13.62 & 12.89 & 92.95 & 48.86 & 1 (2) & & N \\ 
*cEri &  18.48 & 7.64 & 57.39 & 1.69 & 18.43 & 7.23 & 57.39 & 1.69 & 7 (8) & & N \\ 
GJ 3305 &  23.91 & 0.49 & 5.88 & 0.48 & 20.95 & 1.57 & 5.88 & 0.48 & 3 (9) &  & Y \\ 
 HD 22213 &  11.27 & 3.14 & 40.73 & 0.51 & 11.27 & 3.14 & 40.73 & 0.51 & 2 (2) &  & Y \\ 
 HD 21997 &  17.17 & 0.86 & 65.47 & 9.05 & 17.24 & 0.91 & 65.47 & 9.05 & 3 (4) & & N \\ 
 V*AGLep &  25.31 & 0.57 & 23.050 & 4.76 & 25.31 & 0.57 & 23.050 & 4.76 & 4 (5) & & ? \\ 
CD-44 753 &  13.16 & 0.91 & 7.9 & 0.95 & 13.78 & 1.37 & 7.0 & 0.95 & 3 (6) & & N \\ 
 HD 104467 &  11.16 & 2.78 & 25.07 & 2.25 & 11.4 & 2.31 & 25.07 & 2.25 & 6 (8) &  & Y \\ 
 2MASS J12020369-7853012 &  11.17 & 2.91 & 14.97 & 0.71 & 11.17 & 2.91 & 14.97 & 0.71 & 4 (4) & SB1 & Y \\ 
BD-184452A &  \ldots & \ldots & \ldots & \ldots & -19.31 & 2.01 & 8.05 & 4.59 & 0 (2) & & ? \\ 
 GSC 08057-00342 &  \ldots & \ldots & \ldots & \ldots & 13.5 & 5.59 & 5.2 & \ldots & 0 (3) & SB1 & Y \\ 
 2MASS J04470041-513440  & \ldots & \ldots & \ldots & \ldots & 17.92 & 1.98 & 5.1 & \ldots & 0 (2) & & N \\ 
 UCAC3 33-129092 & \ldots & \ldots & \ldots & \ldots & 7.07 & 2.86 & 10.5 & \ldots & 0 (2) & & N \\ 
 UCAC4 110-129613  & \ldots & \ldots & \ldots & \ldots & 3.58 & 6.24 & 25.1 & \ldots & 0 (2) & & N \\ 
 CD-53 544 &  12.62 & 2.90 & 63.45 & 2.18 & 12.56 & 2.55 & 65.47 & 8.26 & 3 (5) & & N \\ 
 TYC8098-414-1 &  \ldots & \ldots & \ldots & \ldots & 19.53 & 8.72 & 11.75 & 9.40 & 0 (6) & & ? \\ 
 HD 207575 &  1.42 & 2.42 & 37.19 & 5.82 & 1.5 & 2.14 & 37.19 & 5.82 & 5 (7) & & ? \\ 
 HD 207964 &  23.46 & 0.2 & 53.86 & 1.52 & 23.26 & 12.65 & 53.86 & 1.52  & 2 (3) & & N \\ 
 TYC 9344-293-1 &  6.16 & 1.01 & 55.37 & 1.43 & 6.95 & 1.57 & 55.35 & 10.0 & 3 (6) & & N \\ 
 UCAC3 92-4597 & \ldots & \ldots & \ldots & \ldots & -5.2 & 9.81 & 4.7 & \ldots & 0 (3) & SB & Y \\ 
HD 3221 &  -2.39 & 3.26 & 68.5 & 5.01 & -2.39 & 3.26 & 68.5 & 5.01 & 3 (3) & & N \\ 
UCAC3 70-2386 & \ldots & \ldots & \ldots & \ldots & 5.65 & 2.33 & 19.2 & \ldots & 0 (2) & SB & Y \\ 
V* CE Ant &  11.7 & 0.06 & 4.87 & 1.75 & 12.4 & 0.32 & 4.87 & 1.76 & 4 (17) & & N \\ 
TWA23 &  10.82 & 0.04 & 9.92 & 3.0 & 7.71 & 2.61 & 9.92 & 3.0 & 2 (16) & SB & Y \\ 
UCAC2 1331888 &  -1.66 & 0.56 & 25.07 & 1.0 & -2.22 & 2.01 & 25.80 & 1.09 & 2 (3) &  & N \\ 
HD 48189 &  36.14 & 0.01 & 16.99 & 1.5 & 33.40 & 2.06 & 17.29 & 0.43 & 2 (3) &  & N \\ 
CD-30 3394 &  12.71 & 2.39 & 37.69 & 0.50 & 14.99 & 2.84 & 37.19 & 0.87 & 4 (5) &  & ? \\ 
CD-30 3394B &  13.94 & 3.21 & 47.79 & 2.71 & 15.09 & 3.24 & 47.29 & 4.07 & 4 (5) &  & ? \\ 
CD-52 9381 &  -13.85 & 2.74 & 39.71 & 1.23 & -13.85 & 2.74 & 39.71 & 1.23 & 4 (4) &  & N \\ 
GSC 08350-01924 & 1.57 &1.45 & 23.05 & 3.0 & 0.21 & 1.46 & 23.05 & 3.0 & 2 (4) &  & N \\ 
V*AFHor & 12.91 & 0.06 & 7.90 & 1.5 & 12.70 & 1.13 & 7.90 & 1.58 & 2 (6) &  & N \\ 
RX J12204-7407 & 14.60 & 1.37 & 39.72 & 1.5 & 14.60 & 1.58 & 39.71 & 1.72 &  4 (4) &  & N \\ 
$[$FLG2003$]$ eps Cha 7 & 13.64 & 1.24 & 23.05 & 0.47 & 13.64 & 1.24 & 23.05 & 0.47 &  3 (3) &  & N \\ 
HD 17250 & 10.51 & 0.54 & 42.24 & 0.82 & 9.73 & 2.92 & 42.24 & 1.01 & 3 (5) & SB & Y \\ 
HD 191089 & -11.69 & 0.47 & 43.75 & 1.23 & -11.18 & 3.13 & 43.75 & 1.42 & 4 (7) &  & ? \\ 
V* AO Men & 16.02 & 0.22 & 16.69 & 0.44 & 16.02 & 1.63 & 16.69 & 0.44 & 8 (10) &  & N \\ 
HD 984 & -2.21 & 1.95 & 39.26 & 1.45 & -2.21 & 2.30 & 39.26 & 1.59 & 6 (8) & & N \\ 
HD 37484 & 21.19 & 0.13 & 52.34 & 1.5 & 21.32 & 2.80 & 52.34 & 1.5 & 2 (3) & & N \\ 
2MASS J01505688-5844032 & \ldots & \ldots & \ldots & \ldots & 9.95 & 1.62 & 10.10 & \ldots & 0 (2) & & N \\ 
UCAC4 137-000439 & \ldots & \ldots & \ldots & \ldots & 7.69 & 2.41 & 11.20 & \ldots & 0 (2) & & ? \\ 
2MASS J12560830-6926539 & \ldots & \ldots & \ldots & \ldots & 11.31 & 3.53 & 16.30 & \ldots & 0 (2) & & Y \\ 
BD-20 1111 & 19.26 & 0.72 & 24.06 & 4.54 & 18.68 & 1.00 & 24.06 & 5.56 & 3 (4) & & ? \\ 
Smethells 165 & 5.98 & 0.72	& 20.02	& 0.47 & 6.04 & 0.69 & 20.02 & 4.04 & 3 (6) & & ? \\ 
\hline\\[-0.3ex]
\end{longtable}
}}

\twocolumn

\section{Accounting for observation sensitivity}
\label{sec:sensitivity}

\begin{figure}[h]
\begin{center}
\includegraphics[width=0.49\textwidth]{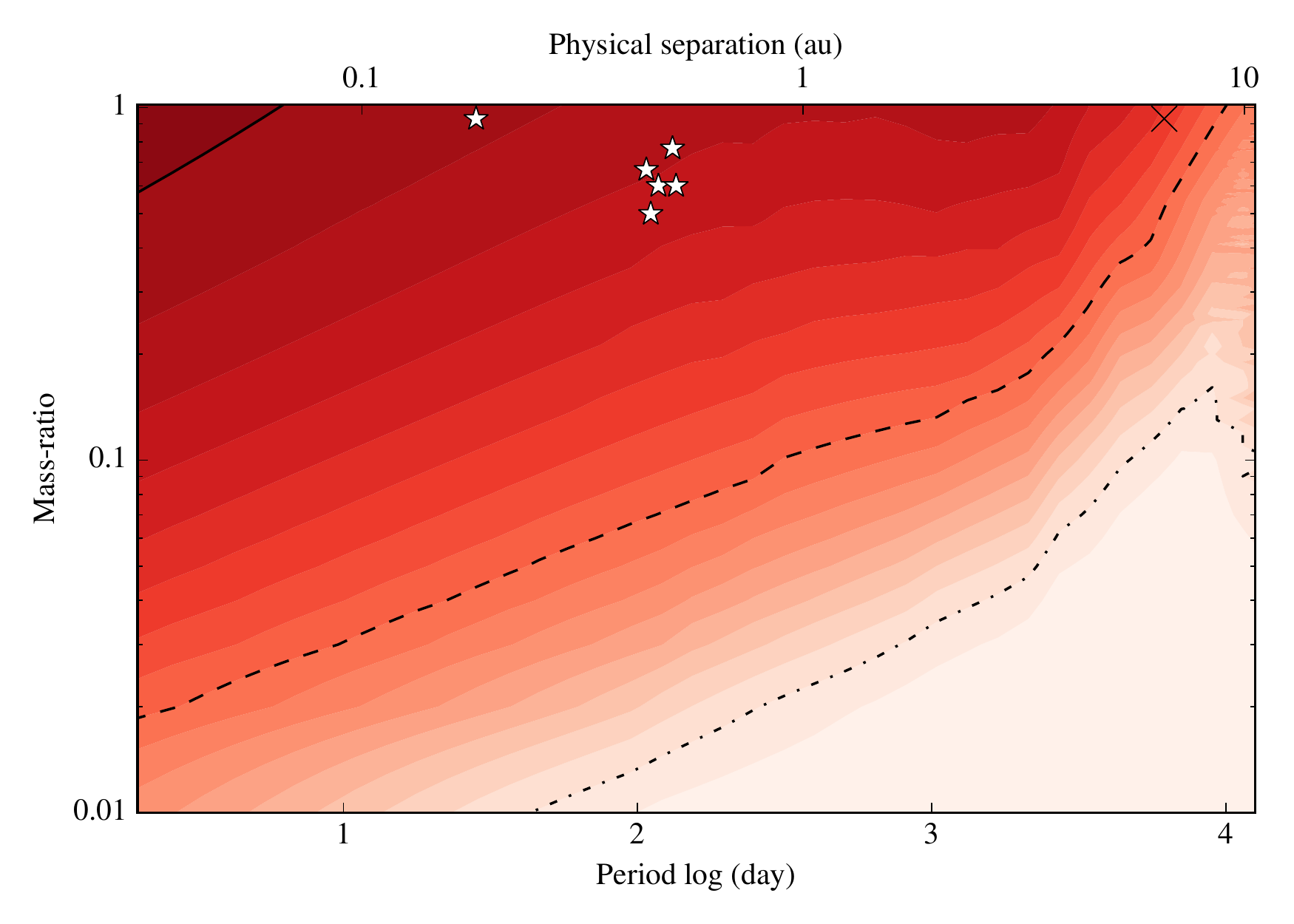}
\includegraphics[width=0.49\textwidth]{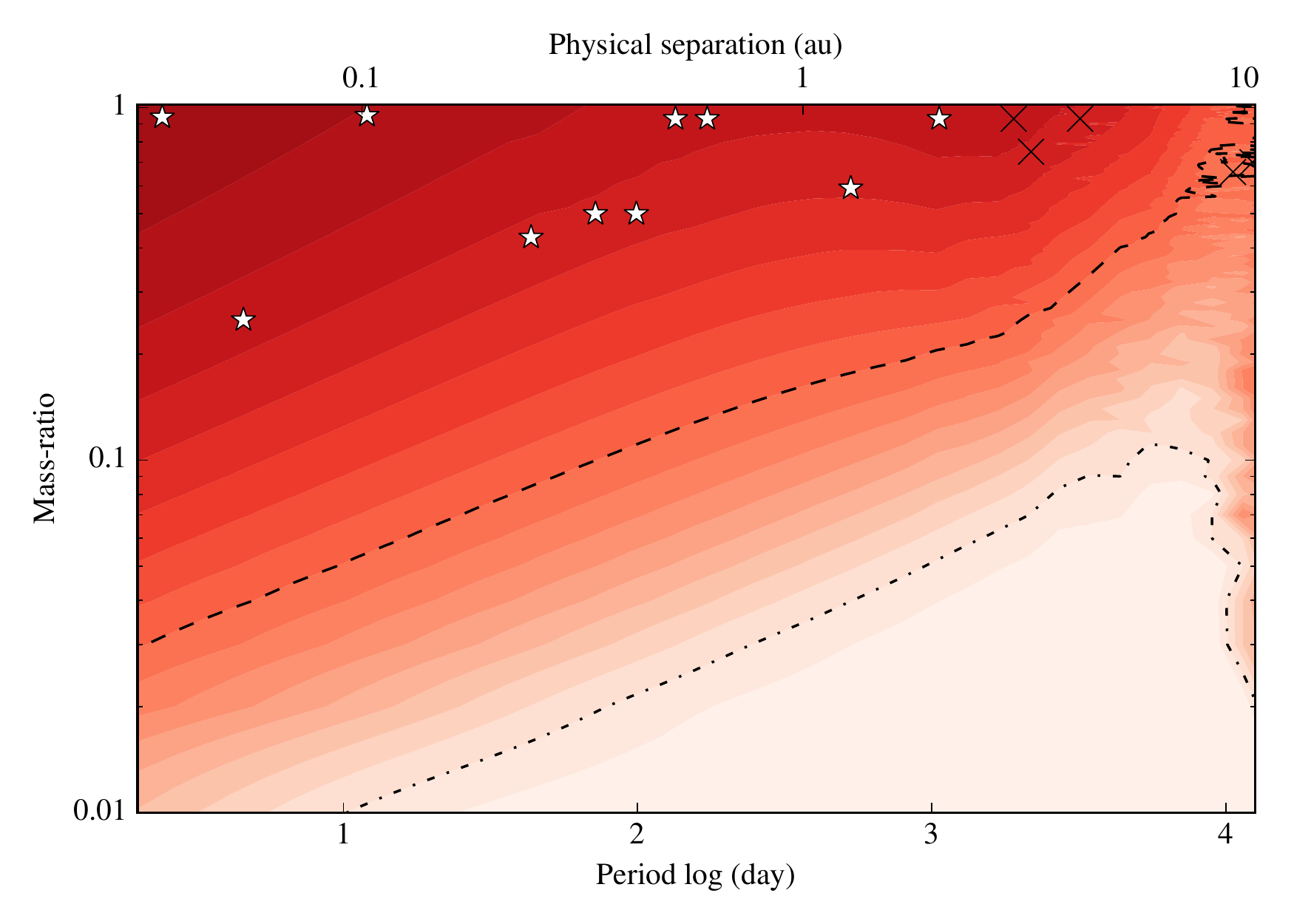}
\vspace{-0.4cm}
\caption{{\it Upper panel}: Average detection probabilities for THA association (contours from red, 100$\%$, to white, 0$\%$), detected spectroscopic companions (white stars) and visual binaries (black crosses) in the physical separation versus mass ratio. The solid, dashed and dash-dotted lines encompass areas with detection probabilities $\geq$ 90$\%$, 50$\%$ and 10 $\%$, respectively. {\it Bottom panel}: Same as upper panel but for BPC association.}
\label{fig:sample_detec_prob}
\end{center}
\end{figure}

As we have seen through this work, tight binaries can be detected in spectroscopic data via identification of double (or multiple) lines, variable RVs (or, unrelated to this work, even unexpected mixes of spectroscopic features). However, our ability to identify these features (multiple lines and variations in RV), can be severely biased by factors such as: the observations strategy (time span $T$ and number of measurements $N_\mathrm{obs}$) and the inherent sensitivity of the spectrographs employed for the observations. These factors have been thoroughly studied and modelled by \cite{Tokovinin2014a}.
The steps incorporated in our analysis to translate this knowledge into detection probability maps were the following:
\begin{enumerate}
    \item We created a set of $10,000$ simulated binaries from the following distributions:
    \begin{itemize}
    \item[$\blacksquare$]  Period ({\it p}): log-normal \citep[$\mu$=5.03, $\sigma$=2.28 log(day),][]{Raghavan2010}
    \item[$\blacksquare$]  mass ratio ({\it q}): uniform \citep[for system between 0.01-1.0\,M$_\odot$;][]{Raghavan2010, Kraus2011, Elliott2015}
    \item[$\blacksquare$]  Eccentricity ({\it e}), two-part:
    \begin{itemize}
        \item{\it p} $\leq$ 12\,days, {\it e}=0
        \item {\it p} $>$12\,days, uniform (for 0 $\leq$ {\it e} $\leq$ 0.6)
    \end{itemize}
    \item[$\blacksquare$]  Initial phase ($\phi_0$), longitude of ascending mode ($\omega$) and inclination ({\it i}): uniform (for 0 $\leq\phi_0\leq$ 1, 0 $\leq\omega\leq$ 2$\pi$, and 0 $\leq{\it i}\leq$ $\pi$, respectively)
    \end{itemize}
    \item From our simulations and using equations 5 to 7 from \cite{Tokovinin2014a}, 
    we calculated a detection probability map for each object characterised by its three detection parameters ($N_\mathrm{obs}$, $T$ and $\sigma_{RV}$). In the case of single epoch data, we assumed the same artificial parameters used by \cite{Tokovinin2014a} (i.e. $T= 100$~days, $N_\mathrm{obs}=3$, and $\sigma_{RV} = 2$\,km~s$^{-1}$), since we are still sensitive to double- and triple-lined multiple systems. 
    \item The detection map of each object was calculated on the same  mass ratio vs. period grid. This ``common-grid" approach, makes it easy to average those maps for objects belonging to the same moving group, yielding an average sensitivity map per association in our sample, see Fig. \ref{fig:sample_detec_prob}. 
    \item These ``association-averaged" probability maps were used to correct our SB fractions from biases induced by the observation strategy and precision. The correction was calculated by taking the mean value in the parameter space $0.1 \leq q  \leq 1$ and $p \leq 10^{\,3.2}$\,days. We excluded mass ratios smaller than 0.1 as very few targets have any meaningful probability of detection in this parameter space (Fig. \ref{fig:sample_detec_prob}, color-scale from red, 100$\%$, to white, 0$\%$).
\end{enumerate}

Note that these corrections are applied across the entire parameter space and do not have assumptions regarding the underlining mass ratio or period distributions (as we have extremely limited information on both).

\section{Updated census of spectroscopic binaries}
\label{sec:results}
\begin{figure}[h]
\begin{center}
\includegraphics[width=0.48\textwidth]{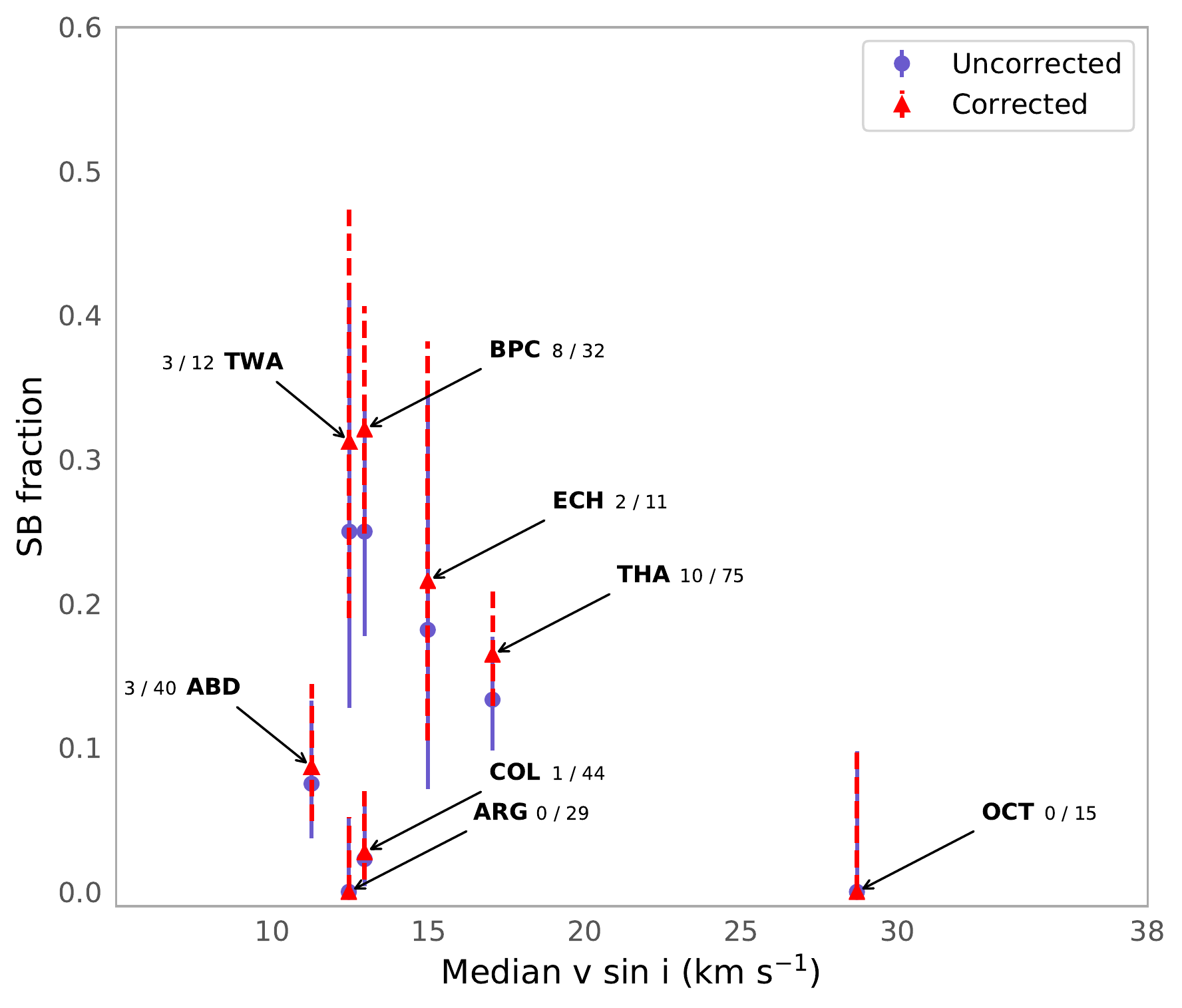}
\vspace{-0.3cm}
\caption{SB fraction as a function of median $v~\mathrm{sin}~i$.  The uncorrected SB fractions are shown in purple and in text next to the name of each association. The corrected SB fractions are shown in red. The primary mass range is $0.6 \leq M \leq 1.5$\, $M_\odot$.}
\label{fig:vsini_sb_frac}
\end{center}
\end{figure}

Building from the previous sections, in Fig.~\ref{fig:vsini_sb_frac}, we present the SB fraction obtained for each associations as a function of the median $v~\mathrm{sin}~i$ of its members. In that figure we present both fractions, the original one that disregards the effects discussed in Sec.~\ref{sec:sensitivity}, and the ``corrected" one (blue and red symbols, respectively). The uncertainties on the derived fractions are calculated from binomial statistics \citep{Burgasser2003}. 

\begin{figure}[h]
\begin{center}
\includegraphics[width=0.48\textwidth]{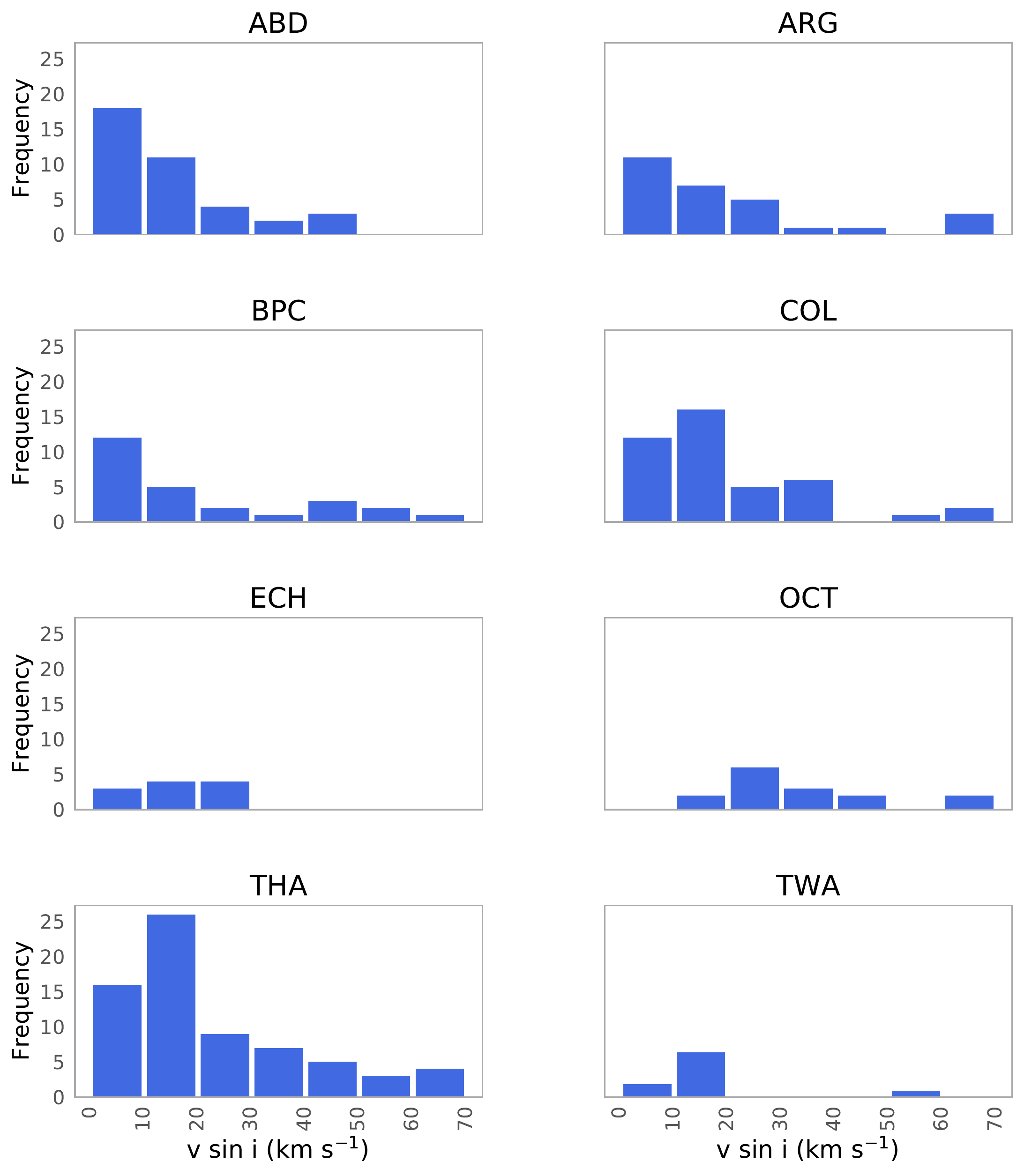}
\vspace{-0.3cm}
\caption{$v\sin\ i$ histogram for each young association from this work with primary mass range $0.6 \leq M \leq 1.5$\, $M_\odot$.}
\label{fig:vsini_histograms}
\end{center}
\end{figure}

As mentioned before, it is extremely difficult to fully account for the effect of $v~\mathrm{sin}~i$ on the sensitivity to identify SBs. Since fast rotators may bias the resulting SB fractions, we opted to look for any relationship between the obtained SB and the median $v~\mathrm{sin}~i$ of the members of each association. No apparent correlation was found between those two quantities, and the distribution of $v~\mathrm{sin}~i$  values for each association are plotted in  Fig.~\ref{fig:vsini_histograms}.  

\begin{figure}[h]
\begin{center}
\includegraphics[width=0.49\textwidth]{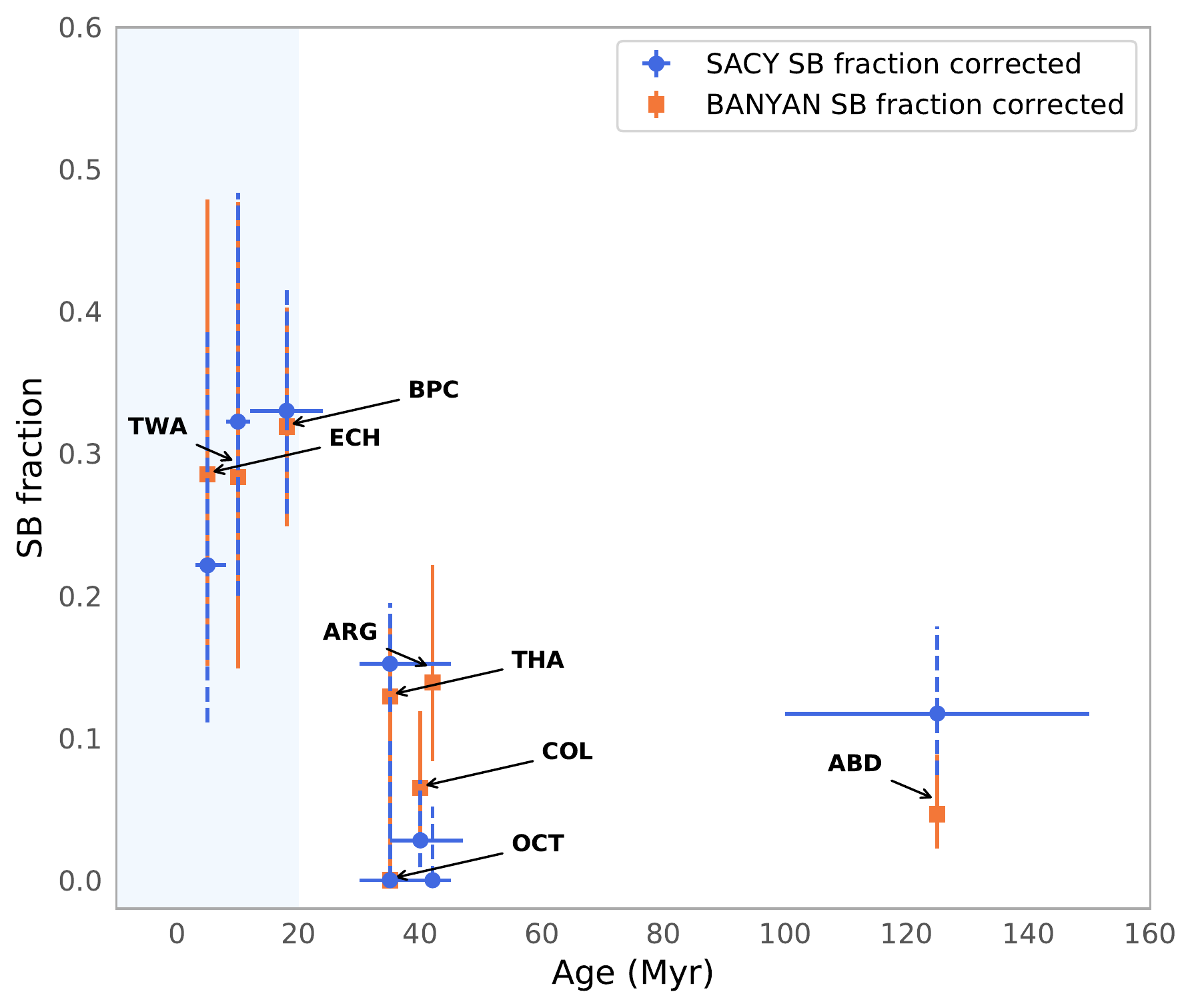}
\vspace{-0.6cm}
\caption{Corrected SB fraction as a function of age (Myr) for membership estimation from our convergence method \citep[blue dots,][]{Torres2006,Torres2008} and BANYAN $\Sigma$ \citep[orange dots,][]{Gagne2018A}. The shaded area highlights the $\leq~20$ Myr zone of the figure. The primary mass range is $0.6 \leq M \leq 1.5$\, $M_\odot$.}
\label{fig:age_sb_frac}
\end{center}
\end{figure}

A striking result from our study is that the SB fraction obtained for the TW Hya association seems to contradict the results from \cite{Elliott2014}. This difference is driven by the discovery of three newly identified SBs in this work, that was possible because of an increase of $30$\% in the amount of data available for this association since \cite{Elliott2014}. To test this result against membership criteria, we compared the fraction estimated using the census obtained from the BANYAN $\Sigma$ tool with that of the convergence method and both figures are fully compatible (see Fig.~\ref{fig:age_sb_frac}). 

Interestingly, the three highest SB fractions are found for the three youngest associations (\textepsilon~Cha $18 \substack{+15 \\ -11}  \%$, TW Hya $22 \substack{+16 \\ -14} \%$ and \textbeta~Pictoris moving group $24 \substack{+9 \\ -8} \%$ prior sensitivity correction, and  $ 22 \substack{+15 \\ -11}  \%$, $ 32 \substack{+16 \\ -12} \%$, and $ 33 \substack{+9 \\ -8} \%$, respectively, when the corrections of Sec.~\ref{sec:sensitivity} are applied).  This is unlikely to result from a lack of sensitivity due to large rotational broadening, as the median $v~\mathrm{sin}~i$ values are relatively low and similar (once the low-number statistics are taken into account) for the three associations (see Fig. \ref{fig:vsini_histograms}). Furthermore, as shown in Fig.~\ref{fig:age_sb_frac}, the higher SB fraction of these associations seems to be insensitive to the membership criteria used, appearing also when the BANYAN $\Sigma$ census is employed. On the other hand, the average SB fraction for the five older associations are $\lesssim 10\%$ (with the possible ``intermediate" case of THA). It must be noted that the confidence interval for this ``dichotomy'' is only 1 to 2\,$\sigma$ given the associated large uncertainties. 

\section{Discussion}
\label{sec:discussion}
The results presented in Section~\ref{sec:results} suggest a counter-intuitive path of evolution for SBs. In this section we compare our results to the literature, discuss whether these results are in fact an artefact produced by our methodology or a physical result; and, in the latter case, if we are really witnessing early SB evolution or the effect of other environmental factors.

\subsection{Comparison with previous results on low density environments}
\label{sec:previous_results}
Figure \ref{fig:age_sb_frac} shows SB fractions ($\approx10\%$) consistent with the field population \citep[$\approx10\%$,][]{Raghavan2010,Tokovinin2014b}, the young clusters Tau-Aur, and Cha I \citep[$\approx7\%$,][]{Nguyen2012} and our previous results from \cite{Elliott2014} for the five older associations ($\gtrsim 20$ Myr) across the mass range of $\sim 0.2-2.0$ M$_{\odot}$. On the other hand, the observed SB fractions for the three youngest associations seem to be larger than those reported for the previously mentioned young regions of Tau-Aur (1\,Myr) and Cha~I (2\,Myr). The estimated distances to these young regions are $\sim$140\,pc and $\sim$160\,pc, respectively \citep{Nguyen2012}, therefore we argue that, given the overall closer distance of our targets, the difference should not rise from a lack of sensitivity or a completeness bias in the SACY sample \citep[see Sec. 5 from][]{Nguyen2012}.

Nevertheless, the relative paucity of SBs in Tau-Aur and Cha~I could be explained by the sample used by \cite{Nguyen2012} which is concentrated on the higher stellar density regions of the clouds. For instance, \cite{Guieu2006} revisited the previously claimed brown dwarf deficit in the same Tau-Aur region, performing a larger scale optical survey including the surroundings of the clouds as well as their densest parts. The authors concluded that the possible deficit was in fact an artefact from target selection rather than a real difference. Interestingly, \cite{VianaAlmeida2012} derived an SB fraction of $\approx 42\%$ for the Rho Ophiuchus star forming region ($\sim~ 0.1-1$ Myr) from targets with mass range of $\sim 0.18-1.4$ M$_{\odot}$ \citep{Natta2006} and a binary fraction of $\approx 71\%$ combining data from different works. These results are more consistent with the SB fraction of our youngest associations and are aligned with the notion that multiplicity is very high at young ages (younger than $\sim$ 1\,Myr). Although the statistical significance in the difference on SB fraction in Fig. \ref{fig:age_sb_frac} is weak, at the level of 1 to 2\,$\sigma$, it is hard to reconcile with the general picture of SB fraction remaining unchanged after $\sim 1$ Myr, and therefore deserves independent confirmation and further characterisation.

\subsection{The impact of the sensitivity correction}
\label{sec:sensitiviy_effect}
In Section \ref{sec:sensitivity} we created sensitivity maps from $10,000$ simulated binaries, to estimate how many binary systems would have been missed because of our observing strategy. The simulated binaries were drawn according to certain priors on the mass ratio, period, and orbital parameters, but those parental distributions were originally estimated from field star surveys \citep{Raghavan2010,Tokovinin2014a}. Those priors may not be representative of the underlying population of binary stars in young associations ($\lesssim\,100$\,Myr). This may have consequences on the sensitivity corrections we obtained which may lead to an artificially large \textit{corrected} SB fraction.

The prior on the period distribution is the most critical one, as it has the most significant effect on the detection probability (shorter periods are easier to detect using spectroscopic observations). Taking this into consideration we created new sensitivity maps using a log-normal period distribution ($\mu= 5.3$, $\sigma=2.28$ log(day), from \citealt{Tobin2016}), representative of Class\,0/I systems ($\lesssim 1$\,Myr). With this period distribution  we obtain an increase of $\sim~2\%$ on the correction factor. This slight increase is not sufficient to explain the difference of $\gtrsim~10-20\%$ between the three younger association with respect to the older ones in our sample. We further tested the impact of the period distribution on the correction factor by taking an even more extreme case. We used a distribution centred  at the smallest separation that a primordial binary system could have \citep[$\approx10$ au from disc fragmentation][]{Vaytet2012}, and even in that almost unrealistic scenario we did not reach a change of sensitivity sufficient to justify the differences of SB fractions between the young and old associations in our sample. The analysis presented here suggests that the differences in SB fractions are not artificially created by our sensitivity correction approach.

\begin{figure*}[h]
\begin{center}
\includegraphics[width=0.85\textwidth]{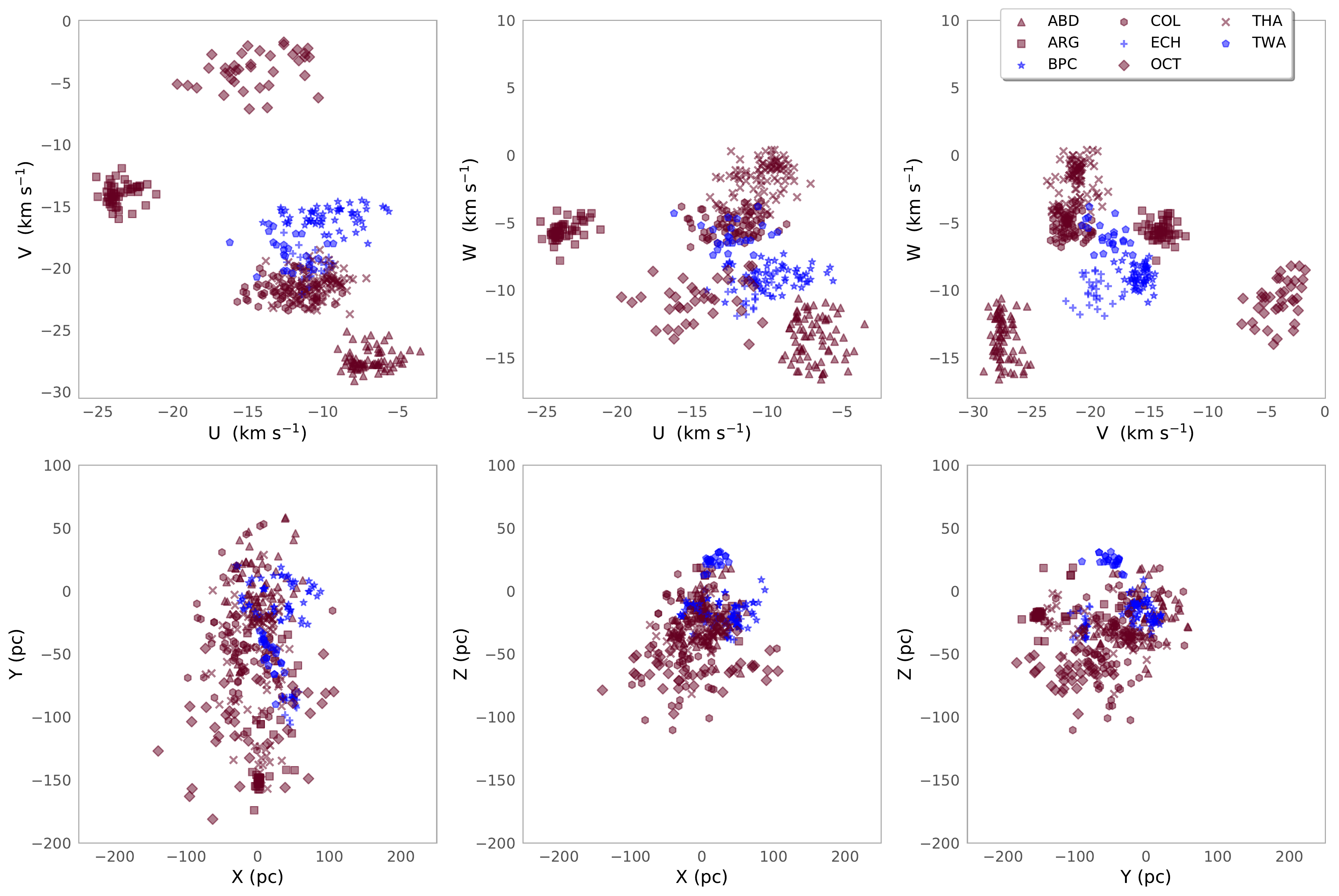}
\vspace{-0.2cm}
\caption{Combinations of the sub-spaces of the UVWXYZ-space for the young associations in the SACY sample. The blue coloured symbols correspond to the three youngest associations (BPC, ECH and TWA). The full membership study and further analysis will be presented in Torres et al. (in prep.).}
\label{fig:sacy_spaces}
\end{center}
\end{figure*}

\subsection{The relation with higher-order multiplicity}
\label{sec:hierarchical_systems}
From the SBs identified in this work, $\sim 77\substack{+8 \\ -7} \% $ are also part of higher-order multiple systems \citep{Elliott2016a,Elliott2016b}. This shows a preference for SBs to be found in triple or higher-order systems, similar to the $63\%$ reported in \cite{Tokovinin2006} for field stars.

There is observational evidence that suggests an overall decrease of binary fraction from pre-MS ages to field ages \citep{Ghez1997,Kouwenhoven2007,Raghavan2010}. \cite{Elliott2016b} suggested that dynamical interactions of triple systems (as proposed by \citealt{Sterzik2002,Reipurth2012}) could explain the population from close (0.1\,au) to very wide (10\,kau) tertiary components where the majority of the wide companions are in the process of being disrupted on timescales of $10-100$\,Myr. The results of \cite{Raghavan2010} also suggest that systems with long periods, or those who have more than two components, tend to lose companions with age due to dynamical evolution. However, these mechanisms that would explain the disruption of wide companions would not necessary explain the SB fraction in this sample. In fact, \cite{Tokovinin2006} suggested that the overall SB fraction seems to remain unchanged after $\sim$1 Myr.

Supporting the dissolution scenario, proposed by \cite{Sterzik2002,Reipurth2012}, $\sim 92\substack{+13 \\ -6} \%$ of SBs in the three youngest associations studied here are part of a triple or high order multiple system in contrast with the $\sim 67\substack{+12 \\ -11} \%$ for the five older associations.

\subsection{The SB fraction evolution with age}
\label{sec:sb_age}
Our results hint that the youngest associations ($\lesssim~20$ Myr) may have a larger SB fraction, even though it remains tentative at the moment. This result suggests a possible decrease of the SB fraction from $\sim5$ to $\sim100$ Myr. A similar result was obtained for the IN-SYNC (INfrared Spectroscopy of Young Nebulous Clusters) sample from high resolution H-band spectra observations of low-mass stars in Orion A, NGC 2264, NGC 1333, IC 348, and the Pleiades \citep{Jaehnig2017}, where the SB fraction of the five pre-MS clusters ($\approx1-10$ Myr) was $\approx20\%-30\%$ in contrast with $\approx5\%-10\%$ found for the Pleiades ($\approx100$ Myr).  \cite{Jaehnig2017} claim that the time sampling of their observations make it more sensitive to the critical $10^2-10^4$ days period range where binary systems are wide enough to be disrupted by dynamical interaction over $\sim$100 Myr timescale in dense environments. However, this scenario is proposed for clusters with typical densities of $\approx 30 M_{\odot}$\,pc$^{-3}$ (at the core radius, \citealt{Piskunov2007}) and may not be compatible with the typical densities of $\approx0.01$ stars\,pc$^{-3}$ for loose associations such as the ones in the SACY sample \citep{Moraux2016}.

\subsection{The role of the environment}
\label{sec:formation_source}
The tentative variations in SB fraction could be related to differences in the primordial multiplicity depending on the formation history and environment of the associations. In Figure\,\ref{fig:sacy_spaces}, we show the sub-spaces of the UVWXYZ-space for all the associations studied in this paper, to search for possible signs of clustering in both velocity and spatial coordinates. Given the proximity of the SACY associations no clear separated groups of points appear for the spatial coordinates \citep{Torres2006}. However, it is more informative to plot the galactic proper motion to trace a possible common origin (UVW: positive toward the Galactic center, Galactic rotation and North Galactic Pole respectively). Qualitatively, we identify possible clustering of points in the UVW sub-spaces (first row of Fig. \ref{fig:sacy_spaces}) for the three youngest associations (blue coloured symbols) that may suggest possible common birth place in the Galactic bars for these associations  compared to the older ones.

Furthermore, previous studies have found evidence that the three associations, \textbeta~Pictoris, TW Hya, and \textepsilon~Cha possibly formed in or near the Sco-Cen giant molecular cloud $5-15$\,Myr ago \citep{Mamajek2001,Torres2008}. Then the difference in the SB fraction presented in this work could arise from different primordial multiplicity instead of being caused by their dynamical evolution. In support of the latter argument, the overall binary fraction in Sco-Cen is $\approx93\%$ among solar-type stars and $\approx75\%$ among low-mass star \citep{Kouwenhoven2006}. These figures are higher than the overall binary fraction for solar-type and low-mass stars in Tau-Aur reported by \cite[][$\sim 66-75$\%, with slightly different binary parameter space explored]{Kraus2011}. In addition, \cite{Cuningham2020} recently claimed that the ABD association may be kinematically linked to a newly discovered ``stellar string'' Theia 301. \cite{Kounkel2019} argue that although they recover Sco-Cen in their kinematic clustering searches, this association is different than the ``typical strings" such as Theia 301.

To summarise, there are hints supporting non-universal multiplicity, however, our current data-set does not allow us to confirm different environmental star formation history among the SACY associations.

\section{Summary and conclusions}
\label{sec:conclusions}
In this work we have presented an update of the SB census for the associations within SACY. The study is based on new observational data (as well as literature and archival data), but also new criteria to identify these tight binaries. We have estimated radial and rotational velocity for $1375$ spectra using CCFs and compiled $\sim 400$ RV measurements from literature (including Gaia DR2, \citealt{GaiaDR2}). Our RVs and $v~\mathrm{sin}~i$ estimates are in good agreement with previously published values, following a 1:1 relation with values in literature (for targets that are not identified as a multiple systems), demonstrating that our CCF analysis is robust. Further robustness is provided by the fact that we have recovered the $84\substack{+11 \\ -8} \% $ of previously known multiple systems.

Besides RV variations as keys to identify SB candidates, we used high-order cross-correlation functions as a complementary diagnostic tool. These features offer a concrete way to quantify the symmetry, curvature and quality of the fitting of the CCFs. More epochs do not only allow to improve the reliability of any RV variation, but it also allows for other statistics to be used when assessing the binary nature of a candidate (see Sec.~\ref{sec:BIS_vs_RV} for instance).

We have calculated the SB fraction for each SACY association and have estimated a correction factor taking into account possible sensitivity issues and biases from the observations (see Sec.~\ref{sec:sensitivity}). The summary of SB candidates can be found in Tables \ref{tab:flagged_variable_sources} and \ref{table:summary_tab}. The analysis and conclusions reached for each target flagged as a candidate can be found in Appendix \ref{sec:individ_sources}.

We find that the three youngest associations have overall higher SB fractions (\textepsilon~Cha $ 22 \substack{+15 \\ -11}  \%$, TW Hya $ 32 \substack{+16 \\ -12} \%$ and \textbeta~Pictoris moving group $ 33 \substack{+9 \\ -8} \%$ when the corrections of Sec.~\ref{sec:sensitivity} are applied) compared with the five oldest associations in the SACY sample ($\sim 35-125$ Myr) which are $\sim 10\%$ or lower. This results seems to be independent of the method used for membership assessment (see Fig. \ref{fig:age_sb_frac}) and not artificially created by the sensitivity correction approach (see Sec. \ref{sec:sensitiviy_effect}). In addition, more than $90\%$ of the SB identified in \textepsilon~Cha, TW Hya and \textbeta~Pictoris are part of a triple or hierarchical system in contrast with $\approx 70\%$ of the five older associations. While the difference in SB fraction remains tentative at the moment, we propose two possible explanations: an evolution effect (previously reported in denser environments), and a primordial non-universal multiplicity. With the data available nowadays we cannot distinguish between the two possibilities.

\begin{acknowledgements}
The authors would like to thank the anonymous referee for constructive comments that helped to improve the content and clarity of this paper. S.Z-F acknowledges financial support from the European Southern Observatory via its studentship program and ANID via PFCHA/Doctorado Nacional/2018-21181044. All the authors acknowledge Dr. M.~Sucerquia and Dr. N.~Cuello for helpful insight and fruitful discussions on multiple systems' evolution. J.O. acknowledges support from Fondecyt (grant 1180395). S.Z-F., A.B., J.O. and C.Z, acknowledge support from Iniciativa Cient\'ifica Milenio via N\'ucleo Milenio de Formaci\'on Planetaria.  A.B acknowledges support from Fondecyt (grant 1190748). N.H. has been funded by the Spanish State Research Agency (AEI) Project No. ESP2017-87676-C5-1-R and No. MDM-2017-0737 Unidad de Excelencia ``Mar\'{\i}a de Maeztu''- Centro de Astrobiología (INTA-CSIC). This research made use of Astropy,\footnote{http://www.astropy.org} a community-developed core Python package for Astronomy \citep{astropy+2013, astropy+2018}. This research has made use of the SIMBAD database and VizieR catalogue access tool, CDS, Strasbourg, France.  The original description of the VizieR service was published in \cite{Vizier2000}.  This research has made use of the services of the ESO Science Archive Facility, based on data products created from observations collected at the European Organisation for Astronomical Research in the Southern Hemisphere under ESO programmes 077.B-0348, 086.A-9014, 088.A-9007, 077.D-0712, 090.D-0061, 091.D-0414, 082.D-0933, 077.C-0138, 078.A-9059, 084.A-9004, 094.A-9012, 077.C-0573, 083.A-9004, 084.B-0029, 077.A-9005, 082.A-9007, 078.C-0378, 079.A-9017, 060.A-9700, 083.A-9003, 077.C-0192, 079.A-9007, 086.A-9006, 078.D-0080, 085.A-9027, 080.A-9006, 090.A-9013, 089.A-9007, 093.A-9029, 084.A-9003, 082.C-0446, 086.D-0460, 087.C-0476, 090.C-0345, 077.D-0478, 086.A-9007, 090.A-9003, 090.A-9010, 089.D-0709, 077.C-0258, 079.A-9002, 078.A-9048, 077.A-9009, 078.A-9058, 079.A-9006, 092.A-9007, 093.C-0343, 095.C-0437 and 097.C-0444. Funding for the TESS mission is provided by the NASA Explorer Program. Funding for the TESS Asteroseismic Science Operations Centre is provided by the Danish National Research Foundation (Grant agreement no.:DNRF106), ESA PRODEX (PEA 4000119301) and Stellar Astrophysics Centre (SAC) at Aarhus University. This work has made use of data from the European Space Agency (ESA) mission Gaia, processed by the Gaia Data Processing and Analysis Consortium (DPAC). Funding for the DPAC has been provided by national institutions, in particular the institutions participating in the Gaia Multilateral Agreement. Data were obtained from the Mikulski Archive for Space Telescopes (MAST). STScI is operated by the Association of Universities for Research in Astronomy, Inc., under NASA contract NAS5- 2655. This research has made use of the Washington Double Star Catalog maintained at the U.S. Naval Observatory.
\end{acknowledgements}

\bibliography{biblio}

\Online

\begin{appendix}

\section{Notes on individual sources}
\label{sec:individ_sources}
\subsection{Sources flagged variable in this work}
\noindent\textbf{CD-46 644}: This target was flagged due to variation in its $v~\mathrm{sin}~i$ value. The CCF profile is somewhat asymmetric however, the evidence is not strong enough to confirm its spectroscopic binary nature.  Therefore,  it was rejected as a spectroscopic binary. \\ 

\noindent\textbf{HD 17332 A}:  This target has two UVES observations and no significant radial velocity variation.  However, its $v~\mathrm{sin}~i$ value was calculated to be $13$ and $4$\,km~s$^{-1}$ in the two epochs.  Closer inspection of the CCF profile shows that the profile is well fitted. However, given we only have two epochs we cannot conclude whether this change is due to a companion or inherent variability of the star.  Therefore at this time we flag the system as a questionable SB, flagged for further investigation. \\ 

\noindent\textbf{CD-56 1032A}: This target has two UVES observations produ-{cing} radial velocity values of $35.99$ and $27.75$\,km/s$^{-1}$.  The target is a relatively fast rotator ($v~\mathrm{sin}~i\approx$ 40\,km~s$^{-1}$) but the rotational profile is well fitted considering.  Therefore we flagged this target as a spectroscopic binary. \\  


\noindent\textbf{CPD-19 878}: This target shows variation in radial velocity.  However, given we only have four epochs we cannot conclude whether this change is due to a companion or inherent variability of the star.  Therefore at this time we flag the system as a questionable SB, flagged for further investigation. \\ 

\noindent\textbf{TYC 7627-2190-1}: This target shows significant variation in radial velocity from both our observations and those including literature values.  Closer inspection of its CCF profile reveals that it is likely a merged double-lined spectroscopic binary.  \\ 


\noindent\textbf{V*PXVir}: This is a known single-lined spectroscopic binary with an orbital solution ($P=216.48\pm0.06$\,day), presented in \cite{Griffin2010}. In this work, when combined with literature values, the system was flagged as variable. \\ 

\noindent\textbf{HD 159911}: This target was flagged as having high $v~\mathrm{sin}~i$ variation. Despite it has a high $v~\mathrm{sin}~i$ value ($\approx 58$\,km~s$^{-1}$) its CCF profile is well fitted and therefore it is flagged as a potential SB1 system. \\  



\noindent\textbf{CD-43 3604}: Its CCF profile has two clear peaks at different depths and the centre of the single Gaussian fit moves significantly from epoch to epoch. The target's rotational broadening is poorly constrained due to the merged double-peak nature of the profile. This target is likely a merged double-lined spectroscopic binary.  \\ 

\noindent\textbf{V* 379 Vel, TYC 8594-58-1, HD 37484}: These targets were flagged due to variation in its radial velocity when  a literature value was included. Given that the variation come only for one extra epoch, there is not enough evidence to establish the origin of this variation.  Therefore, these targets are rejected as a spectroscopic binary for now. \\ 

\noindent\textbf{2MASS J12203437-7539286}:  This target only has three observations (two presented here, the other from \citealt{Torres2006}).  However, given its low $v~\mathrm{sin}~i$ value ($\approx 8$\,km~s$^{-1}$) the difference in radial velocities ($0.6$ and $4.8$\,km~s$^{-1}$) is significant.    \\ 

\noindent\textbf{HD 129496}: This target was initially flagged as having potentially variable radial velocity, however it has a very high  $v~\mathrm{sin}~i$ value ($\approx 67$\,km~s$^{-1}$).  It's CCF profile is poorly fitted and therefore it is rejected as a spectroscopic binary.  \\ 

\noindent\textbf{CD-52 9381}:    This target has a high $v~\mathrm{sin}~i$ value ($\approx 40$\,km~s$^{-1}$) and was flagged due to radial velocity variation ($\sigma_\mathrm{\subrv} = 2.75$\,km~s$^{-1}$).  A closer inspection of its CCF reveals that the profile is asymmetric however, there are not two distinguishable peaks.  At this time we reject this target as a spectroscopic binary.\\

\noindent\textbf{V*AFLep}: This target was flagged due to variation in its $v~\mathrm{sin}~i$ value from 3 measurements. The CCF profile is somewhat asymmetric however, the evidence is not strong enough to confirm its spectroscopic binary nature.  Therefore,  it was rejected as a spectroscopic binary.\\ 

\noindent\textbf{HD 139084}: This is a known single-lined spectroscopic and close visual binary.  The orbital solution of this system was recently presented in \cite{Nielsen2016}.  The period of the system is 4.576\,yr putting it on the limit of detectability, see Fig.~\ref{fig:sample_detec_prob}.    \\ 

\noindent\textbf{HD 139084 B}: This target is a fast rotator ($v~\mathrm{sin}~i\geq50$\,km~s$^{-1}$) and only has two observations (one presented here and other from \citealt{Torres2006}).  For that reason, there is not enough evidence to establish the origin of the variation.  Therefore, this target is rejected as a spectroscopic binary.  \\ 

\noindent\textbf{HD 164249 B}:  This target was flagged for potential variable $v~\mathrm{sin}~i$ values.  However its CCF profiles are poorly fitted and therefore it was rejected as a spectroscopic binary.  \\ 

\noindent\textbf{CD-31 16041}:  This target was flagged due to variation in its $v~\mathrm{sin}~i$ value from 3 measurements. The CCF profile is somewhat asymmetric however, the evidence is not strong enough to confirm its spectroscopic binary nature.  Therefore,  it was rejected as a spectroscopic binary.\\ 

\noindent\textbf{V*PZTel}:  This target was flagged due to variation in its $v~\mathrm{sin}~i$ value.  However,  it is a very fast rotator ($v~\mathrm{sin}~i$ 64\,km~s$^{-1}$) and its CCF profile is poorly fitted, there it was rejected as a spectroscopic binary. \\ 

\noindent\textbf{HD 191089}: From our measurements alone this target would not be flagged as variable.  However, with the inclusion of literature values it's radial velocity significantly changes.  There are two separate measurements (\cite{Gontcharov2006}: -5.9\,km~s$^{-1}$ and \cite{Desidera2015}: -6.4\,km~s$^{-1}$).  The values calculated from our 3 UVES observations are -12.18, -12.14 and -11.24\,km~s$^{-1}$. In \cite{Grandjean2020} analysis this source was flagged as a variable due to stellar pulsations. Therefore at this time we flag the system as a questionable SB, flagged for further investigation. \\ 


\noindent\textbf{HD 199143}: This target is a fast rotator and has been flagged for both variable $v~\mathrm{sin}~i$ value and radial velocity. The value calculated in this work is $v~\mathrm{sin}~i\approx$58\,km~s$^{-1}$, compared to that of \cite{Torres2006}, 128\,km~s$^{-1}$.  Closer inspection of its CCF shows that our fit of rotational broadening is most likely underestimated due to the velocity span of the CCF fit (-180 -- +180\,km~s$^{-1}$).  Therefore the value of 58\,km~s$^{-1}$ should be treated as a conservative lower limit. Additionally the profile is extremely noisy and poorly fitted by both a Gaussian for its radial velocity value and the rotational broadening profiles.  Given these limitations the system was rejected as a spectroscopic binary.\\ 

\noindent\textbf{*cEri}: This target is a very fast rotator ($v~\mathrm{sin}~i\approx$57\,km~s$^{-1}$).  Additionally, its CCF is very noisy and poorly fitted.  Therefore, it is likely that the apparent radial velocity variation is not physical and the result of a poorly constrained profile.  This system is rejected as a spectroscopic binary.\\ 

\noindent\textbf{GJ 3305}: Given its low $v~\mathrm{sin}~i$ value ($\approx$5\,km~s$^{-1}$) its radial velocity variation ($\sigma_\subrv\approx$1.6\,km~s$^{-1}$) is well above the threshold for identifying it as a spectroscopic binary. \\ 


\noindent\textbf{HD 22213}: This target has two UVES observations producing radial velocity values of 8.13 and 14.41\,km~s$^{-1}$.  The target is a relatively fast rotator ($v~\mathrm{sin}~i\approx$41\,km~s$^{-1}$) but the rotational profile is well fitted considering.  Therefore we flagged this target as a spectroscopic binary.  \\ 

\noindent\textbf{V*AGLep}: This target has three UVES observations and no significant radial velocity variation.  However, its $v~\mathrm{sin}~i$ value was calculated to be $\sim$ 23 and 33\,km~s$^{-1}$ between the three epochs.  Closer inspection of the CCF profile shows that firstly, for a relatively fast rotator the profile is well fitted.  However, the shape changes significantly between the two epochs (the bisector slope, curvature and bisector inverse slope change dramatically).  However, given we only have two epochs we cannot conclude whether this change is due to a companion or inherent variability of the star.  Therefore at this time we flag the system as a questionable SB, flagged for further investigation. \\ 

\noindent\textbf{HD 21997}: This target was flagged as having variable $v~\mathrm{sin}~i$, however, given the associated uncertainty and high $v~\mathrm{sin}~i$ value this variation is not significant. \\ 


\noindent\textbf{CD-44 753}: This target were flagged due to variation in its radial velocity when  a literature value was included. Given that the variation come only for one extra epoch, there is not enough evidence to establish the origin of this variation.  Therefore, this targets is rejected as a spectroscopic binary for the moment. \\ 

\noindent\textbf{HD 104467}: This target was flagged due to significant radial velocity variation.  The $v~\mathrm{sin}~i$ value of the target is $\approx$25\,km~s$^{-1}$, and the profile is well fitted.  Therefore this system is flagged as a spectroscopic binary. \\  

\noindent\textbf{2MASS J12020369-7853012}: This target was flagged due to significant radial velocity variation.  The $v~\mathrm{sin}~i$ value of the target is $\approx$15\,km~s$^{-1}$, and the profile is well fitted.   This target was previously flagged as a single-lined spectroscopic binary in \cite{Elliott2014}.  Therefore this system is flagged as a spectroscopic binary.\\  

\noindent\textbf{BD-20 1111}: We have 3 UVES observations of this target and it has been flagged as having a variable $v~\mathrm{sin}~i$ value.  The shape of the profile significantly changes between 2 epochs resulting in the different $v~\mathrm{sin}~i$ values of 25 and 15\,km~s$^{-1}$.  Given that we only have 3 epochs currently we cannot assess whether this asymmetry is a result of the star's changing surface or of a physically bound companion.  Therefore at this time we flag the target as a questionable SB system.\\ 


\noindent\textbf{CD-66 395}: This target is a very fast rotator ($v~\mathrm{sin}~i\approx$60\,km~s$^{-1}$).  Additionally, its CCF is very noisy and poorly fitted.  Therefore, it is likely that the apparent radial velocity variation is not physical and the result of a poorly constrained profile.  This system is rejected as a spectroscopic binary. \\ 

\noindent\textbf{BD-184452A}: This target only has two  $v~\mathrm{sin}~i$ observations from \citealt{Torres2006} and one RV value from Gaia DR2.  Therefore is not enough evidence yet to establish the origin of the variation. At this time we flag the target as a questionable SB system. \\ 

\noindent\textbf{GSC 08057-00342}: This target has 3 radial velocity values in the literature from \cite{Rodriguez2013}, \cite{Malo2014}, and \cite{Kraus2014}. Given its low $v~\mathrm{sin}~i$ value ($\approx$5\,km~s$^{-1}$) its large radial velocity variation ($\sigma_\subrv\approx$5\,km~s$^{-1}$) is well above the threshold for identifying it as a spectroscopic binary. This object was also independently identified as a SB by \cite{Flagg2020}. \\ 



\noindent\textbf{HD 17250}: This target has 3 RV from UVES observations and 2 from literature \citep{Gontcharov2006,GaiaDR2}. This object is the main star of a quadruple system with two visual companions and was flagged as an SB by \citep{Tokovinin2016}. \\ 

\noindent\textbf{2MASS J04470041-5134405, UCAC3 33-129092, UCAC4 110-129613}: These targets only has two observations (one from GDR2 and other from \citealt{Kraus2014}). There is not enough evidence yet to establish the origin of the variation.  Therefore these target are rejected as a spectroscopic binary.\\ 

\noindent\textbf{CD-53 544}: This target was flagged due to variation in RV and $v~\mathrm{sin}~i$ values. The CCF profile is somewhat asymmetric however the evidence is not strong enough to confirm its spectroscopic binary nature. \\ 


\noindent\textbf{TYC8098-414-1}:  There are 6 available radial velocity measurements for this system.  Five of these six measurements would give an RV $\sim$ 19.60 \,km~s$^{-1}$, which would not be flagged as SB candidate. However, the inclusion of one value from \cite{Kraus2014} of -1.60\,km~s$^{-1}$ makes the apparent variation significant. It is difficult to assess these individual values given the available information.  At this time, the system is flagged as a potential SB for further investigation. \\ 

\noindent\textbf{HD 207575}:  This target shows variation in radial velocity and $v~\mathrm{sin}~i$ value. The CCF profile shows that the shape change between the epochs (the bisector slope, curvature and bisector inverse slope).  However, given we only have five epochs we cannot conclude whether this change is due to a companion or inherent variability of the star.  Recently, \cite{Grandjean2020} flag this source as a variable due to pulsations from HARPS observations. Therefore, this target is rejected as a spectroscopic binary. \\ 

\noindent\textbf{HD 207964}: This targets only has three observations (one from GDR2 and two from our work). Given that there is not enough evidence to establish the origin of the variation.  Therefore, this target is rejected as a spectroscopic binary. \\ 

\noindent\textbf{TYC 9344-293-1}: This object has a variable number of $v~\mathrm{sin}~i$ values. The values are 61\,km~s$^{-1}$ \citep{Torres2006}, 59.5, 65.4 and 67.5\,km~s$^{-1}$ \citep{Malo2014} and 55, 55, and 58\,km~s$^{-1}$ (this work). The most different was the value of 33.1 \,km~s$^{-1}$ published in \cite{Kraus2014}. This system was tagged as a rotational variable but for the moment rejected as potential spectroscopic binary.\\ 

\noindent\textbf{UCAC3 92-4597}: This target was previously flagged as a SB in \citep{Malo2014}. In this work, the system was flagged as a variable using the literature values.  \\ 

\noindent\textbf{CD-30 3394, CD-30 3394B}: These objects was flagged due to RV variation. The CCF profile shows that the shape change between the epochs (the bisector slope, curvature and bisector inverse slope).  However, given we only have four epochs we cannot conclude whether this change is due to a companion or inherent variability of the star. At this time, the systems are flagged as a potential SBs for further investigation. \\ %

\noindent\textbf{HD 3221}: This target is a very fast rotator ($v~\mathrm{sin}~i\geq68$\,km~s$^{-1}$) and its profile is extremely noisy and poorly fitted.  For that reason the radial velocity variation is likely non physical.  Therefore, this target is rejected as a spectroscopic binary.  \\ 

\noindent\textbf{SCRJ0103-5515}: This target was previously flagged as a double or multiple star in WDS. In this work, the system was flagged as a variable using the literature values from \citep{Malo2014} and \citep{Kraus2014}. \\ 


\noindent\textbf{V* CE Ant }: This target was flagged due to variation in its $v~\mathrm{sin}~i$ value from our measurements. The CCF profile is somewhat asymmetric however, the evidence is not strong enough to confirm its spectroscopic binary nature.  Therefore,  it was rejected as a spectroscopic binary. \\ 

\noindent\textbf{TWA23}: This target has 16 individual radial velocity measurements (the majority from \citealt{Bailey2012}) and shows significant radial velocity variation.  Although we only have one observation, from UVES, the profile is consistent as resulting from a merged SB2 system.  There is a significant asymmetry at approximately half the depth of the profile, causing a large bisector slope. Therefore this target is flagged as an SB2 system. \\ 

\noindent\textbf{V* AO Men}: This target was flagged due to variation in its radial velocity when a Gaia DR2 value was included . Given that the variation come only for one extra epoch, there is not enough evidence to establish the origin of this variation.  On the other hand, \cite{Grandjean2020} estimated that the variation was due to stellar activity (spots). Therefore, this target is rejected as a spectroscopic binary. \\ 

\noindent\textbf{HD 984}:  This target was flagged due to variation in its radial velocity when a Gaia DR2 value was included. \cite{Johnson_Groh2017} calculated the orbit of this system as $\sim$ 70\,yr, which is outside outside the region where a visual binary can be detected through radial velocity variation given $\sim$10\,yr measurements. Therefore, although this object is a visual binary it cannot be flagged as a spectroscopic binary. \\ 

\noindent\textbf{2MASS J01505688-5844032, UCAC4 137-000439}:  These targets were flagged due to variation in its radial velocity from two literature values \citep{Kraus2014,GaiaDR2}. \cite{Shan2017} did not find sign of companion from adaptive optics observations conducted on the 6.5 m Magellan Clay Telescope for these objects. UCAC4 137-000439 was noted as potential tight binary in \cite{Janson2017} with an estimated separation of $\sim 0.01\arcsec$. 2MASS J01505688-5844032 is rejected as a spectroscopic binary for the moment and UCAC4 137-000439 is flagged as a potential SB for further investigation. \\ 

\noindent\textbf{2MASS J12560830-6926539}: This target only has two observations (one from \citealt{Torres2006} and another from  Gaia DR2). \cite{Elliott2015} probed binarity in this object by high-resolution
imaging with an estimated angular separation of $0.1\arcsec$, physical separation of 13.1 AU and mass ratio of $0.55$. This object is at the boundaries of the region where a visual binary can be detected through radial velocity variation given $\sim$10\,yr measurements.  At this time we flag the target as a questionable SB system. \\ 

\noindent\textbf{Smethells 165}: This target was previously flagged as a double or multiple star in WDS. In this work, the system was flagged as a variable using the $v~\mathrm{sin}~i$ values from literature. The variation came from one $v~\mathrm{sin}~i$ measurement from \cite{Kraus2014}. At this time we flag the target as a potential SB for further investigation. \\ 



\subsection{Sources previously flagged as spectroscopic multiple systems not recovered in this work}

\noindent\textbf{CD-29 4446}: This is a known binary system with an orbital solution presented in \cite{Rodet2018}. In this work, the system was flagged as a variable using the literature values. \\ 

\noindent\textbf{V* V1005 Ori}:  This target was flagged as an SB1 system in \cite{Elliott2014}. The compilation of further radial velocities do not show significant radial velocity variation caused by a companion.  \\ 

\noindent\textbf{HD 98800A}:  \cite{Torres1995b} calculated the orbit of this SB1 system as 262\,day. In the results presented here we only have 2 radial velocity values which are 4 days apart and therefore did not detect any significant change in velocity.  This is one of the few clear spectroscopic systems missed by our analysis.\\ 

\noindent\textbf{CD-33 7795}: This target is a known triple system with companions at $\approx$0.06\arcsec \citep{Macintosh2001} and 2\arcsec \citep{Webb1999}.  \cite{Konopacky2007} calculated the orbit of the inner system as 5.94$\pm$0.09\,yr, which puts it in the approximate region where a visual binary can be detected through radial velocity variation given $\sim$10\,yr measurements.  However, this object is a fast rotator ($v~\mathrm{sin}~i\approx50$\,km~s$^{-1}$) and only has 2 epochs of radial velocity data which do not show significant variation. Therefore, although this object is a visual binary it cannot be flagged as a spectroscopic binary. \\ 

\noindent\textbf{HD 13183}: This target was flagged as a potential SB1 system in the CORAVEL database \citep{Nordstrom1996}.  Furthermore, \cite{Cutispoto2002} found evidence for significant radial velocity variation.  From our compilation of values this system does not exhibit significant variation given its rotational velocity ($v~\mathrm{sin}~i\approx24$\,km~s$^{-1}$), however it does have an asymmetrical CCF profile.  Given the previous notes in multiple other works this system is flagged as a spectroscopic binary.\\ 


\subsection{Double- and triple-lined spectroscopic binaries}

Double and triple-lined spectroscopic multiple systems can be identified from a single epoch of data, and are essentially confirmed as multiple systems with one detection.  For that reason the notes below on each system are brief, with references to their original discovery where applicable. \\

\noindent\textbf{HD 67945}: This target was flagged as a potential SB2 system in \cite{Torres2006}.  However, given its extremely fast rotation $v~\mathrm{sin}~i\geq58$\,km~s$^{-1}$ and extremely noisy CCF profile we do not find sufficient evidence to confirm that.  Additionally it does not have significant radial velocity variation.  Therefore, it was rejected as a spectroscopic binary.\\ 

\noindent\textbf{HD 155177}   There are 3 individual radial velocity values for this target with uncertainties $<$3\,km~s$^{-1}$, two of which are calculated in this work.  Both the shape ($b_b, c_b$ and $BIS$) and the peak of the CCF profile change significantly in the 2 observations.    Therefore, this system is flagged as a spectroscopic binary. \\  

\noindent\textbf{GSC 06513-00291}: \cite{Malo2014} flag this system as an SB2 and quote values of 12.1, 21.6 and 2.4 for $v~\mathrm{sin}~i$ of this target from three observations.  Interestingly the RV values from the three epochs 22 and 23.9 and 22.8 do not vary significantly.  This target has a companion at $\approx$0.1\arcsec.  Therefore, it is likely an SB3 system. The companion at 0.1\arcsec (3\,au using a trigonometric distance of 29.4\,pc, \citealt{Riedel2014}) would have a period $>$1000\,day. Such a period would not typically induce a large RV difference unless the orbit was extremely eccentric. This system is therefore flagged as an SB3. \\ 

\noindent\textbf{V4046 Sgr}:  This target is a well known SB2 system, the orbital solution was presented in \cite{Quast2000}. We recover both components of this system in all CCF profiles.  \\ 

\noindent\textbf{LP 476-207 A}:   This is a known SB2 system whose orbital solution was presented in \cite{Delfosse1999}.  We recover both components of this system in all CCF profiles. \\  

\noindent\textbf{Barta 161 12}: We do not have our own observations of this target and therefore cannot further investigate the spectroscopic binary-nature of this object with our measurements.  However, \cite{Malo2014} reported this target as an SB2 system. There are multiple radial velocity measurement that show apparent variation, however, it was not recovered in our analysis as the majority of measurements have uncertainties larger than 3\,km~s$^{-1}$.  This target is therefore flagged as a spectroscopic binary.  \\ 

\noindent\textbf{HD 217379A}: This is a previously discovered SB3 system \citep{Elliott2014}.  More recently \cite{Tokovinin2016b} presented an orbital solution for both the inner and outer system.  We recover all three components of this system in our CCF profiles. \\ 


\noindent\textbf{TWA 3A}: This target was flagged as an SB2 system in \cite{Malo2014}  We do not have further observations from UVES, FEROS or HARPS.  However, from our compilation of radial velocities this system has significant radial velocity variation. \\ 

\noindent\textbf{UCAC3 112-6119, UCAC3 92-4597}:  \cite{Kraus2014} flagged these two targets as an SB2 systems.  We do not have further observations from UVES, FEROS or HARPS.  However, from our compilation of radial velocities these systems have significant radial velocity variation. \\ 

\noindent\textbf{HD 309751, HD 33999}:   These two systems were previously reported in \cite{Elliott2014} and recovered in this   analysis.\\ 

\noindent\textbf{HD 36329}:   This SB2 system was previously reported in \cite{Torres2006} recovered in this analysis. \\ 

\noindent\textbf{TYC 8098-414-1}:     \cite{Kraus2014} noted this target as an SB2 system, however, we do not recover the component in our analysis. Most likely the companion is not detected as its flux ratio is to low in our optical spectra. \cite{Malo2014} also noted that their $v~\mathrm{sin}~i$ value did not agree with the literature values and mentioned that this could be an unresolved spectroscopic binary.  Given this information the system is flagged as an SB2 in our analysis. \\ 

\noindent\textbf{HD 199058}: \cite{Chauvin2015} noted this object as a binary or multiple system. In this work we flagged this target as an SB2. \\

\noindent\textbf{TYC 6872-1011-1, BD-20 951, GSC 08077-01788, UCAC3 116-474938, V* V1215 Cen,  HD 36329}: To the best of our knowledge these systems have not previously been reported in the literature.  All are newly discovered SB2 systems. 

\section{Measurements of v~{sin}~i}
\label{sec:vsini_calib}

\subsection{Calibrating using CCF width}
\label{sec:calib_sigma_0}
In the case of slow rotators ($v~\mathrm{sin}~i\lesssim20$\,km~s$^{-1}$) there is a significant contribution to the width ($\sigma_\mathrm{obs}$) of the cross correlation function (CCF) from non-rotation related broadening mechanisms which can either be inherent to the star (effective temperature and turbulence) or from the instrument that is used for the observation. The width of the CCF profile is described by:

\begin{equation}
\sigma_\mathrm{obs}^2 = \sigma_\mathrm{rot}^2 - \sigma_\mathrm{0}^2
\end{equation}

where $\sigma_\mathrm{obs}$ is the width of the resultant CCF profile, $\sigma_\mathrm{rot}$ is the rotational broadening of the star and $\sigma_\mathrm{0}$ is the width of a non-rotating star, which can be very well expressed as a function of colour.

Beyond $\approx$20\,km~s$^{-1}$ the width of the CCF profile is dominated by the rotation of the star and therefore these effects become small or negligible.  Note that within our sample of objects there are very few measurements with FEROS or HARPS with $v~\mathrm{sin}~i$ values $\geq$20\,km~s$^{-1}$.

The $v~\mathrm{sin}~i$ value can be expressed as \citep{Queloz1998}:

\begin{equation}
\label{eq:vsini}
v~\mathrm{sin}~i = A \sqrt{\sigma_\mathrm{obs}^2 - \sigma_\mathrm{0}^2}
\end{equation}

where $A$ is the coupling constant, calibrating one set of CCF measurements to previously calibrated $v~\mathrm{sin}~i$ values.

\begin{figure}[h]
\begin{center}
\includegraphics[width=0.499\textwidth]{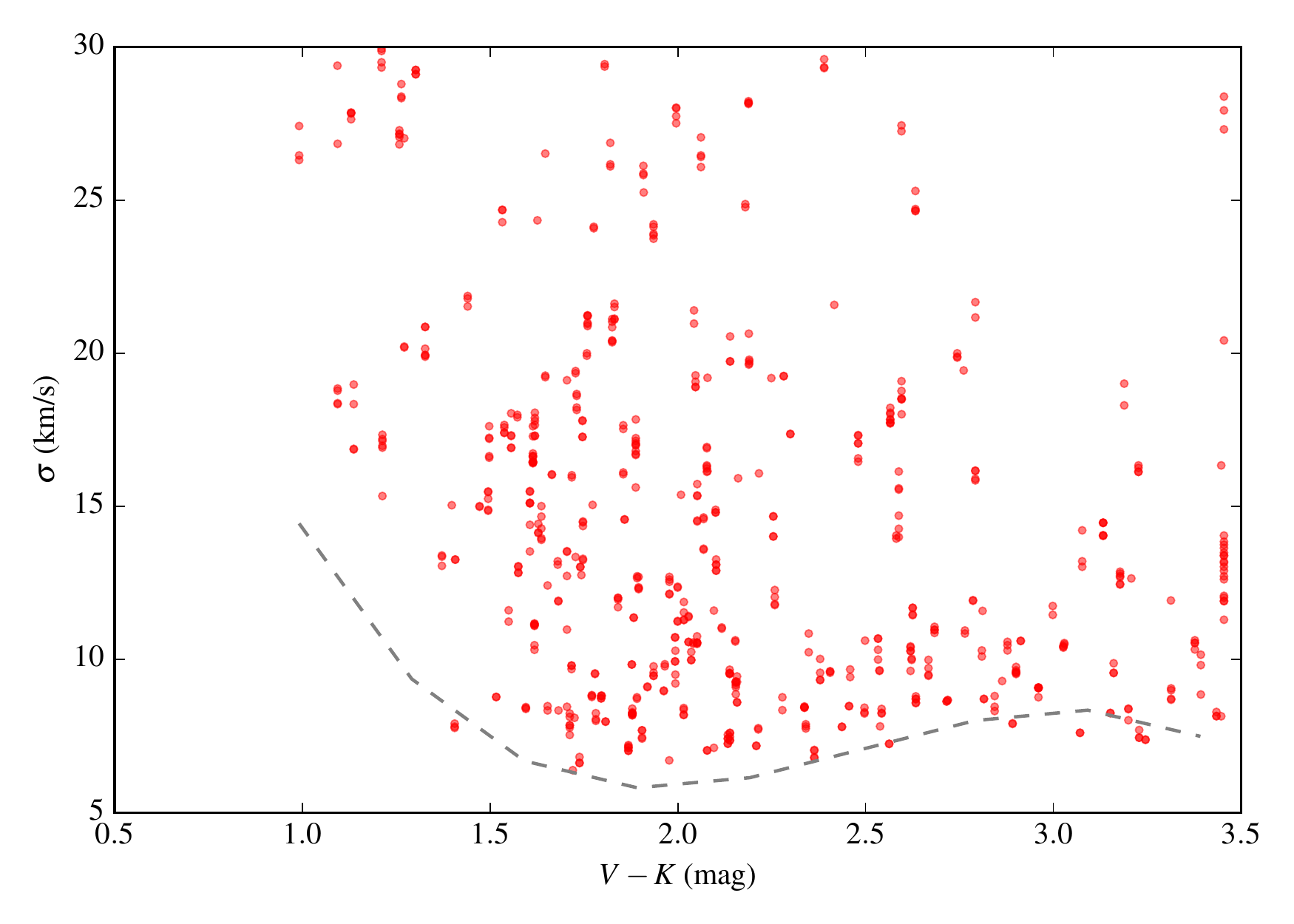}
\vspace{-0.3cm}
\caption{$V - K$ colour versus $\sigma$ (the observed width of the CCF profile) for all individual UVES observations.  The dotted line represents a polynomial fitted to the lower envelope of these measurements.}
\label{fig:uves_sigma_obs}
\end{center}
\end{figure}

First, to determine the value of $\sigma_\mathrm{0}$ we computed the lower envelope of points in a $V - K$ versus $\sigma_\mathrm{obs}$ diagram, see Fig.~\ref{fig:uves_sigma_obs} for an example using UVES observations.  The envelope was fitted with a polynomial and is shown as the dotted line.  This is similar to the technique used in \cite{Melo2001} and \cite{Boisse2010}. We used this $\sigma_0$ value for each star with its respective $V - K$ colour and found the slope (and offset) between published $v~\mathrm{sin}~i$ values and our calculated {\small $A \sqrt{\sigma_\mathrm{obs}^2 - \sigma_\mathrm{0}^2}$} values.  Note that in this analysis we used CCF profiles with low fit residuals in order to better constrain the results.

\begin{figure}[h]
\begin{center}
\includegraphics[width=0.499\textwidth]{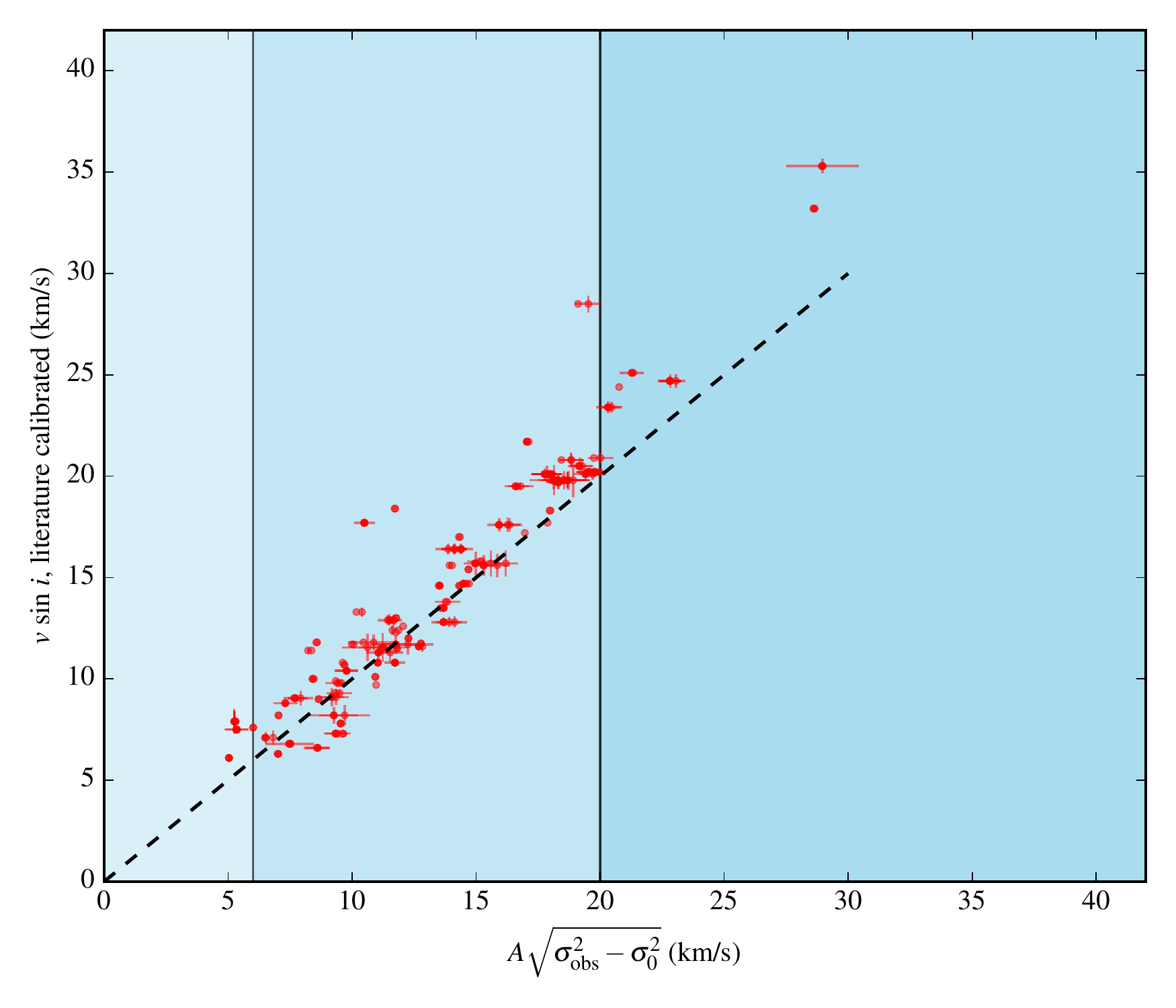}
\vspace{-0.1cm}
\caption{{\small $A \sqrt{\sigma_\mathrm{obs}^2 - \sigma_\mathrm{0}^2}$} versus  literature $v~\mathrm{sin}~i$ values for UVES observations. Three regions are highlighted.  From left to right: Our lower limit on reliable  $v~\mathrm{sin}~i$ values (6\,km~s$^{-1}$), the intermediate range (6-20\,km~s$^{-1}$) where the 1:1 relation should hold and the fast rotator range ($> 20$\,km~s$^{-1}$).  The dotted line represents the 1:1 relation between the two sets of values.}
\label{fig:uves_sigma_0}
\end{center}
\end{figure}

 Figure~\ref{fig:uves_sigma_0} shows the resultant relation for observations using UVES.  We have highlighted 3 regions of the Fig. to guide the reader's eye.  Below $\approx$6\,km~s$^{-1}$, in the case of UVES, $\sigma_0\approx\sigma_\mathrm{obs}$ and therefore this is our reliable lower limit on  $v~\mathrm{sin}~i$ values.  Between $\approx$6-20\,km~s$^{-1}$ the 1:1 linear relation sufficiently describes the majority of our data.

Figure~\ref{fig:uves_sigma_0} shows that, at least in the case of UVES observations this calibration is relatively successful as the literature  $v~\mathrm{sin}~i$ values match the {\small $A \sqrt{\sigma_\mathrm{obs}^2 - \sigma_\mathrm{0}^2}$} value.  However, in the case of FEROS and HARPS we were unable to perform the same analysis successfully.  Due to the smaller number of objects an accurate calculation of $\sigma_0$ was severely inhibited.  With this in mind, below we outline an alternative approach to  $v~\mathrm{sin}~i$ calculation.

\subsection{Calibrating using rotational profiles}
\label{sec:rotational_calib}
\begin{figure}[h]
\begin{center}
\includegraphics[width=0.499\textwidth]{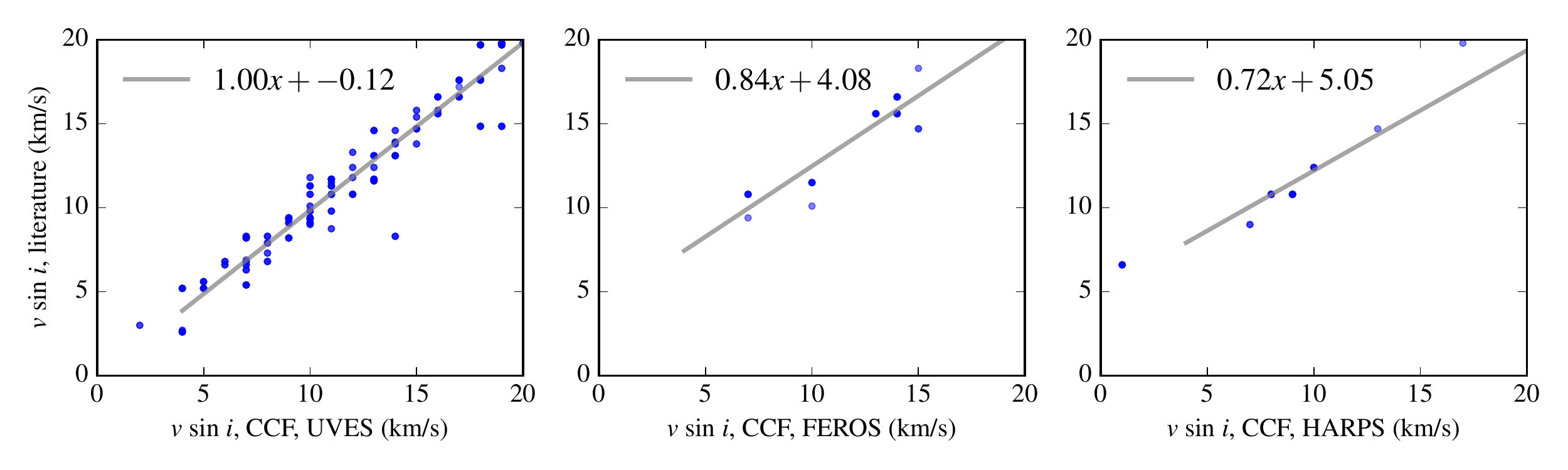}
\vspace{-0.2cm}
\caption{$v~\mathrm{sin}~i$ values from fitted rotational profiles versus literature  $v~\mathrm{sin}~i$  values.  The left, middle and right panels show measurements for UVES, FEROS and HARPS observations.  The linear relation ($y=mx+c$) is shown for each set of measurements.}
\label{fig:all_instruments_calib_vsini}
\end{center}
\end{figure}

We directly compared our calculated values using rotational profiles to published values. We used $v~\mathrm{sin}~i$ with published uncertainties $< 3$\,km~s$^{-1}$ in this analysis.  Figure~\ref{fig:all_instruments_calib_vsini} shows the results for UVES, FEROS and HARPS in the left, middle and right panels, respectively.  A linear relation ($y=mx+c$) was fitted to each set of points and was used to calibrate our values.

\begin{figure}[h]
\begin{center}
\includegraphics[width=0.499\textwidth]{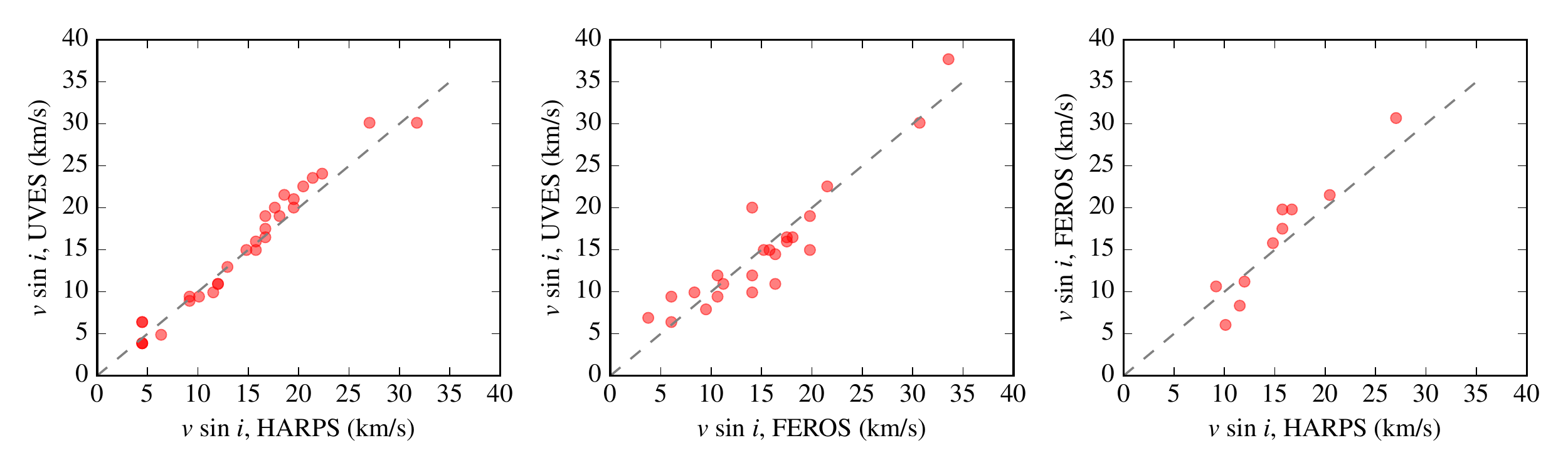}
\vspace{-0.3cm}
\caption{$v~\mathrm{sin}~i$ values calculated in this work for each pair of instruments.  Left, middle and right panels are HARPS versus UVES, FEROS versus UVES and HARPS versus FEROS, respectively.  The 1:1 relation in each case is plotted as the dotted line.}
\label{fig:instrument_comparison}
\end{center}
\end{figure}

To verify this relationship we performed an internal check by comparing $v~\mathrm{sin}~i$ values for objects that were observed with at least 2 of the 3 instruments. Figure~\ref{fig:instrument_comparison} shows the results of this comparison for each pair of instruments.  Given typical uncertainties on $v~\mathrm{sin}~i$  values are 1-2\,km~s$^{-1}$ \citep{Melo2001, Malo2014} the resultant 1:1 relationships adequately describe our data.  The advantage of this calibration technique is that the linear relation can be applied to all stars in our sample.  However, in the case of the technique described in Section~\ref{sec:calib_sigma_0}, a $V - K$ value is needed and some stars in our sample do not have reliable $V$ magnitudes.  Additionally, our stars cover the age range $\approx$5-150\,Myr and therefore can be at very different evolutionary stages, which could hinder a robust $\sigma_0$ calculation.


\subsection{{\it v}~sin~{\it i} lower limit }

From our calibration of $v~\mathrm{sin}~i$ values described in the previous section we arrive at lower limits of 0.83, 4.47 and 8.36\,km~s$^{-1}$ using a star rotating with a projected rotational velocity of 1\,km~s$^{-1}$ for UVES, HARPS and FEROS, respectively. However, as highlighted in Section~\ref{sec:calib_sigma_0}, a more realistic lower limit on $v~\mathrm{sin}~i$ values for UVES is 6\,km~s$^{-1}$, where $\sigma_0 \approx \sigma_\mathrm{obs}$.

\subsection{Limitations on {\it v}~sin~{\it i} measurements of extremely fast rotators}
\label{sec:vsini_fast_rotators}

In the case of very large rotational broadening ($v~\mathrm{sin}~i\geq$60\,km~s$^{-1}$), some stars' $v~\mathrm{sin}~i$ values can be underestimated.  This is due to the width of the profile approaching the width of the velocity span used in the CCF calculation.  This causes a lack of continuum and when the profile is fitted the outer wings of the profile can be wrongly ignored.  For fast rotators in our sample ($v~\mathrm{sin}~i\geq$50\,km~s$^{-1}$) we reran our CCF calculation using a wider velocity window of -250 to +250\,km~s$^{-1}$. Even with this broader window some star's CCF profile widths were still underestimated.  In these cases we use our calculated value as a lower limit.

\subsection{Measurement uncertainties on {\it v}~sin~{\it i} values}
\label{sec:vsini_uncertainties}

\begin{figure}[h]
\begin{center}
\includegraphics[width=0.49\textwidth]{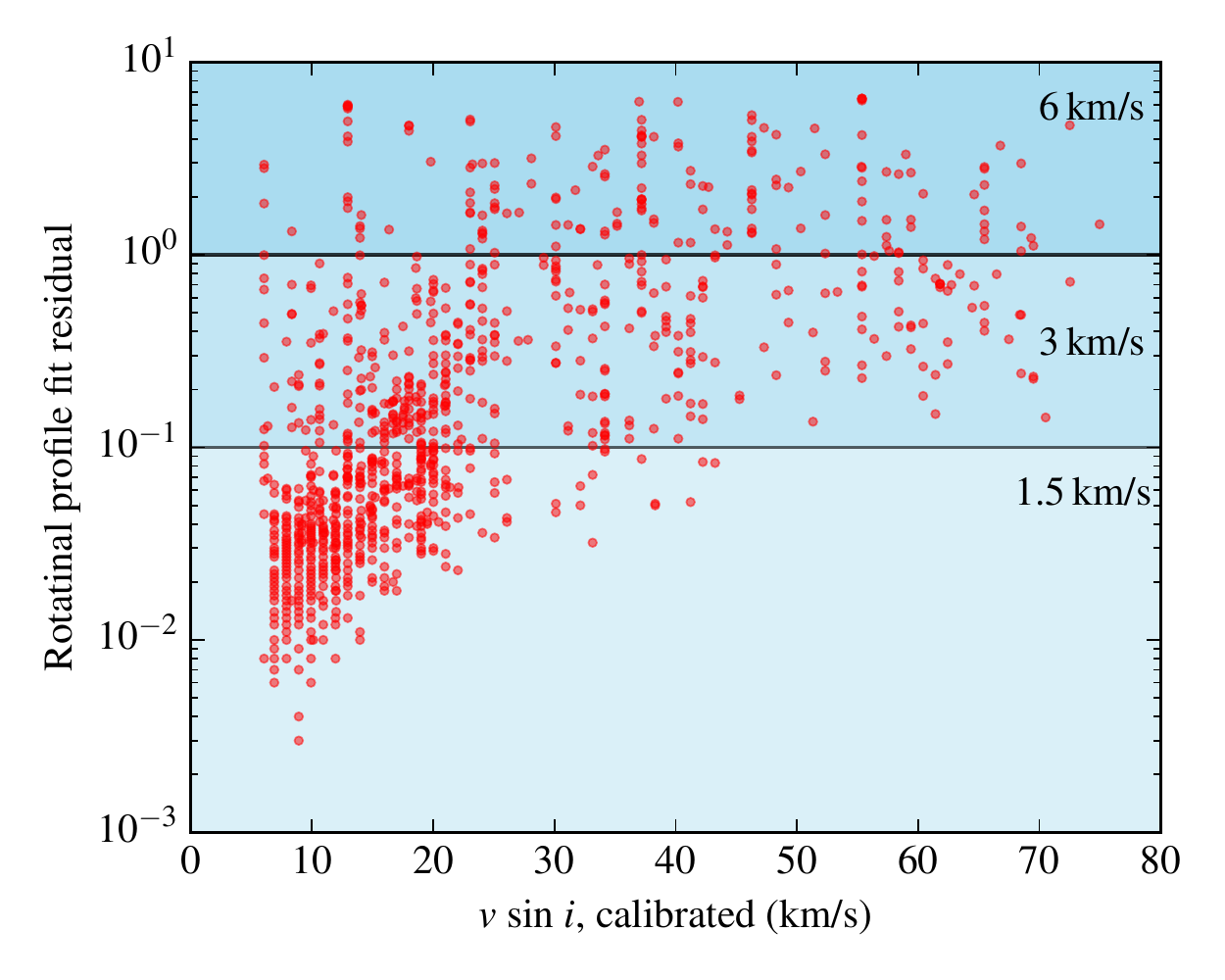}
\vspace{-0.7cm}
\caption{The rotational profile fit residual as a function of calibrated $v~\mathrm{sin}~i$ values. The $v~\mathrm{sin}~i$ uncertainties value is defined depending on the range of fit residual values.}
\label{fig:vsini_uncertainties}
\end{center}
\end{figure}

We compared our calibrated $v~\mathrm{sin}~i$ values with the fitted linear relation (see Section \ref{sec:rotational_calib}) and calculate the quadratic sum of the error as tracer of uncertainties. We set three uncertainties values based on three order of magnitude from residuals. These values were selected from the mean uncertainty value from the errors between the calibrated $v~\mathrm{sin}~i$ and the fitted linear relation on each range of profile fit residuals (see Fig. \ref{fig:vsini_uncertainties}).

\section{Sensitivity maps}
\label{sec:sensitivy_appendix}
Average detection probability maps (contours from red, 100$\%$, to white, 0$\%$) computed for the population of binaries described in Sec.~\ref{sec:sensitivity}. Detected spectroscopic companions (white stars) and visual binaries (black crosses) in the physical separation versus mass ratio. The solid, dashed and dash-dotted lines encompass areas with detection probabilities $\geq$ 90$\%$, 50$\%$ and 10 $\%$, respectively. For THA and BPC association see Fig. \ref{fig:sample_detec_prob}

\begin{figure}[H]
\begin{center}
\includegraphics[width=0.45\textwidth]{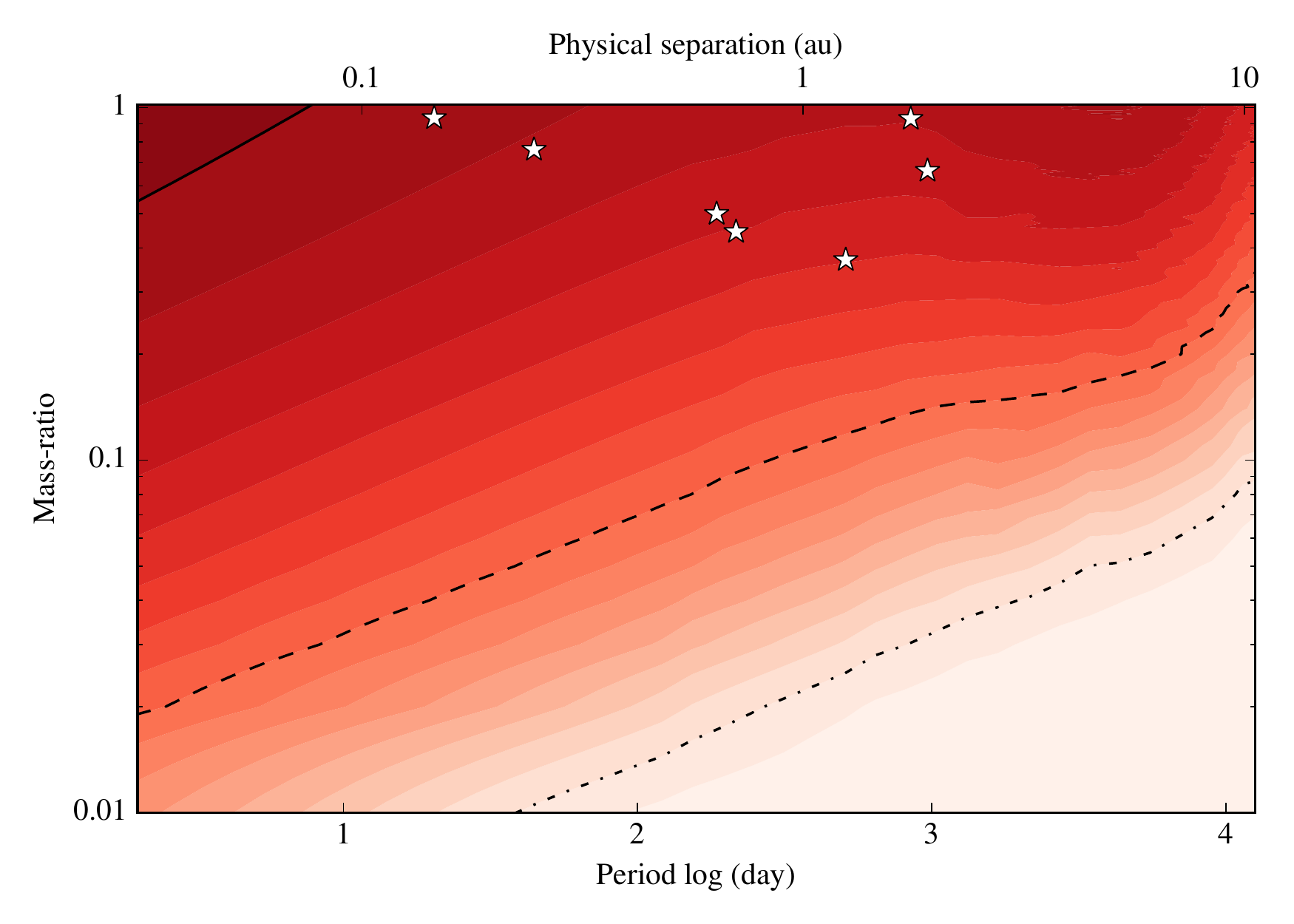}
\vspace{-0.4cm}
\caption{Average detection probabilities for ABD association.}
\end{center}
\end{figure}
\vspace{-0.4cm}
\begin{figure}[H]
\begin{center}
\includegraphics[width=0.45\textwidth]{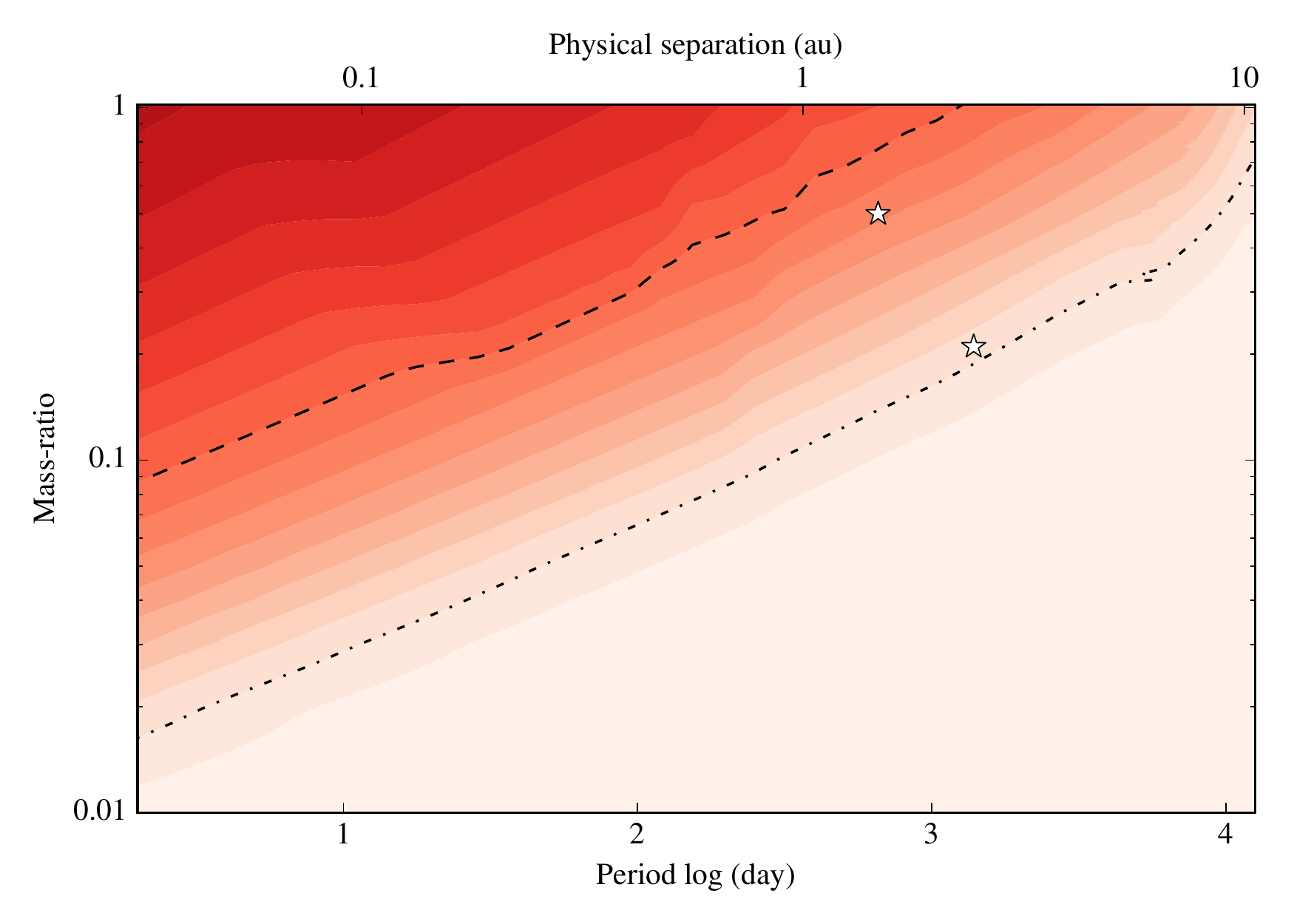}
\vspace{-0.4cm}
\caption{Average detection probabilities for ARG association.}
\end{center}
\end{figure}
\vspace{-0.4cm}
\begin{figure}[H]
\begin{center}
\includegraphics[width=0.45\textwidth]{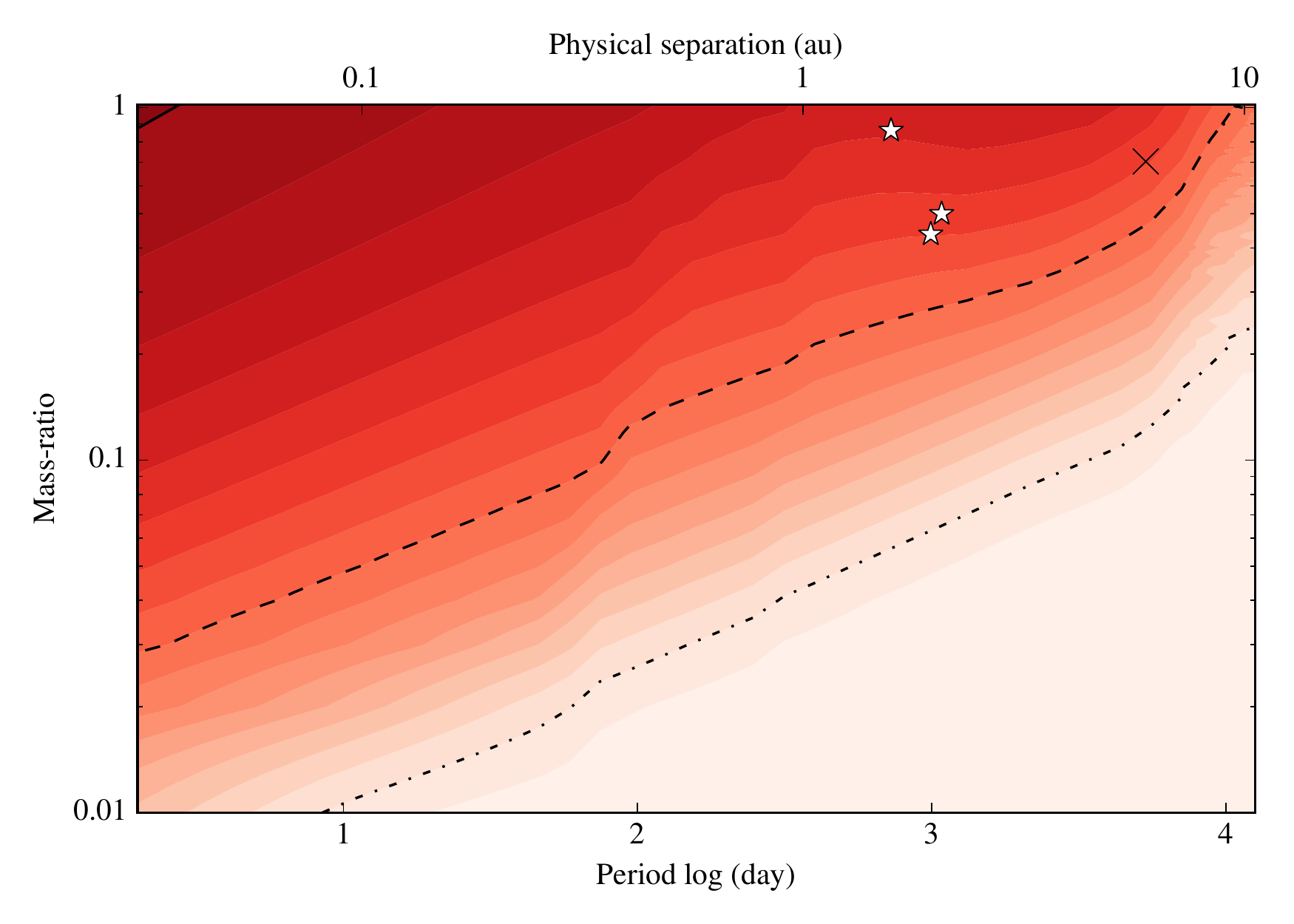}
\vspace{-0.4cm}
\caption{Average detection probabilities for COL association.}
\end{center}
\end{figure}

\begin{figure}[H]
\begin{center}
\includegraphics[width=0.49\textwidth]{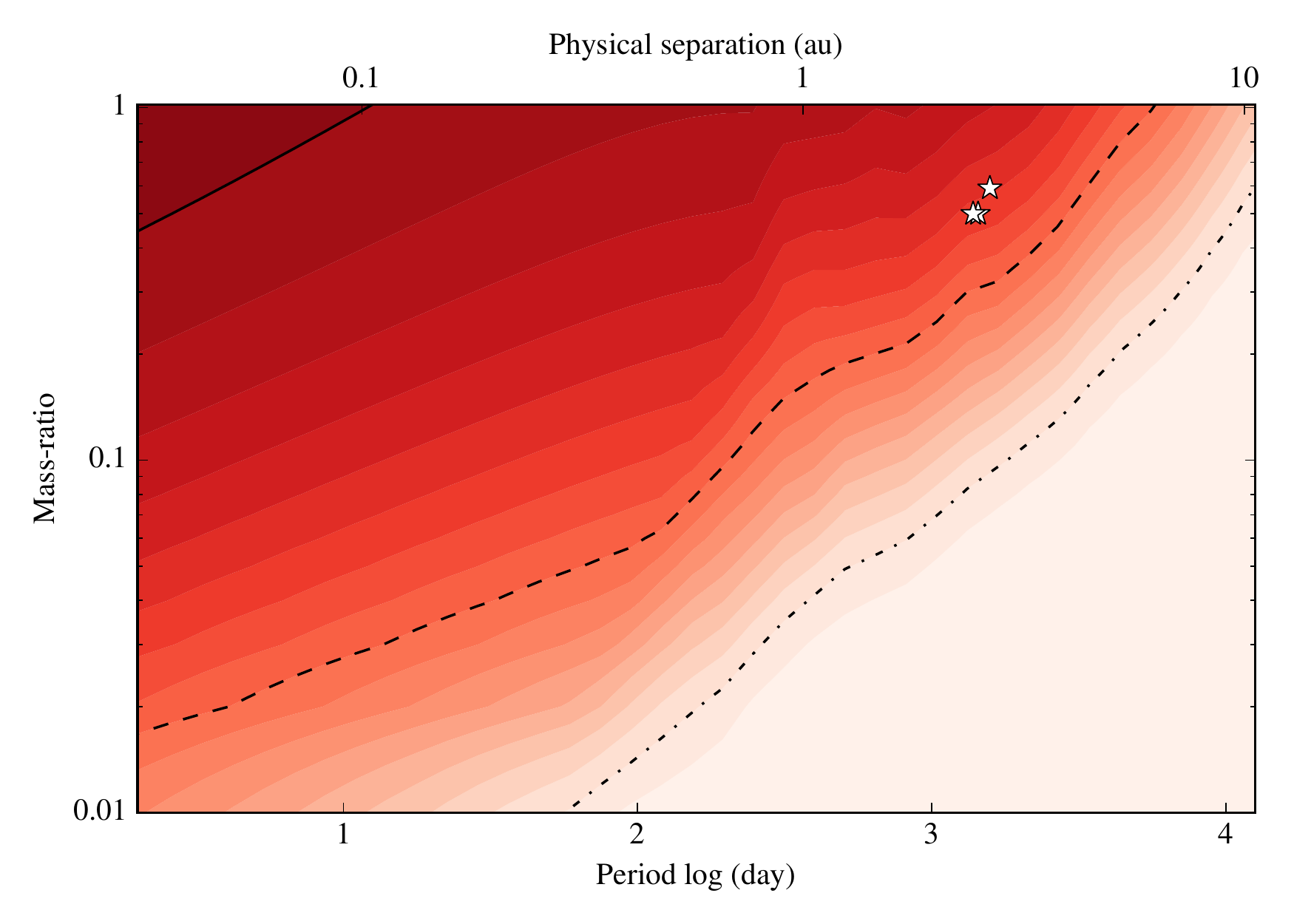}
\vspace{-0.4cm}
\caption{Average detection probabilities for ECH association.}
\end{center}
\end{figure}

\begin{figure}[H]
\begin{center}
\includegraphics[width=0.49\textwidth]{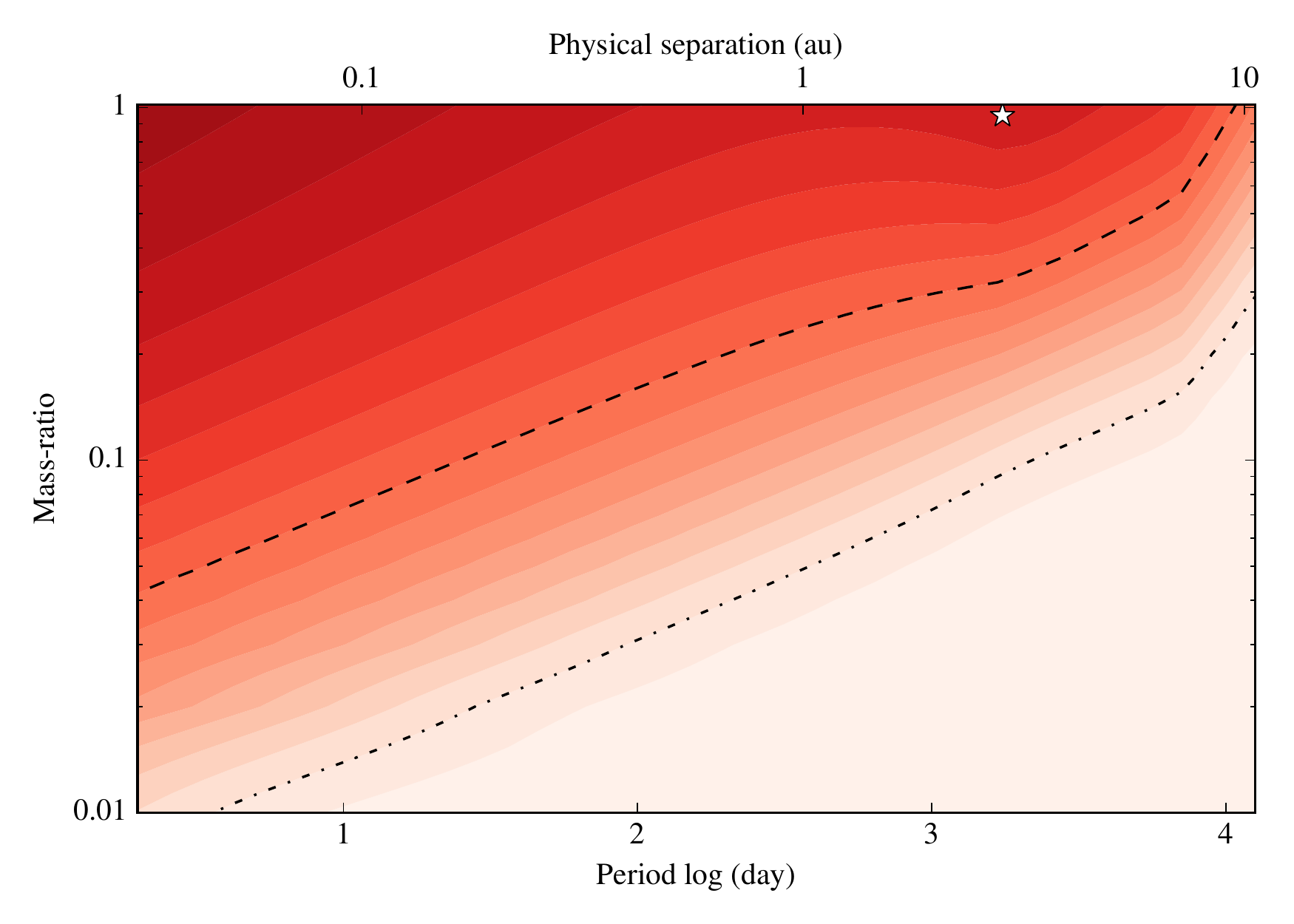}
\vspace{-0.4cm}
\caption{Average detection probabilities for OCT association.}
\end{center}
\end{figure}

\begin{figure}[H]
\begin{center}
\includegraphics[width=0.49\textwidth]{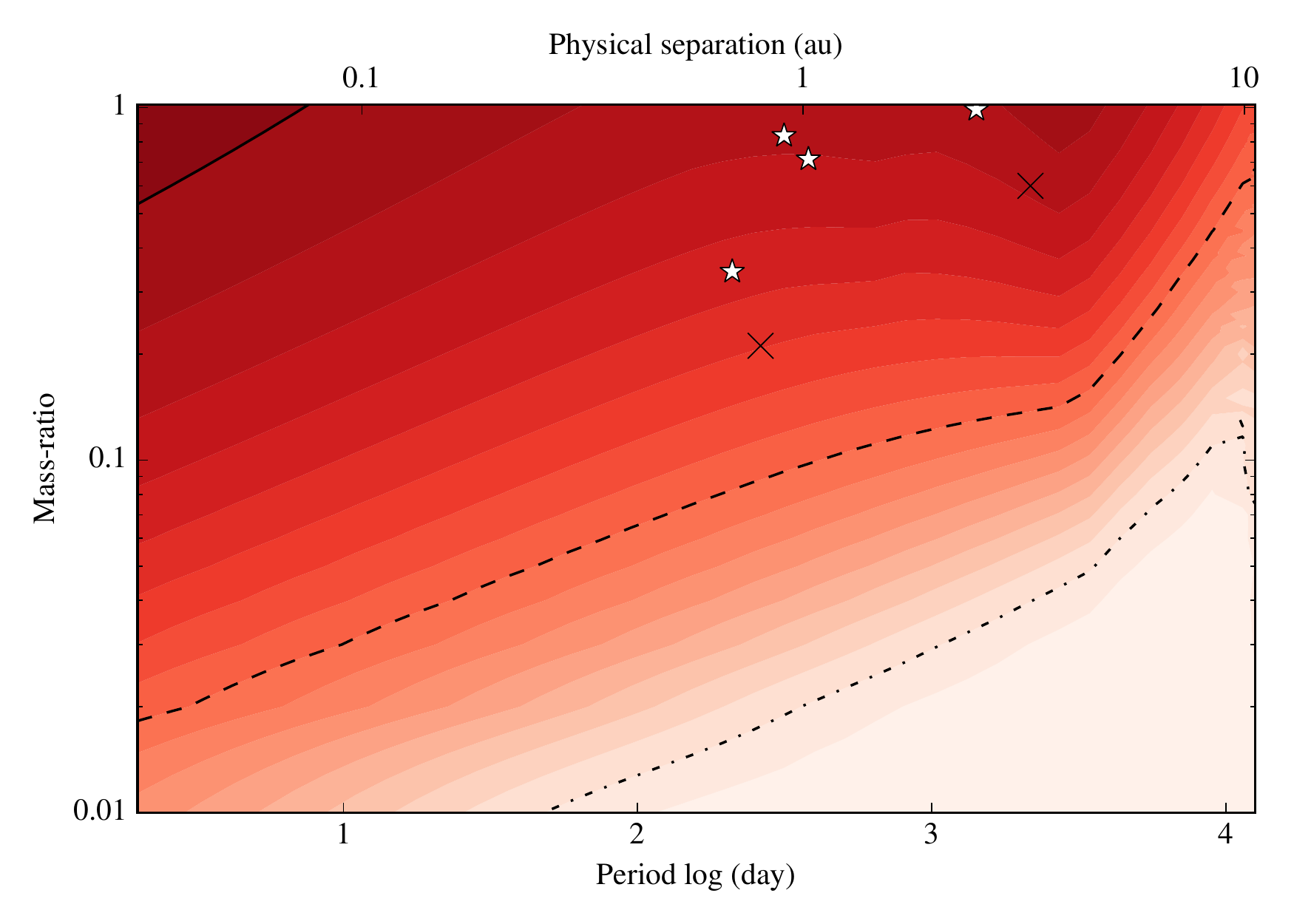}
\vspace{-0.4cm}
\caption{Average detection probabilities for TWA association.}
\end{center}
\end{figure}

\section{SB1 systems identified in this work}
\begin{figure}[h]
\begin{center}
\includegraphics[width=0.49\textwidth]{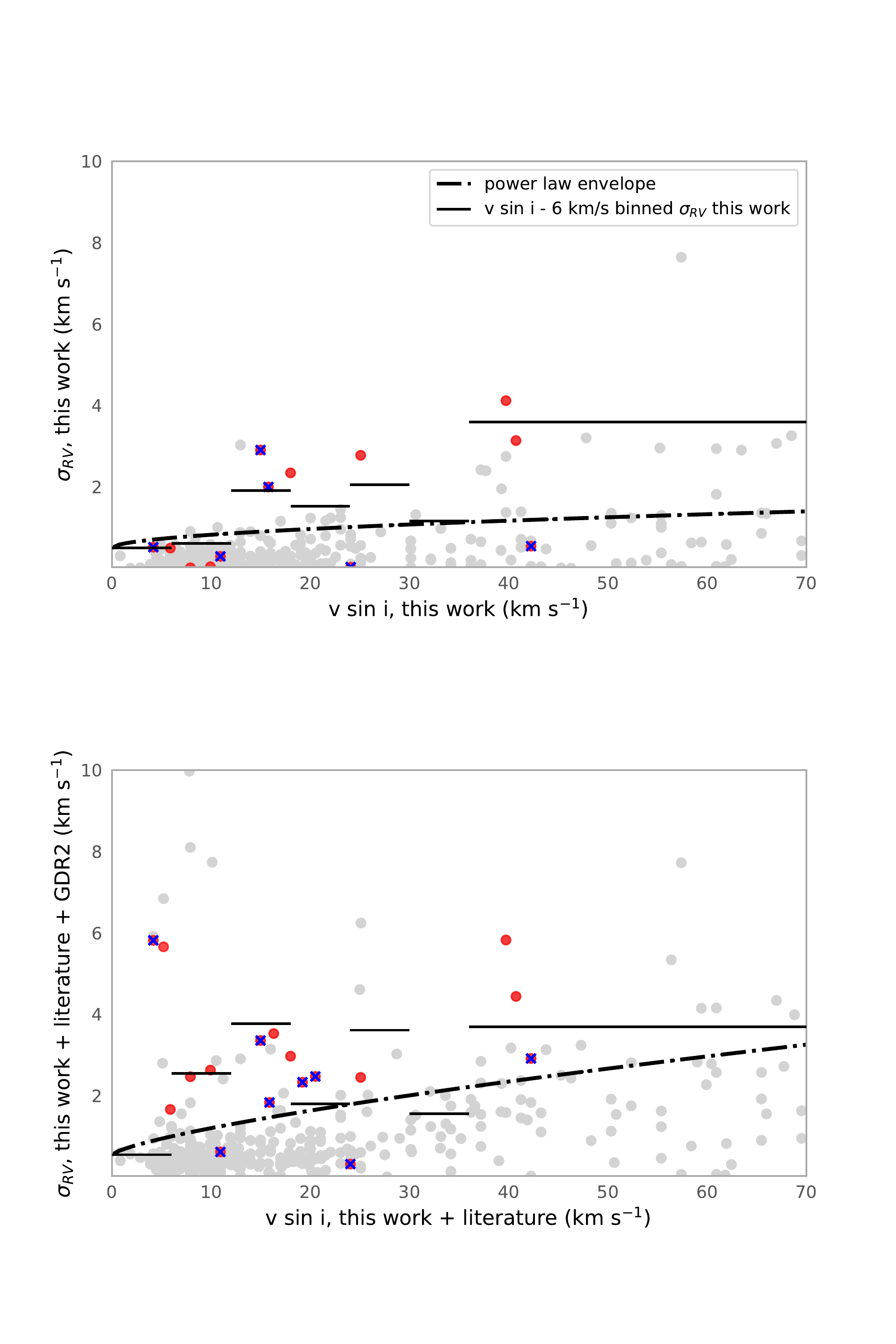}
\vspace{-0.2cm}
\caption{{\it Upper panel}: The standard deviation in RV as a function of  $v~\mathrm{sin}~i$ for measurements calculated in this work. The 3\,$\sigma$ value from binning in $6$\,km\,s$^{-1}$ bins are represented by the solid lines. The power law envelope is represented by dash-dotted line. The SB1s identified in this work are plotted as a red dots and the previously identified SB1s from literature are represented as a blue crosses. {\it Bottom panel}: Same as upper panel but including values from literature and Gaia DR2. Some SB1 were confirmed only when literature values were included (red dots under the $3\sigma$ envelope in upper panel). Details on each candidate can be found in Appendix \ref{sec:individ_sources}.}
\label{fig:appendix_rv_std_vs_vsini}
\end{center}
\end{figure}

\newpage
\section{Gaia DR2}

Example of the sanity checks performed regarding the correct identification of the Gaia DR2 counterparts to the SACY members.

\begin{figure}[H]
\begin{center}
\includegraphics[width=0.49\textwidth]{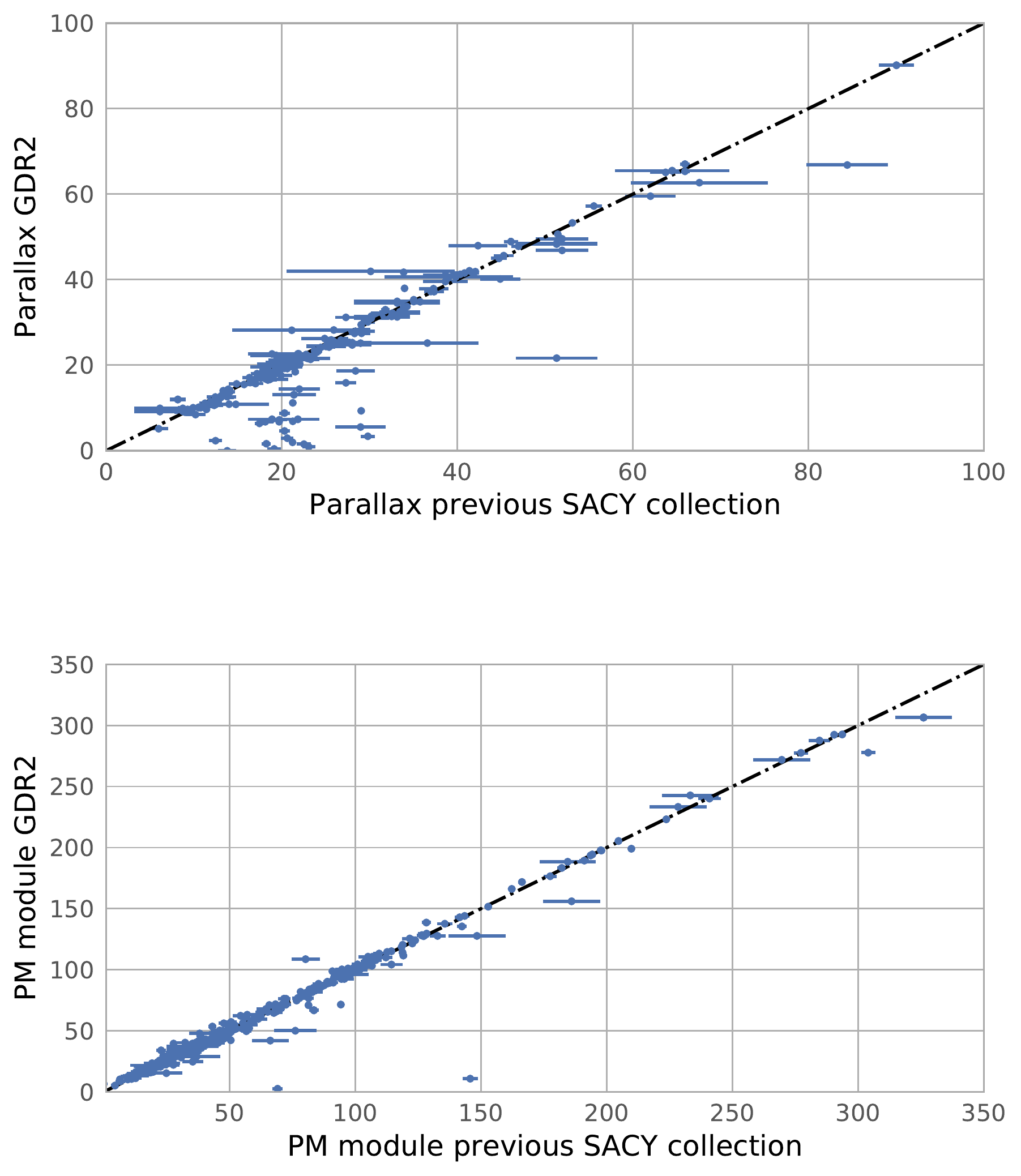}
\vspace{-0.2cm}
\caption{Possible mismatched results was visually inspected and crosschecked to avoid false positives. The dotted-dashed line represent the 1:1 relation.}
\label{fig:GDR2appendix}
\end{center}
\end{figure}

\section{Rotational periods from light curves}
\label{sec:LC_qflag}

\begin{figure}[h]
\begin{center}
\includegraphics[width=0.49\textwidth]{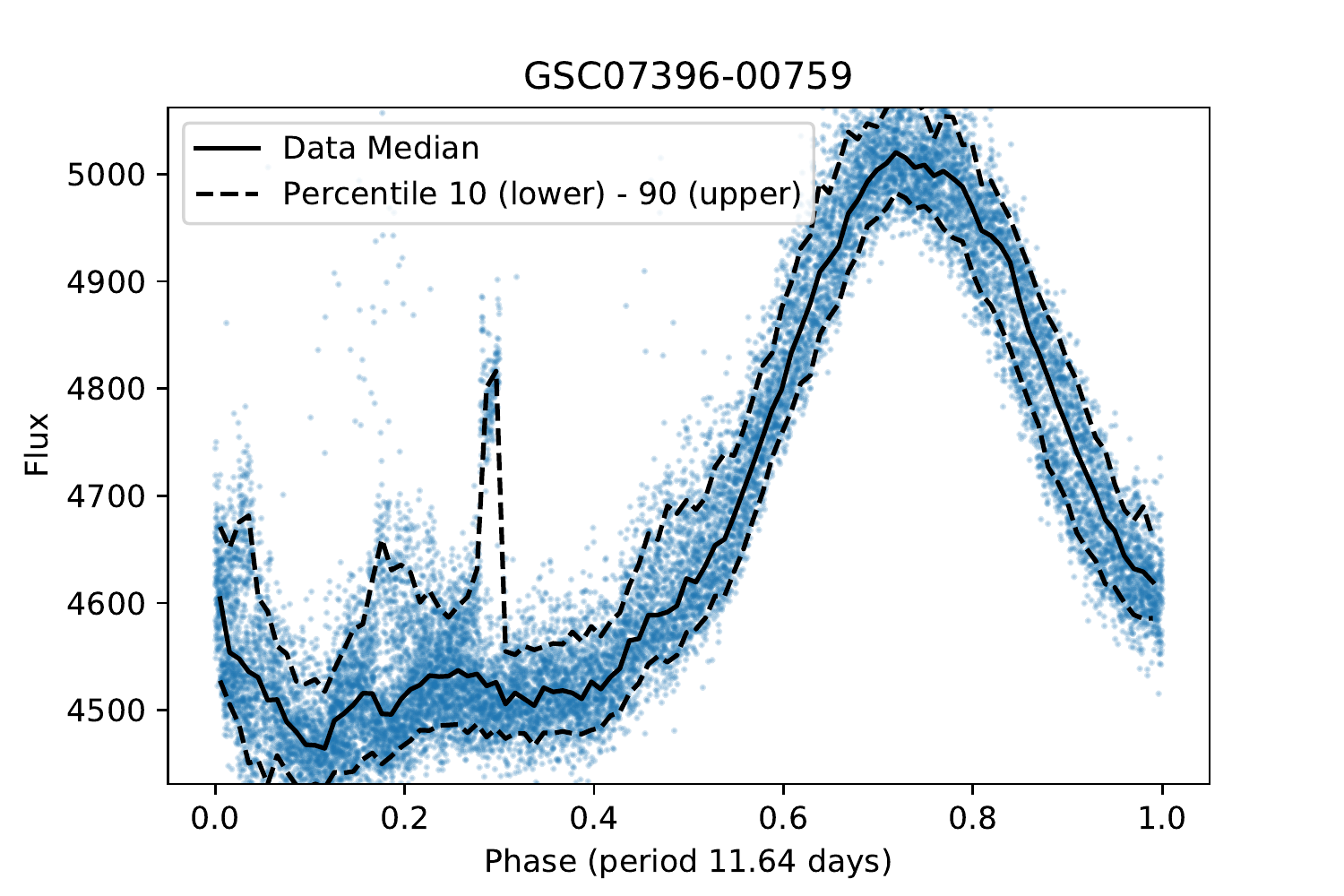}
\includegraphics[width=0.49\textwidth]{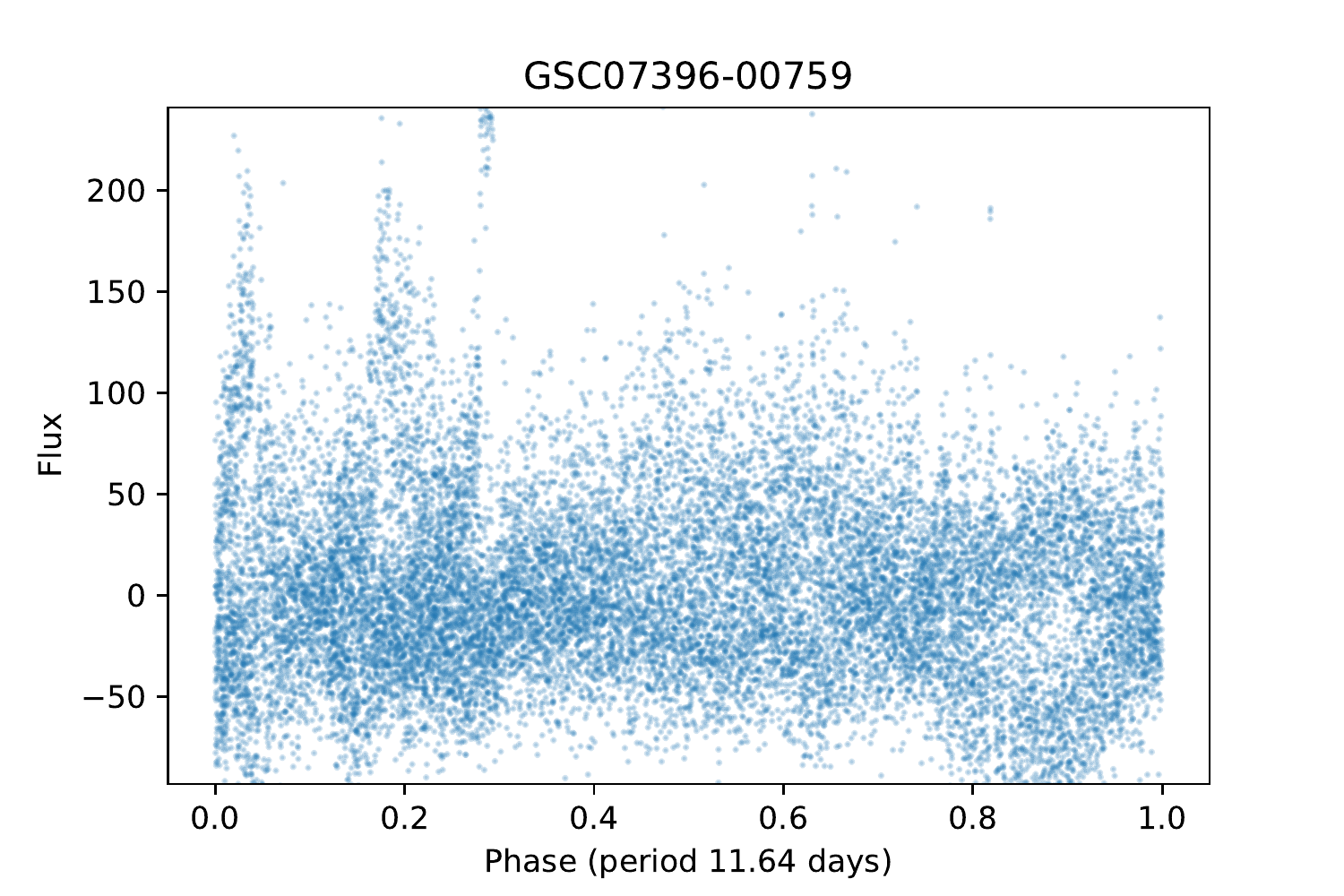}
\vspace{-0.4cm}
\caption{\textit{Upper panel:} Phased light curve for GSC 07396-00759. The solid line represent the median calculated by binning the phased curve in $100$ bins. The MAD for the phased curve for this object is $652.67$. \textit{Bottom panel:} Residuals from subtracting light curve values from the ``median model" (solid line). The MAD of the residuals is $121.84$.}
\label{fig:LC_residuals}
\end{center}
\end{figure}

In Fig.~\ref{fig:LC_residuals} we show an example of the TESS light curve folded to the period estimated in this work for GSC 07396-00759. The lower panel shows the residuals obtained after subtraction of the binned / smoothed phased light curve to be used to asses the reliability of the period. We can see that despite possible flares in the data-set, our procedure offers a simple but robust diagnostic.

\begin{figure}[H]
\begin{center}
\includegraphics[width=0.42\textwidth]{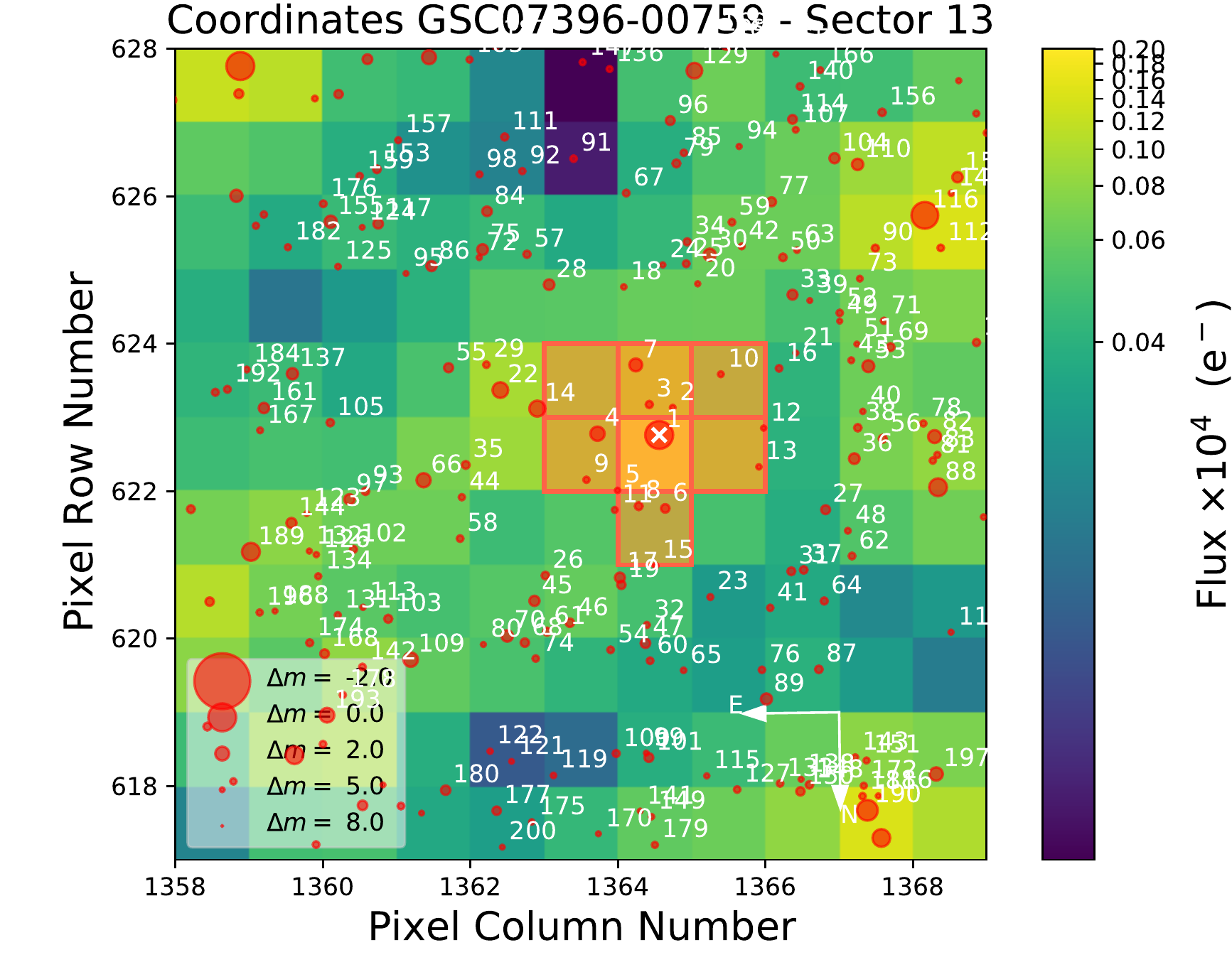}
\vspace{-0.3cm}
\caption{Output figure for GSC 07396-00759 from the package \texttt{tpfplotter} \citep{Aller2020}. We count the number of Gaia sources within a $\Delta$G mag $\leq 5$ of the science target that fall in the pipeline aperture of TESS and save the minimum $\Delta\mathrm{Gmag}$ value to assess the quality of the rotational period.}
\label{fig:LC_contaminats}
\end{center}
\end{figure}

On the other hand, as it is evident from Fig.~\ref{fig:LC_contaminats}, the aperture used to derive the TESS light curve is contaminated by similar brightness objects and therefore, we cannot assure that the reported value is the rotational period of this particular source.

\newpage
\onecolumn{
\section{Individual and summary tables}
\label{appendix:rv_sini_collect}

\begin{table}[H]
\caption{Table of all individual radial velocity values calculated in this work and compiled from literature/Gaia DR2 (first 10 rows). The full table (2048 RV values) is published online in the machine-readable format. The high order features (BIS, $b_b$, $c_b$) are available for all our CCF calculations. The reference code in \textit{Ref.} column correspond to: ZF20: this work or updated value of \cite{Elliott2014}, SC12: \cite{Schlieder2012}, SH12: \cite{Shkolnik2012}, TO06: \cite{Torres2006}, LO06: \cite{Lopez2006}, RO13: \cite{Rodriguez2013}, MA10: \cite{Maldonado2010}, MO13: \cite{Moor2013}, RE09: \cite{Reiners2009}, GO06: \cite{Gontcharov2006}, MA14: \cite{Malo2014}, KR14: \cite{Kraus2014}, MO01b: \cite{Montes2001b}, MO02: \cite{Mochnacki2002}, BA12: \cite{Bailey2012}, DE15: \cite{Desidera2015} and GDR2: Gaia DR2, \cite{GaiaDR2}. The MJD and instrument information is not available for all rows in the table, more details in Sec. \ref{sec:sample}.}
\centering
\tiny{
\begin{tabular}{lllrrlrrrll}
\toprule
 SIMBAD ID & RA J2000 (deg) & DEC J2000 (deg) &     RV &    RV err &      MJD &    BIS &     $b_b$ &    $c_b$ & Instrument &   Ref. \\
\midrule
 BD-202977 &     144.964005 &      -21.571400 &  18.87 &  0.532750 &    53906 & -0.404 &  -8.353 & -0.150 &      FEROS &  ZF20 \\
 BD-202977 &     144.964005 &      -21.571400 &  17.73 &  0.532750 &  54240.1 & -0.103 &  -2.283 & -0.138 &       UVES &  ZF20 \\
 BD-202977 &     144.964005 &      -21.571400 &  17.75 &  0.532750 &  54240.1 & -0.089 &  -1.893 & -0.133 &       UVES &  ZF20 \\
   HD99827 &     171.324005 &      -84.954399 &  20.01 &  1.390170 &  54906.3 & -0.430 & -73.848 & -0.676 &       UVES &  ZF20 \\
   HD99827 &     171.324005 &      -84.954399 &  19.94 &  1.390170 &  54906.3 & -0.747 & -60.319 & -0.830 &       UVES &  ZF20 \\
   HD99827 &     171.324005 &      -84.954399 &  16.30 &  1.390170 &  55371.1 &  0.894 &   7.246 & -1.184 &       UVES &  ZF20 \\
   HD99827 &     171.324005 &      -84.954399 &  19.42 &  1.390170 &  56734.3 &  1.609 &  81.204 & -0.746 &       UVES &  ZF20 \\
   HD99827 &     171.324005 &      -84.954399 &  19.55 &  1.390170 &  56748.1 &  0.810 &   9.514 & -2.268 &       UVES &  ZF20 \\
 CD-691055 &     194.606995 &      -70.480301 &  13.70 &  0.894763 &    54577 & -1.860 & -90.714 &  4.322 &      FEROS &  ZF20 \\
 CD-691055 &     194.606995 &      -70.480301 &  12.53 &  0.894763 &  55978.4 &  1.439 &  13.629 &  2.944 &       UVES &  ZF20 \\
\bottomrule
\end{tabular}
    \label{tab:RV_indiv}
    }
\end{table}

\begin{table}[H]
\caption{Table of all individual rotational velocity values calculated in this work and compiled from literature  (first 10 rows). The full table (1480  $v~\mathrm{sin}~i$ values) is published online in the machine-readable format. The reference code in \textit{Ref.} column correspond to: ZF20: this work, SC12: \cite{Schlieder2012}, TO06: \cite{Torres2006}, MA14: \cite{Malo2014}, BA12: \cite{Bailey2012} and DE15: \cite{Desidera2015}.}
\centering
\tiny{
\begin{tabular}{lllrrl}
\toprule
 SIMBAD ID & RA J2000 (deg) & DEC J2000 (deg) &  vsini &  vsini err & Ref. \\
\midrule
 BD-202977 &     144.964005 &      -21.571400 &  13.49 &        1.5 &  ZF20 \\
 BD-202977 &     144.964005 &      -21.571400 &   9.92 &        1.5 &  ZF20 \\
 BD-202977 &     144.964005 &      -21.571400 &   9.92 &        1.5 &  ZF20 \\
   HD99827 &     171.324005 &      -84.954399 &  41.23 &        3.0 &  ZF20 \\
   HD99827 &     171.324005 &      -84.954399 &  40.22 &        3.0 &  ZF20 \\
   HD99827 &     171.324005 &      -84.954399 &  39.21 &        3.0 &  ZF20 \\
   HD99827 &     171.324005 &      -84.954399 &  41.23 &        3.0 &  ZF20 \\
   HD99827 &     171.324005 &      -84.954399 &  41.23 &        6.0 &  ZF20 \\
 CD-691055 &     194.606995 &      -70.480301 &  15.20 &        6.0 &  ZF20 \\
 CD-691055 &     194.606995 &      -70.480301 &  28.10 &        6.0 &  ZF20 \\
\bottomrule
\end{tabular}
    \label{tab:vsini_indiv}
    }
\end{table}

\begin{table}[H]
    \caption{Component radial velocity values for SB2 systems estimated in this work.}
    \centering
    \tiny{
\begin{tabular}{lllrrrrc}
\toprule
       SIMBAD ID &    RA J2000 (deg) &    DEC J2000 (deg) &      RV1 &   RV1 err &       RV2 &   RV2 err &     MJD \\
\midrule
  GSC08077-01788 &   72.970802 &  -46.791901 & -21.2927 &  1.618303 &  70.72530 &  0.694135 &  56735.1 \\
  GSC08077-01788 &   72.970802 &  -46.791901 & -15.4776 &  1.312729 &  65.67440 &  1.338559 &  56738.1 \\
        HD199058 &  313.588013 &    9.040000 & -30.0904 &  1.000865 & -11.77740 &  1.547928 &  56828.4 \\
        HD199058 &  313.588013 &    9.040000 & -24.9204 &  0.972239 & -13.04740 &  1.352694 &  56836.3 \\
        HD199058 &  313.588013 &    9.040000 & -24.5381 &  0.860092 & -15.96410 &  1.037438 &  57275.1 \\
        HD199058 &  313.588013 &    9.040000 & -25.0000 &  0.904851 & -13.90000 &  1.964319 &  54783.0 \\
         HD36329 &   82.350403 &  -34.515598 &  23.8599 &  0.979925 &  23.85990 &  0.904718 &  57271.4 \\
         HD36329 &   82.350403 &  -34.515598 & -44.8610 &  1.290879 &  90.64900 &  1.382289 &  57276.4 \\
         HD36329 &   82.350403 &  -34.515598 & -19.5175 &  1.143967 &  68.21940 &  1.918715 &  57295.3 \\
         HD51062 &  103.447998 &  -43.114201 &  14.6000 &  0.912295 &  38.90000 &  0.997951 &  55522.3 \\
         HD99827 &  171.324005 &  -84.954399 &   1.7000 &  1.269730 &  33.50000 &  1.642376 &  54169.2 \\
 UCAC3116-474938 &  299.011993 &  -32.121899 & -29.8203 &  1.363111 &  15.90560 &  1.102911 &  57255.3 \\
 UCAC3116-474938 &  299.011993 &  -32.121899 & -66.4405 &  1.185148 &  54.73050 &  1.044546 &  57272.1 \\
 UCAC3116-474938 &  299.011993 &  -32.121899 & -40.6756 &  1.003758 &  28.52240 &  1.520679 &  57275.1 \\
 UCAC3116-474938 &  299.011993 &  -32.121899 & -14.4882 &  1.280286 &   2.47079 &  0.857250 &  57292.2 \\
\bottomrule
\end{tabular}
}
    \label{tab:my_label}
\end{table}
}

{\centering
\begin{table}[h]
\caption{Summary table of the sample presented in this work. This table is available only in electronic format.}
\begin{center}
\begin{tabular}{lll}
\hline
\hline
Label & Units & Description \\
\hline
Simbad ID                           &                    & Simbad identifier        \\
RA J2000                            & degrees            & Right ascension at J2000     \\
DEC J2000                           & degrees            & Declination at J2000     \\
$\mathrm{RV}_{\mathrm{median}}$ CCF & km~s$^{-1}$ & Median RV from our CCF calculation    \\
$\sigma_{\mathrm{RV}}$ CCF & km~s$^{-1}$ & Standard deviation in RV from our CCF calculation             \\
$\mathrm{vsini}_{\mathrm{median}}$ CCF & km~s$^{-1}$ & Median $v~\mathrm{sin}~i$ from our CCF calculation \\
$\sigma_{v~\mathrm{sin}~i}$ CCF & km~s$^{-1}$ & Standard deviation in $v~\mathrm{sin}~i$ from our CCF calculation  \\
$\mathrm{N}_{\mathrm{obs}}$  CCF   &       & Number of observation from our CCF calculation      \\
$\mathrm{RV}_{\mathrm{median}}$   & km~s$^{-1}$ & Median RV from our work + literature \\
$\sigma_{\mathrm{RV}}$  & km~s$^{-1}$          & Standard deviation in RV from our work + literature    \\
$\mathrm{N}_{\mathrm{obs}}$ RV   &       & Number of RV observations from our work + literature  \\
$\mathrm{vsini}_{\mathrm{median}}$ & km~s$^{-1}$ & Median $v~\mathrm{sin}~i$ from our work + literature \\
$\sigma_{v~\mathrm{sin}~i}$ & km~s$^{-1}$ & Standard deviation in $v~\mathrm{sin}~i$ from our work + literature  \\
$\mathrm{N}_{\mathrm{obs}}$ $\mathrm{vsini}$   &       & Number of $\mathrm{vsini}$ observations from our work + literature  \\
Period                     & days    & Period from light curves                             \\
$\sigma_{\mathrm{Period}}$      & days    & Period uncertainty             \\
FAP                &     & False alarm probability            \\
Phased-MAD          &   & MAD on phased light curve \\
Residual-MAD        &   & MAD on residuals of phased light curve \\
P-MAD/R-MAD         &   &  Ratio between phased-MAD and residuals-MAD    \\
INSTR.                   &    &  Instrument that has measured the light curve \\
TESS sector           &   &  TESS sector       \\
TESS/K2 ID    &  & TESS or K2 identifier    \\
$\mathrm{N}_{\mathrm{sources}}$ TESS & & number of sources in TESS aperture with $\Delta Gmag < 5$  \\
$\mathrm{Min}_{\Delta Gmag}$ TESS & mag & Minimum $\Delta Gmag$ in TESS aperture  \\
LC notes           &   & Light curves notes on the object    \\
LC $\mathrm{q}_{\mathrm{flag}}$       &  & Light curve quality flag (Good, Caution or Bad)         \\
GaiaDR2 ID                &   & Gaia DR2 source identification        \\
mass    & $M_{\odot}$    & Stellar mass                 \\
Spt & & Spectral type \\
$\mathrm{SACY}_{MG}$              &                & Best MG match from SACY convergence method        \\
$\mathrm{SACY}_{P}$               &                & SACY membership probability    \\
$\mathrm{BAN}_{MG}$              &                & Best MG match from BANYAN$\Sigma$        \\
$\mathrm{BAN}_{P}$               &                & BANYAN$\Sigma$ membership probability    \\
Notes           &   &  Notes on SB candidates \\
\hline
\end{tabular}
\label{table:summary_tab}
\end{center}
\end{table}}

\end{appendix}

\end{document}